\documentclass[%
 aip,
 amsmath,amssymb,
 reprint,%
]{revtex4-1}

\usepackage{graphicx}
\usepackage{dcolumn}
\usepackage{bm}
\usepackage{adjustbox}
\usepackage[utf8]{inputenc}
\usepackage[T1]{fontenc}
\usepackage{mathptmx}
\usepackage{xcolor}
\usepackage{hyperref}
\hypersetup{
	colorlinks   = true,
	citecolor    = blue,
	linkcolor = blue,
    urlcolor = blue
}

\begin{document}

\preprint{AIP/123-QED}

\title{A first-principles Quantum Monte Carlo study of two-dimensional (2D) GaSe}

\author{Daniel Wines}
\affiliation{%
Department of Physics, University of Maryland Baltimore County, Baltimore MD 21250
}%

\author{Kayahan Saritas}%

\affiliation{ 
Department of Applied Physics, Yale University, New Haven CT 06520
}%

\author{Can Ataca}
 \email{ataca@umbc.edu}
\affiliation{%
Department of Physics, University of Maryland Baltimore County, Baltimore MD 21250
}%

\date{\today}

\begin{abstract}

Two-dimensional (2D) post-transition metal chalcogenides (PTMC) have attracted attention due to their suitable band gaps and lower exciton binding energies, making them more appropriate for electronic, optical and water-splitting devices than graphene and monolayer transition metal dichalcogenides (TMDs). Of the predicted 2D PTMCs, GaSe has been reliably synthesized and experimentally characterized. Despite this fact, quantities such as lattice parameters and band character vary significantly depending on which density functional theory (DFT) functional  is used. Although many-body perturbation theory (GW approximation) has been used to correct the electronic structure and obtain the excited state properties of 2D GaSe, and solving the Bethe-Salpeter equation (BSE) has been used to find the optical gap, we find that the results depend strongly on the starting wavefunction. In attempt to correct these discrepancies, we employed the many-body Diffusion Monte Carlo (DMC) method to calculate the ground and excited state properties of GaSe because DMC has a weaker dependence on the trial wavefunction. We benchmark these results with available experimental data, DFT [local-density approximation, Perdew-Burke-Ernzerhof
(PBE), strongly constrained and appropriately normed (SCAN) meta-GGA, and hybrid (HSE06) functionals] and GW-BSE (using PBE and SCAN wavefunctions) results. Our findings confirm monolayer GaSe is an indirect gap semiconductor ($\Gamma$-M) with a quasiparticle electronic gap in close agreement with experiment and low exciton binding energy. We also benchmark the optimal lattice parameter, cohesive energy and ground state charge density with DMC and various DFT methods. We aim to present a terminal theoretical benchmark for pristine monolayer GaSe, which will aide in the further study of 2D PTMCs using DMC methods. 
\end{abstract}

\maketitle
\section{\label{sec:intro}Introduction}

It has been reported that 2D post-transition metal chalcogenides (PTMCs), which have an MX stoichiometry (M is a group IIIA-IVA post transition metal atom, X is a chalcogen atom) can possess desirable properties such as strong second harmonic generation \cite{doi:10.1002/anie.201409837,gase-layered,gase-jacs}, unusual band renormalization effects \cite{NI201310,PhysRevB.84.085314,PhysRevB.87.195403,Pozo_Zamudio_2015}, and lower exciton binding energies than transition metal dichalcogenides (TMDs) \cite{NI201310,PhysRevB.84.085314,PhysRevB.87.195403}, which can additionally have applications for water-splitting \cite{water-splitting-mx}.
Since the synthesis and characterization of graphene \cite{Novoselov666,RevModPhys.81.109}, which lacks a native band gap, researchers have investigated the properties of other layered two-dimensional (2D) materials such as TMDs and PTMCs. Several experimental and theoretical studies have demonstrated that 2D TMDs such as MoS$_2$ or WS$_2$, which have an MX$_2$ stoichiometry, possess larger band gaps than their bulk counterparts and certain indirect to direct band gap transitions can occur with decreasing thickness, making them suitable candidates for optoelectronic devices \cite{mos2-photo,PhysRevLett.105.136805,mos2-nature,valley,tmddesign,nature-mose2}. Due to the fact that PTMCs possess a suitable band gap for photovoltaics and transistors \cite{NI201310,PhysRevB.84.085314,PhysRevB.87.195403,Pozo_Zamudio_2015,gase-atomic-layers,C3CP50233C,doi:10.1002/adma.201201361,bulk-gase,C8CP04723E,JAPPOR2018251}, excellent thermal transport \cite{doi:10.1063/1.5094663,C8CP04723E}, inherent flexibility \cite{PhysRevApplied.11.024012,PhysRevB.94.245407,doi:10.1142/S0217984915500499} and smaller exciton binding energies \cite{NI201310,PhysRevB.84.085314,PhysRevB.87.195403} make them viable substitutes for TMDs in device applications. In addition, it has been reported that applying strain \cite{PhysRevApplied.11.024012,C5NR08692B,doi:10.1142/S0217984915500499}, creating heterostructures \cite{Lie1501882,C8CP03740J,gate-mos2,JAPPOR2017109} and chemical functionalization \cite{o-func,C5CP00397K,pyradine} can effectively tune the electronic and optical properties of monolayer GaSe and it has been reported that GaSe can be used as a suitable substrate for other 2D materials \cite{D0CP00357C,PhysRevMaterials.2.104002}.

\begin{figure}
\begin{center}
\includegraphics[width=6.5cm]{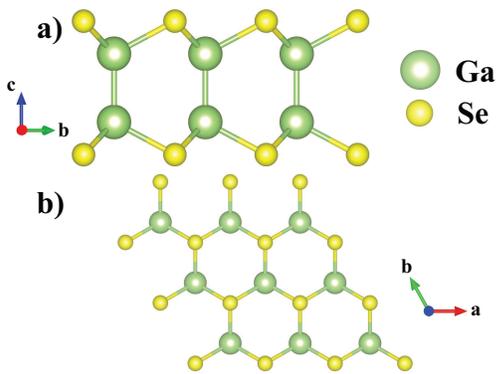}
\caption{The a) side and b) top view of monolayer GaSe. Green spheres represent Ga atoms and yellow spheres represent Se atoms.}
\label{structure}
\end{center}
\end{figure}

Although GaSe has been reliably synthesized and the lattice constant, quasiparticle gap and optical gap have been experimentally characterized \cite{PhysRevB.96.035407,doi:10.1063/1.4973918,GaSe-optical,C8NR01065J,GaSe-nature-synthesis,Rahaman_2018,Rahaman_2017,doi:10.1002/adma.201601184,doi:10.1063/1.5094663}, results obtained from different computational methods vary slightly for these values. Since most of the predicted PTMCs other than GaSe have not been synthesized, a careful benchmark made for GaSe can provide a pathway to better analyze the other predicted PTMCs that have not yet been synthesized. Due to quantum confinement of the c-direction (see Fig. \ref{structure}), monolayer GaSe possesses a much larger indirect quasiparticle band gap \cite{gase-atomic-layers,C3CP50233C,doi:10.1002/adma.201201361,bulk-gase,C8CP04723E} (compared to bulk GaSe which has a band gap of 2.0 eV), which has been measured to be 3.5 eV on top of a graphene substrate from angle-resolved photoemission spectroscopy (ARPES) \cite{PhysRevB.96.035407}. In addition, the optical band gap of GaSe has been measured to be 3.3 eV from cathodoluminescence (CL) \cite{GaSe-optical}, which implies that the exciton binding energy is smaller than 0.2 eV (much smaller than that of the 2D TMDs \cite{2dexciton}). In addition, the lattice constant of synthesized 2D GaSe has been measured to be $a=b=3.74$ \AA \cite{GaSe-nature-synthesis}. Despite these experimental findings, the theoretical predictions of the optimal lattice constant and electronic structure can vary significantly based on which density functional \cite{PhysRev.136.B864, PhysRevLett.77.3865,PhysRevLett.115.036402,doi:10.1063/1.1564060,doi:10.1063/1.2404663} is used. The largest discrepancy comes from the location of the conduction band edge at each high symmetry point. Since the energy difference between each high symmetry point is so slight ($\sim$ 0.2 - 0.3 eV), different functionals can predict the indirect gap to have different values, be at various locations in reciprocal space, and even incorrectly predict a direct gap for the material. In addition, when the electronic structure is corrected using many-body perturbation methods such as the GW approximation \cite{PhysRevB.34.5390,PhysRev.139.A796} and the Bethe-Salpeter equation (BSE) \cite{RevModPhys.74.601}, the results depend significantly on which functional is used to generate the starting wavefunction. These discrepancies are evident from our DFT and GW-BSE calculations (presented in the following section) and previous computational studies \cite{PhysRevB.87.195403,water-splitting-mx,C3CP50233C,PhysRevApplied.11.024012,doi:10.1063/1.5094663,C8CP04723E,PhysRevB.94.245407}.  

To obtain estimates for optimal lattice constant, cohesive energy and electronic and optical band gaps, we employed the Quantum Monte Carlo (QMC) method, using Variational Monte Carlo (VMC) and Diffusion Monte Carlo (DMC) for ground and excited state calculations. Although more computationally demanding than DFT, QMC has been shown to produce more accurate results for ground and excited state energies in condensed matter systems than DFT and GW  \cite{ataca_qmc,PhysRevLett.115.115501,bilayer-phos,doi:10.1063/1.5026120,PhysRevB.95.081301,PhysRevB.96.119902,PhysRevB.96.075431,PhysRevB.99.081118,PhysRevB.101.205115,PhysRevMaterials.3.124414,PhysRevLett.109.053001,nanopart,PhysRevX.9.011018,shin_krogel_kent_benali_heinonen_2020,PhysRevB.82.115108,doi:10.1063/1.4937421,doi:10.1142/S0217979203020533,doi:10.1063/1.4934262,PhysRevB.92.235209,PhysRevMaterials.2.085801,formation-qmc,PhysRevB.95.121108,Luo_2016,C6CP02067D,PhysRevLett.114.176401,PhysRevMaterials.1.065408,PhysRevB.95.075209,doi:10.1063/1.4919242,PhysRevMaterials.1.073603,PhysRevMaterials.2.075001,PhysRevB.98.155130,phosphors,LiNiO2,PhysRevMaterials.1.065408}. In addition to low dimensional studies involving nanoclusters \cite{PhysRevLett.109.053001} and nanoparticles \cite{nanopart}, there have been few studies involving the study of monolayer \cite{PhysRevB.95.081301,PhysRevB.96.119902,PhysRevB.96.075431,PhysRevX.9.011018,shin_krogel_kent_benali_heinonen_2020} and bilayer \cite{PhysRevLett.115.115501,bilayer-phos,doi:10.1063/1.5026120} materials. Specifically, the cohesive energies and band gaps have been accurately determined at the DMC level for monolayer phosphorene \cite{PhysRevX.9.011018} and monolayer GeSe \cite{shin_krogel_kent_benali_heinonen_2020}. By calculating the ground and excited state properties of 2D GaSe with DMC, we aim to prove that our method is a significant improvement over DFT and GW-BSE methods, relying weakly on the starting wavefunction and exchange-correlation functional. In demonstrating that this DMC method works well for 2D GaSe, we hope that this will influence other works in pursuing other PTMCs that have not been synthesized or characterized.

In section \ref{sec:methods} we outline our DMC approach and the different convergence criteria for our DMC, DFT, GW and BSE calculations. Section \ref{sec:geo} outlines our procedure of calculating the equation of state with DMC by applying in-plane biaxial strain and presents the results for the optimal lattice constant of GaSe. In section \ref{sec:cohesive} we calculate the cohesive energy with DMC and compare with results from various DFT functionals and van der Waals (vdW) corrections. We then calculate the electronic and optical band gaps for different electronic transitions using DMC and benchmark with DFT and GW-BSE results in section \ref{sec:gap}. In section \ref{sec:charge} we present the total ground state charge density of 2D GaSe calculated with DMC and DFT and contrast the results. Finally, we provide some concluding remarks and future perspectives in section \ref{sec:conclusion}.

\section{\label{sec:methods}Computational Methods}

The ground and excited state energies of monolayer GaSe were calculated using DFT, GW and DMC. Using these methods, we benchmarked cohesive energies, band gaps and quasiparticle energies. The DFT calculations were performed using the VASP code with projector augmented wave (PAW) potentials \cite{PhysRevB.54.11169,PhysRevB.59.1758}. Local-density approximation (LDA) \cite{PhysRev.136.B864}, Perdew-Burke-Ernzerhof (PBE) \cite{PhysRevLett.77.3865}, strongly constrained and appropriately normed (SCAN) meta-GGA \cite{PhysRevLett.115.036402}, and hybrid HSE06 \cite{doi:10.1063/1.1564060,doi:10.1063/1.2404663} functionals were used for benchmarking. A reciprocal grid of 6x6x1 was used in addition to a kinetic energy cutoff of 350 eV and at least 20 \AA\space of vacuum between periodic layers of GaSe in the \textit{c}-direction. The DFT calculations using VASP were mainly used for benchmarking. Using VASP for these several reference calculations is more advantageous because the PAW potentials have lower kinetic energy cutoff energies, which increases the computational efficiency. In addition, the DFT wavefunctions calculated in VASP can easily be stored and used in GW and BSE calculations.

We employed the GW method to obtain quasiparticle energies and BSE to obtain optical properties using the VASP code. We applied the GW method perturbatively using the G$_0$W$_0$ "single shot" method. This was used to obtain first order corrections to the Kohn-Sham eigenvalues and wavefunctions obtained from DFT, using the PBE and SCAN functionals. The GW and BSE calculations were performed using the the same reciprocal grid and kinetic energy cutoff as DFT. For the GW calculations the number of empty bands was converged using at least ten times the number of electrons in the simulation cell. For the BSE calculations, wavefunctions and quasiparticle energies from GW were used as an input and the Tamm-Dancoff approximation \cite{PhysRev.78.382} with 24 occupied and 24 unoccupied bands was used. 

The VMC and DMC \cite{RevModPhys.73.33,Needs_2009} calculations were carried out using the QMCPACK \cite{Kim_2018,doi:10.1063/5.0004860} code, where the DFT-VMC-DMC workflow was generated using the Nexus \cite{nexus} software suite. For all DMC calculations, the trial wavefunction was constructed from DFT using the PBE functional (except for the optimal geometry calculation, where trial wavefunctions created with PBE, LDA and SCAN were used for DMC and the results obtained with each method were compared). The Quantum Espresso (QE) \cite{Giannozzi_2009} DFT code was used to generate the single determinant wavefunction and our wavefunctions in DMC were of the Slater-Jastrow form \cite{PhysRev.34.1293,PhysRev.98.1479}. Terms up to three-body Jastrow correlation functions \cite{PhysRevB.70.235119} were included and were parameterized in terms of radial blip-splines for one-body and two-body functions and in terms of low-order polynomials for three-body functions. The Jastrow parameters were optimized with VMC variance and energy minimization respectively using the linear method \cite{PhysRevLett.98.110201}. The cost function of the energy minimization is split as 95 $\%$ energy minimization and 5 $\%$ variance minimization, which has been shown to improve the variance for DMC calculations \cite{PhysRevLett.94.150201}. The goal of optimizing the trial wavefunction in the Slater-Jastrow form is to achieve a more accurate ground state energy with smaller localization error \cite{doi:10.1063/1.460849} and reduced variance \cite{RevModPhys.73.33}. 

Our QMC simulations were performed at a minimum supercell size of 8 formula units (16 atoms) and a maximum supercell size of 36 formula units (72 atoms) for finite-size extrapolation. For calculations of quasiparticle and optical gap, the supercell sizes and shapes (tiling matrices) were advantageously chosen to contain all of the necessary high symmetry points ($\Gamma$, M, and K) of 2D GaSe. Jackknife fitting was used to obtain a linear fit of the calculated data and extrapolate to the infinite-size limit for cohesive energy and band gap. The locality approximation \cite{doi:10.1063/1.460849} was used to evaluate the nonlocal pseudopotentials in DMC. For Ga and Se we used energy-consistent Hartree-Fock Burkatzki-Filippi-Dolg pseudopotentials (BFD potentials). \cite{doi:10.1063/1.2741534,doi:10.1063/1.2987872} We also benchmarked a newly developed set of effective core potentials from correlated calculations (ccECP) \cite{doi:10.1063/1.5121006} at the DFT level along with BFD potentials. For pseudopotential validation and testing, refer to the discussion and Table S1 in the Supplementary Information (SI). For these pseudopotentials a kinetic energy cutoff of 120 Ry was used (see Fig. S1). After convergence testing, we decided to use a supercell reciprocal twist of 6x6x1 (Fig. S2) and a timestep of 0.02 Ha$^{-1}$ (Fig. S3) for all DMC calculations.

In DMC, we calculated the quasiparticle gap using: $\textrm{E}^{QP}=\textrm{E}_{N+1}+\textrm{E}_{N-1}-2\textrm{E}_N$, where N is the number of electrons in the neutral system, $\textrm{E}_N$ is the ground state energy of the neutral cell, and $\textrm{E}_{N+1}$ and $\textrm{E}_{N-1}$ are the ground state energies of the negatively and positively charged cells respectively. This is equivalent to the difference between the electron affinity and the ionization potential. An optical excitation is produced by annihilating an electron at the valence band maxima (VBM) and creating another at the conduction band minima (CBM). Excitations at specific high symmetry points between the valence and conduction bands are also considered at the DMC level in order to determine the lowest energy excitation which might not be predicted correctly by DFT. The optical gap, $\textrm{E}^{opt}$, is calculated by $\textrm{E}^{opt}=\textrm{E}^{ex}-\textrm{E}_N$, where $\textrm{E}^{ex}$ is the energy of the created excited state. For all optical and quasiparticle band gap calculations, the ground state trial wavefunction (single determinant PBE) was used and electrons were either added/removed (quasiparticle excitations) or swapped between the conduction band and valence band (optical excitations) at the DMC level. We calculated the quasiparticle and optical gaps for various transitions between $\Gamma$, M and K wavevectors in the first BZ. In order to avoid finite size effects, all supertwists used in the finite size extrapolation are optimized to simultaneously accommodate these three wavevectors ($\Gamma$, M and K). The workflow to get DMC band gaps was implemented in NEXUS, where the primitive cell was standardized using SPGLIB \cite{togo2018textttspglib} and the irreducible BZ path was obtained from using SeeK-PATH \cite{HINUMA2017140}. Therefore, the procedure is seamless and general for any material.

\section{\label{sec:results}Results and Discussion}

\subsection{\label{sec:geo}Optimal Geometry}

\begin{figure}
\begin{center}
\includegraphics[width=7.5cm]{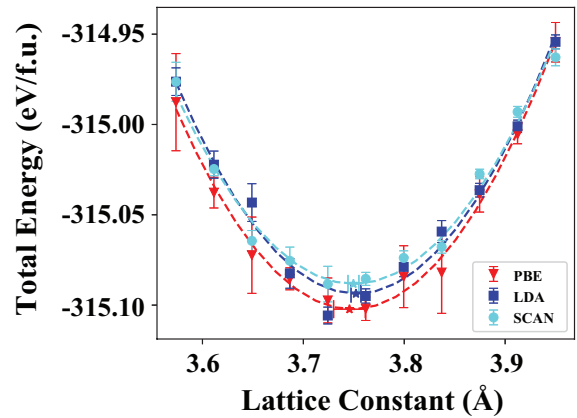}
\caption{The total energy per formula unit calculated with DMC with starting wavefunctions from PBE (red), LDA (blue), and SCAN (teal) versus the isotropically scaled lattice constant (in the \textit{x} and \textit{y}-direction). The dotted lines represent the fitted curves and the star represents the minimum energy point on the curve: $a=b=3.74(2)$ \AA\space for PBE, $a=b=3.75(1)$ \AA\space for LDA, $a=b=3.75(1)$ \AA\space for SCAN. }
\label{geo}
\end{center}
\end{figure}

Although the experimental in-plane lattice parameter of monolayer GaSe is well characterized (3.74 \AA \cite{GaSe-nature-synthesis}), further details of the monolayer GaSe geometry are not yet known. Therefore, computational methods can be used to obtain a relaxed geometry and benchmarked using the experimental lattice parameters. Although full geometry optimization (atomic coordinates and lattice constants) in DMC is possible, it is computationally rather demanding \cite{doi:10.1063/1.5040584,doi:10.1063/1.3516208}. On the other hand, various DFT functionals can provide a feasible way to obtain the geometry of GaSe. We went on to benchmark PBE, LDA and SCAN lattice parameters, which were 3.81, 3.71 and 3.76 \AA\space respectively. We found that SCAN is in better agreement with experiment, which is in accordance with recent reports that SCAN yields lattice constants closer to experiment for 2D materials \cite{scan-2d}.

Starting from the DFT-SCAN optimized geometry, we obtained the equation of state with DMC by applying various values of in-plane uniform biaxial strain, depicted in Fig. \ref{geo}. The biaxial strain is applied to a supercell with 36 atoms and using a 6x6x1 reciprocal grid to minimize finite size effects. The energy and lattice parameter values are then normalized for a single unit cell (per formula unit). The energy vs. lattice parameter values were then fitted with a quadratic equation (dotted lines in Fig. \ref{geo}). 

To assess whether the starting wavefunction had a significant impact on the total energy and optimized lattice constant, we performed the same DMC calculations using PBE, LDA and SCAN starting wavefunctions (starting from the DFT-SCAN optimized geometry and applying various in-plane uniform biaxial strain). This fit yields in an in-plane lattice parameter of 3.74(2) \AA\space for DMC-PBE, 3.75(1) \AA\space for DMC-LDA and 3.75(1) \AA\space for DMC-SCAN (the optimal lattice constants are marked with a star in the figure). These values are in very close agreement to the experimental value of $3.74$ \AA \cite{GaSe-nature-synthesis}. In addition, the total energies obtained by running DMC on top of each DFT method are within 15 meV of each other (see Fig. \ref{geo}). By recovering the same total energy and optimal lattice constant by running DMC with PBE, LDA and SCAN wavefunctions, we demonstrate that DMC has a weaker dependence on different flavors of the single determinant trial wavefunctions.

\subsection{\label{sec:cohesive}Cohesive and Interlayer Binding Energies}

An important fundamental quantity of 2D GaSe is the cohesive energy.  To our knowledge, the cohesive energy of monolayer and bulk GaSe has not been reported experimentally. The cohesive energy is such an important quantity because it can give us insight on how 2D GaSe is held together in nature. Bulk GaSe consists of quasi-2D layers weakly bound by vdW forces. Due to the limitations of DFT, this layer-layer interaction in bulk or few-layer materials is often times difficult to model and more sophisticated semi-empirical methods can be used \cite{PhysRevLett.115.115501,bilayer-phos,doi:10.1063/1.5026120}. Although vdW interactions are more prevalent in bulk materials, it has been reported that it is important to take vdW effects into account for monolayers with a larger thickness \cite{D0CP00357C}. Unlike graphene, 2D GaSe has a thickness of about 4.75 \AA. The quasi-2D structure shown in Fig. \ref{structure} consists of two sub-layers held together by Ga-Ga bonds parallel to the c-direction, where certain long range interactions are present due to the thickness. In addition to 2D GaSe, there are other 2D materials that possess these long range interactions, and DFT fails to model certain properties (such as phosphorene and GeSe \cite{shin_krogel_kent_benali_heinonen_2020,scan-2d,PhysRevX.9.011018,bilayer-phos}). Thus by calculating the cohesive energy of 2D GaSe with DMC, we can capture these complicated weak interactions and provide an intuitive theoretical benchmark that does not have any empirical correction or inherent dependence on a specific DFT functional.

\begin{figure}
\begin{center}
\includegraphics[width=7.5cm]{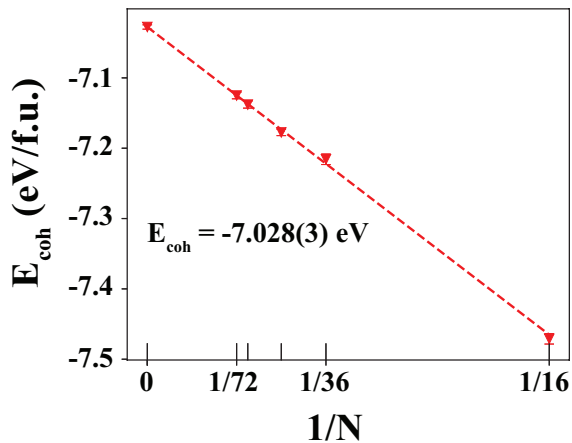}
\caption{The finite-size scaling of the DMC calculated cohesive energy (E$_{coh}$). DMC simulations were done at N$=16,36,48,64,72$ atoms in the simulation cell and the value as N $\rightarrow \infty$ was extrapolated from the calculated data.}
\label{coh}
\end{center}
\end{figure}

Similar to other 2D semiconducting materials \cite{PhysRevX.9.011018}, GaSe has moderately large cohesion. Cohesive energy of GaSe is defined as: $\textrm{E}_{coh}=\textrm{E}_{GaSe}-\textrm{E}_{Ga}-\textrm{E}_{Se}$ where $\textrm{E}_{GaSe}$ is the total energy of GaSe (per formula unit), and $\textrm{E}_{Ga}$ and $\textrm{E}_{Se}$ are the total energies of isolated Ga and Se atoms respectively. The DMC calculations of the single Ga and Se atoms were also performed using a single determinant trial wavefunction and BFD pseudopotentials. The timestep extrapolated total energies of each atom are given in Fig. S4. From this we obtain an energy of -2.034(1) eV for Ga and -9.280(1) eV for Se. This is in good agreement with previous accurate total atomic energy calculations \cite{atomcalcs}. Fig. \ref{coh} depicts the DMC finite-size scaling of the cohesive energy performed at supercell sizes of N$=16,36,48,64$ and $72$ atoms, where N is the number of atoms in the simulation cell. The cohesive energy per formula unit extrapolated from the infinite size limit (1/N$=0$, N $\rightarrow \infty$) is found to be -7.028(3) eV.

\begin{table}
\caption{\label{tab:tablecoh}%
The cohesive energy per formula unit (E$_{coh}$) and the interlayer binding energy (E$_b$) for 2D GaSe calculated with various DFT methods and DMC. Pseudopotential choice is indicated in parenthesis. }
\begin{tabular*}{0.48\textwidth}{l @{\extracolsep{\fill}} cc}
\hline
\hline
 Method   & E$_{coh}$ (eV/f.u.)  & E$_{b}$ (eV) \\
 \hline
 \hline
PBE (PAW)   & -6.636 & 0.004    \\
PBE+D2 (PAW)   & -7.086 & -0.087     \\
PBE (BFD)  & -7.096 & -0.026  \\
PBE+D2 (BFD)  & -7.552 & -0.111  \\
PBE (ccECP)  &  -6.965 & 0.007  \\
PBE+D2 (ccECP)  & -7.422 & -0.077   \\

SCAN (PAW)      & -7.429 & -0.025   \\
SCAN+rVV10 (PAW)      & -7.365 & -0.079   \\

SCAN (BFD)  &  & -0.026  \\
SCAN+rVV10 (BFD)  &  & -0.097  \\
SCAN (ccECP)  &  & -0.022   \\
SCAN+rVV10 (ccECP)  &  & -0.062   \\
HSE06 (PAW)  & -6.300 &     \\
DMC (BFD)    & -7.028(3) &   \\

\hline
\hline
\end{tabular*}
\label{table:coh}
\end{table}

Before moving on the the discussion about benchmarking DMC results with different DFT functionals and pseudopotentials, we wanted to stress that although these PAW potentials are meant to reproduce the all-electron results (as opposed to norm-conserving Hartree-Fock potentials that provide a many-body valence interaction), the total energy calculated with PAW cannot be directly compared to the total energy calculated with these Hartree-Fock potentials. Despite this fact, a meaningful comparison and benchmark can be made between PAW potential results and Hartree-Fock potential results for quantities that involve energy differences (cohesive energy, interlayer binding, band gap). 

To benchmark our DMC results, we calculated the cohesive energy using PBE (PAW, BFD, ccECP), SCAN (PAW) and HSE06 (PAW). In order to gain further insight of the weak interaction, we also benchmarked the cohesive energy using semi-empirical vdW corrections such as PBE+D2 \cite{doi:10.1002/jcc.20495} (method of Grimme) and SCAN+rVV10 \cite{PhysRevX.6.041005}. We also used these vdW corrected functionals to calculate the interlayer binding energy E$_b$ (which we define as the difference between the total energy of bulk GaSe and monolayer GaSe per formula unit) and compare to PBE and SCAN (see Table \ref{table:coh}).

As expected, the interlayer interaction calculated with regular PBE fails to correctly capture the binding in bulk GaSe. Although there is a slight discrepancy in the binding of PBE calculations (when comparing PAW, BFD and ccECP), the interlayer binding energies are in much better agreement with each other when SCAN and vdW corrections are added (see Table \ref{table:coh}) for each pseudopotential, with vdW corrected energies having stronger binding than regular SCAN. Due to the semi-empirical nature of these vdW corrections, it is possible that the long-range weaker interactions in the monolayer (due to thickness) can be overestimated, which is why our DMC results are so crucial since they have no semi-empirical correction and weaker dependence on the starting wavefunction. Despite our attempts, we were not able to converge isolated atom calculations with SCAN and SCAN+rvv10 using the BFD and ccECP potentials due to numerical instabilities associated with meta-GGA functionals. Hence, we do not provide cohesive energies for SCAN and SCAN+rvv10 in Table \ref{table:coh}. It is also important to note that the DFT and DMC results in Table \ref{table:coh} are calculated using a fixed geometry from SCAN using PAW potentials (same as in Section \ref{sec:geo}) but we provide the cohesive energies and interlayer binding energy obtained by relaxing the structure with PBE and PBE+D2 (Table S2), where there are minimal changes in the final results.

The fact that two fundamentally different vdW functionals (PBE+D2 and SCAN+rVV10) result in nearly identical interlayer binding energies (difference of 0.008 eV/f.u. for PAW, 0.014 for BFD and 0.015 for ccECP) gives us reason to believe that there should not be a huge discrepancy in interlayer binding energy between other DFT vdW functionals. For this reason we did not calculate the interlayer binding energy with DMC, but rather used our PBE+D2 and SCAN+rVV10 results as theoretical benchmarks to further explain our cohesive energy calculations with DMC. Since various vdW functionals are finding the interlayer binding energy of bulk GaSe to be low (on the order of 100 meV/f.u.), this can imply that each quasi-2D layer of GaSe is bonded together almost as strongly as the entire bulk material. This means that theoretically the cohesive energies of bulk and monolayer GaSe should be close in value and the interlayer interaction should account for a very small portion of the cohesive energy of bulk GaSe. As mentioned earlier, more complicated long range interactions can become important in thicker monolayers such as 2D GaSe, and DMC can recover these interactions without any semi-empirical corrections. Our DMC extrapolated cohesive energy is therefore important because it provides an important theoretical benchmark to aid future computational studies of 2D GaSe and other lesser studied 2D PTMCs.

\subsection{\label{sec:gap}Electronic and Optical Gaps}

In attempt to resolve any discrepancies in the electronic structure of 2D GaSe, we calculated the band gaps with DMC. Monolayer GaSe is an indirect semiconductor with a 3.5 eV quasiparticle gap (on graphene) \cite{PhysRevB.96.035407} and a 3.3 eV optical gap (on SiO$_x$/Si) \cite{GaSe-optical}. The results of these experiments indicate that the exciton binding energy (the difference of the quasiparticle and optical band gap) can be estimated as 0.2 eV, although the substrate of the material can slightly effect this value. We and others \cite{PhysRevB.87.195403,water-splitting-mx,C3CP50233C,PhysRevApplied.11.024012,doi:10.1063/1.5094663,C8CP04723E,PhysRevB.94.245407} have reported that the energy difference at the conduction band edge of each high symmetry point for the first BZ ($\Gamma$-M-K-$\Gamma$) is very small (on the order of 0.2 - 0.3 eV with respect to the $\Gamma$ point).

\begin{figure}
\begin{center}
\includegraphics[width=8.5cm]{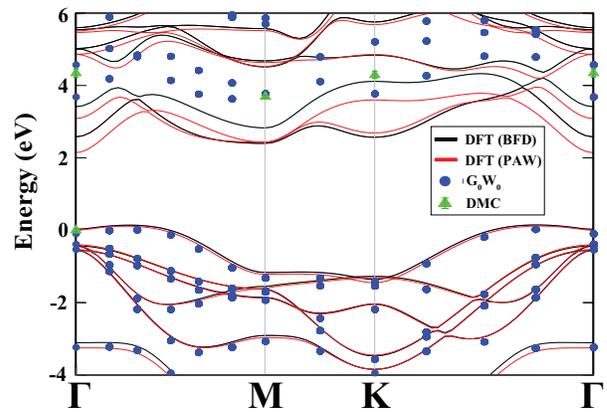}
\caption{The electronic band structure of monolayer GaSe calculated with PBE using BFD potentials (black), VASP PAW potentials (red), and G$_0$W$_0$ using PBE wavefunctions and VASP PAW potentials (blue). In addition the DMC excitation energies and associated error bars (with respect to the $\Gamma$ point) are given in green at each high symmetry point, with the error bars also given in green. }
\label{band}
\end{center}
\end{figure}

Since the energy differences at the conduction band edges are so small, this can result in incorrect predictions of the band gap location and the band gap value. This is observed in Fig. \ref{band}, which depicts the PBE calculated band structure of 2D GaSe calculated with BFD potentials (black) and PAW potentials (red) in addition to the quasiparticle band structure calculated with G$_0$W$_0$ using PAW potentials and PBE wavefunctions (blue) and the DMC excitation energies and error bars (green). This is apparent at the DFT level, where we observe that using the PAW potentials yield in a near-direct/direct band gap close to the $\Gamma$ point while using the BFD potentials yield in an indirect band gap from the $\Gamma$ to M point. It is important to note that the character of the conduction band is nearly identical at all points in the BZ except for around the $\Gamma$ point, which is resulting in the discrepancy in band gap prediction. This makes sense since because DFT is a ground state theory, any functional will modify the conduction band significantly. In addition, the quasiparticle band gap between $\Gamma$-$\Gamma$, $\Gamma$-M and $\Gamma$-K is almost indistinguishable at the G$_0$W$_0$ level (using PBE wavefunctions from PAW potentials). Due to the fact that the band diagram of 2D GaSe critically depends on the intricacies of the density functional used, it is crucial to use more accurate methodologies. To ensure this discrepancy cannot be resolved by soley examining DFT and GW results, we looked at the band decomposed charge densities (BDCD) of the conduction band at the $\Gamma$ point for PAW, BFD and ccECP calculations. We observe that the BDCD of each functional/pseudopotential is similar, having contributions from $p$-orbitals of Ga and Se. This also mandates a higher accuracy method to calculate the electronic structure of GaSe.

\begin{figure}
\begin{center}
\includegraphics[width=8.5cm]{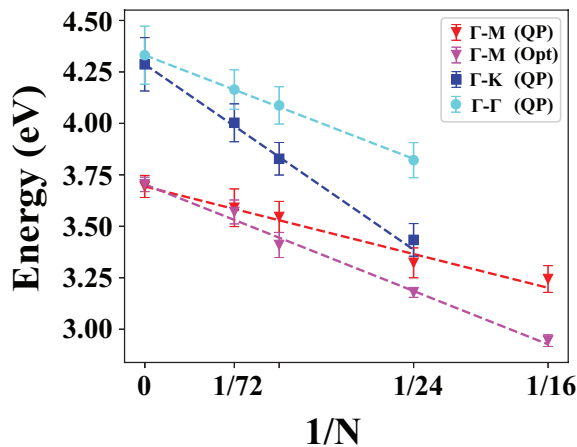}
\caption{The finite-size scaling of the quasiparticle (QP, fundamental) band gap for the $\Gamma$-M, $\Gamma$-K, and $\Gamma$-$\Gamma$ transitions (labeled appropriately) and the optical gap (Opt) for $\Gamma$-M. Simulations were performed at N $=72,48,24$ atoms for each transition (and N $=16$ for $\Gamma$-M) and the gaps at N $\rightarrow \infty$ were extrapolated from the calculated data. }
\label{gap}
\end{center}
\end{figure}

For freestanding GaSe we used the DMC method to calculate excited state energies at each transition of interest ($\Gamma$-$\Gamma$, $\Gamma$-M and $\Gamma$-K) due to the fact that there is weaker dependence on the starting wavefunction . We performed simulations at supercell sizes of 72, 48 and 24 atoms (and an additional simulation of 16 atoms for the $\Gamma$-M transition). Fig. \ref{gap} depicts the finite size scaling of the quasiparticle gap of each electronic transition. At the infinite size limit (N $\rightarrow \infty$), the extrapolated band gap values were determined to be: 3.69(5) eV for $\Gamma$-M, 4.34(14) eV for $\Gamma$-$\Gamma$ and 4.29(13) eV for $\Gamma$-K. 

These DMC results in Fig. \ref{gap} and tabulated energy differences at each high symmetry point in Table \ref{tab:table5} indicate that GaSe is in fact an indirect material with an indirect quasiparticle gap that ranges from the $\Gamma$ to M point and is in good agreement with experiment \cite{PhysRevB.96.035407}. The small difference between our calculated indirect gap value of 3.69(5) eV and the measured value of 3.5 eV from ARPES \cite{PhysRevB.96.035407} can be attributed to the change in dielectric environment of the synthesized GaSe on top of a graphene substrate. Our results resolve the previously mentioned discrepancy of DFT and GW calculations, where we saw competing energy minima at the conduction band at the $\Gamma$, M and K points. In order to fully benchmark these DMC results, we calculated the electronic gap of these transition using PBE (PAW, BFD, ccECP), SCAN (PAW, BFD, ccECP), and HSE06 (PAW) in addition to G$_0$W$_0$ (PAW and PBE/SCAN starting wavefunctions). As expected, PBE and SCAN both underestimate the band gaps significantly while HSE06 only slightly underestimates. A significant discrepancy arises when the electronic structure is calculated with PAW potentials vs. the BFD and ccECP potentials. With PAW potentials, the transition with the smallest energy is $\Gamma$-$\Gamma$, which means that the gap is near-direct/direct. This is also present in the G$_0$W$_0$ calculated quasiparticle band structure. Although this discrepancy is present in the GW results, the GW calculated quasiparticle gaps (using PBE and SCAN starting wavefunction) are closer to experiment and closer to our DMC results. In contrast, the transition with the smallest energy for the BFD and ccECP potentials is $\Gamma$-M, which means the gap is indirect. A summary of the DFT and GW benchmarked band gaps, in addition to the DMC calculated quasiparticle gaps, can be found in Table \ref{tab:table5}. Again it is important to note that all results in Table \ref{tab:table5} are calculated using a fixed geometry obtained from SCAN using PAW potentials (same as in Section \ref{sec:geo} and \ref{sec:cohesive}) but we provide the Kohn-Sham gaps obtained by relaxing the structure with PBE (Table S3), where as expected, there are minimal changes in the final results.

\begin{table}

\caption{\label{tab:table5}%
The Kohn-Sham electronic gap for 2D GaSe calculated with various DFT methods, quasiparticle gaps calculated with GW and DMC, and optical gaps calculated with BSE and DMC for the $\Gamma$-M, $\Gamma$-$\Gamma$ and $\Gamma$-K transitions in eV. Pseudopotential choice is indicated in parenthesis.}
\begin{tabular*}{0.48\textwidth}{l @{\extracolsep{\fill}} ccc}
\hline
\hline
  & $\Gamma$-M & $\Gamma$-$\Gamma$ & $\Gamma$-K \\
 \hline
 \hline
 \multicolumn{4}{l}{\textit{Kohn-Sham eigenstates}} \\
PBE (PAW)    & 2.430 & 2.143 & 2.687    \\
PBE (BFD)   & 2.373 & 2.555 & 2.541  \\
PBE (ccECP)   & 2.412 & 2.621 & 2.570   \\
SCAN (PAW)       & 2.700 & 2.531 & 2.902 \\
SCAN (BFD)   & 2.782 & 3.018 & 3.220  \\
SCAN (ccECP)   & 2.695 & 3.072 & 2.814  \\
HSE06 (PAW)   & 3.203 &  3.114 & 3.406    \\
\hline
\multicolumn{4}{l}{\textit{Quasiparticle gaps}} \\
G$_0$W$_0$-PBE (PAW)   & 3.893 & 3.805 & 3.892   \\
G$_0$W$_0$-SCAN (PAW)   & 4.054 & 3.139 & 4.541    \\
DMC (BFD)     & 3.69(5) & 4.34(14) & 4.29(13)   \\
\hline
\multicolumn{4}{l}{\textit{Optical gaps}} \\
BSE-PBE (PAW) &  & 3.42  & \\
BSE-SCAN (PAW) & & 3.21 & \\
 DMC (BFD) & 3.70(4) & 4.35(4) & 3.98(5)\\
\hline
\hline
\end{tabular*}
\label{table:bulk}

\end{table}

In addition to the quasiparticle band gap, we also calculated the optical band gap using DMC to be consistent with CL measurements \cite{GaSe-optical}. From our infinite-size extrapolated results (see Fig. \ref{gap}), we calculated the DMC optical gap for the $\Gamma$-M transition to be 3.70(4) eV. This is almost identical to the DMC quasiparticle gap for $\Gamma$-M since both quantities are identical within the uncertainty. This implies the upper bound on the exciton binding energy is 80 meV (when the exciton binding energy is defined as the difference between the indirect quasiparticle gap and the indirect optical gap at for the $\Gamma$-M transition). This is lower than the previously determined experimental value of 0.2 eV deduced from separate CL and ARPES measurements, which again can be attributed to the change in dielectric environment due to substrate effects for each sample. From a theoretical standpoint, this confirms the assertion that PTMCs such as GaSe have a lower exciton binding energy than TMDs (typical exciton binding energies of 2D TMDs are on the order of 0.5 - 1.0 eV) \cite{NI201310,PhysRevB.84.085314,PhysRevB.87.195403,2dexciton}, which can be advantageous for applications such as water-splitting \cite{water-splitting-mx,JAPPOR2018251}. Since the exciton binding of the $\Gamma$-M transition is low, we do not expect different results for the $\Gamma$-$\Gamma$ and $\Gamma$-K transitions. For these transitions we included the finite size extrapolation for the DMC optical gap and the respective DMC ground and excited state energies (Fig. S5 and S6 respectively) and from these results we calculated a $\Gamma$-$\Gamma$ optical gap of 4.35(4) eV and a $\Gamma$-K optical gap of 3.98(5) eV.

As an alternative method to calculate the optical band gap, we performed BSE calculations using GW as a starting point and calculated the frequency dependent dielectric function (see Fig. S7). From the first peak of the dielectric function, we obtain an optical gap of 3.42 eV for PBE wavefunctions and 3.21 eV for SCAN wavefunctions (both using PAW potentials). To further understand the excitonic properties obtained with the GW-BSE method, we also calculated the electron-hole coupling strength of the first bright exciton for PBE and SCAN wavefunctions and projected them on the GW calculated band structure (further details in the SI and Fig. S8). These results indicate that the strongest electron-hole coupling strength comes from the $\Gamma$-$\Gamma$ point transition. The PBE wavefunction used for GW-BSE simulations significantly overestimates the exciton binding of the $\Gamma$-$\Gamma$ transition for monolayer GaSe to be 0.39 eV when the optical gap is subtracted from the $\Gamma$-$\Gamma$ quasiparticle gap (see Table \ref{tab:table5}).  The SCAN wavefunction used for the GW-BSE simulations result in a negative exciton binding energy of $\sim 70$ meV, which is unphysical. 
Since there is such a disagreement in quantities calculated with different density functionals (even one instance resulting in an unphysical result), this gives justification for why DMC is advantageous for this system, closely matching the experiments and having weaker dependence on the starting wavefunction and functional. This conclusion is, however, only valid for single determinant methods, we expect that ground state DMC can be further reduced using multideterminant wavefunctions. Our band gap results obtained with various DFT methods, GW, BSE and DMC serve as a terminal theoretical benchmark to aid in further characterization of 2D GaSe and other predicted PTMCs.

\subsection{\label{sec:charge}Charge Density}

\begin{figure}
\begin{center}
\includegraphics[width=7.5cm]{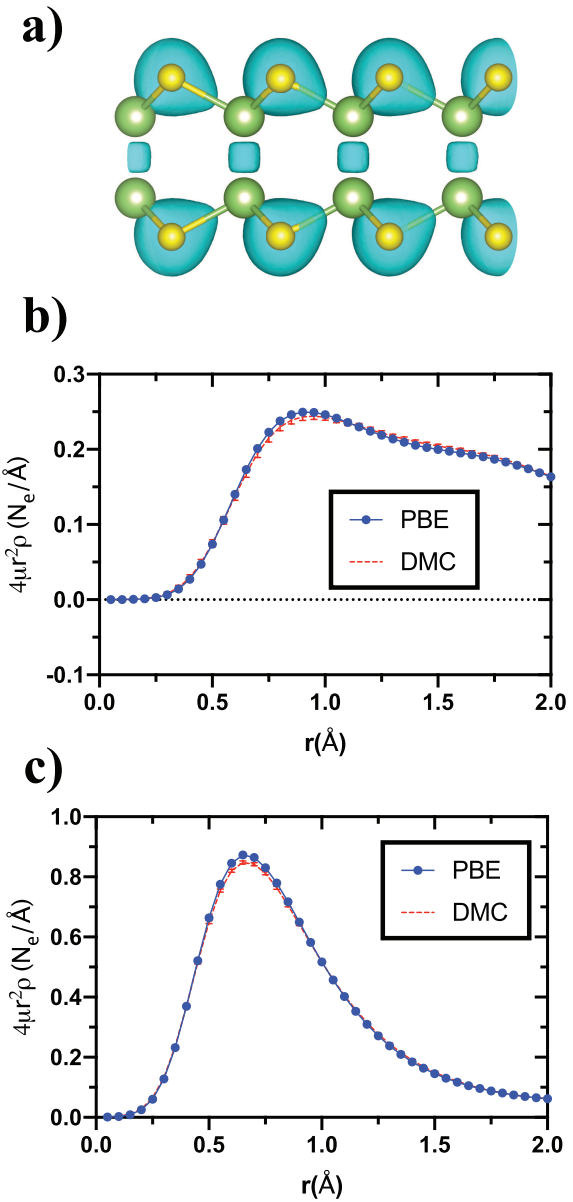}
\caption{a) The total ground state charge density of monolayer GaSe calculated with DMC using BFD pseudopotentials. The isosurface value is set to 0.05 e/\AA$^3$. Depicted are the radial charge densities of b) Ga and c) Se in 2D GaSe calculated with PBE (blue) and DMC (red). }
\label{charge}
\end{center}
\end{figure}

We additionally benchmarked the DFT and DMC calculated charge densities to understand if the significant discrepancies in the band gaps are related to a potential discrepancy in charge density. There has been various reported instances where DFT functionals and DMC predict charge densities with significant differences \cite{bilayer-phos,LiNiO2,PhysRevMaterials.1.065408}, but there have been other reported instances where these differences are quite small \cite{phosphors}. Although it is possible to analyze the contribution of the charge density at each point in the BZ at specific bands using DFT (band decomposed charge densities), in DMC we are restricted to analyzing the total charge density. We used an extrapolation scheme on the DMC charge densities to eliminate the bias that arises from using a mixed estimator. Since the charge density estimator does not commute with the Hamiltonian (at the fixed-node DMC level), the DMC charge density we calculated is a mixed estimator between the pure fixed-node DMC and VMC densities. These extrapolation formula can be used for such \cite{RevModPhys.73.33}:

\begin{equation} \label{rho1}
\rho_1 =2\rho_{\textrm{DMC}}-\rho_{\textrm{VMC}}+\mathcal{O}[(\Phi-\Psi_{\textrm{T}})^2]
\end{equation}
where $\rho_{\textrm{DMC}}$ and $\rho_{\textrm{VMC}}$ are respectively the DMC and VMC charge densities. Additionally $\Phi$ is the trial wavefunction from the DMC Hamiltonian and $\Psi_{\textrm{T}}$ is the trial wavefunction from VMC.

Fig. \ref{charge} depicts the total ground state charge density calculated with DMC while figure S9 depicts the DMC and DFT (PBE) calculated densities side by side, both using the BFD pseudopotentials. As seen in the figure, the total ground state density calculated with DFT and DMC are very similar. For both methods, we observe more density surrounding Se atoms and a small bit of density between Ga atoms of each quasi-2D layer. This charge distribution is due to the ionic character of the Ga-Se bonds in monolayer GaSe and the covalent Ga-Ga bonds, where Ga has only 3 valence electrons while Se has 6. From the isosurfaces we observe that the charge density around Se is slightly more spread out for PBE than DMC and there is more charge density around the Ga-Ga covalent bonds for PBE. To examine these densities quantitatively, we calculated the radial density around the Ga and Se atoms from our DFT and extrapolated DMC results (see Fig. \ref{charge}). From these results we can see that the charge density around Ga is nearly identical for PBE and DMC and around Se the density is slightly larger for PBE.

\section{\label{sec:conclusion}Conclusion}

In attempt to resolve the discrepancies in DFT, GW and BSE calculations that depend heavily on the density functional used, we have employed the DMC method to calculate the optimal lattice constant, cohesive energy, quasiparticle band gap and optical band gap, and total ground state charge density of 2D GaSe. Our DMC calculated optimal lattice constant (3.74(2) \AA\space for DMC-PBE, 3.75(1) \AA\space for DMC-LDA, and 3.75(1) \AA\space for DMC-SCAN) is in close agreement with the experimental value of 3.74 \AA. This also has demonstrated that DMC has a weaker dependence on the starting single determinant trial wavefunction. In attempt to understand our DMC calculated cohesive energy of -7.028(3) eV, we calculated the cohesive energy and interlayer binding energy with various DFT methods, including semi-empirical vdW corrections. Using a DFT benchmark, we find that the interlayer binding energy of the bulk structure should be small, but negative. Using DMC, we resolve the discrepancy of the conduction band edge energies in the electronic structure and confirm that monolayer GaSe is an indirect gap material ($\Gamma$-M) with a quasiparticle gap of 3.69(5) eV, which is in close agreement with the experimental quasiparticle gap of 3.5 eV (on graphene substrate). We find an upper bound of 80 meV for the exciton binding energy using DMC, which confirms that monolayer GaSe is a 2D material with low exciton binding energy, hence suitable for water-splitting applications. Finally, we calculated the total ground state charge density with DMC and found that it is in very close agreement with PBE. By presenting these DMC, DFT, GW and BSE results, we present a terminal benchmark for pristine 2D GaSe. We hope that these results will aide in the future investigation of other 2D PTMCs and other 2D materials using theoretical methods. 

\section*{Supplementary Material}

See the supplementary information for additional details and convergence tests of DFT and DMC calculations and additional supporting electronic and optical property results using DFT, GW-BSE and DMC.

\begin{acknowledgments}
This work was supported by the National Science Foundation through the Division of Materials Research under NSF DMR-1726213.
\end{acknowledgments}

\section*{Data Availability}

The data that support the findings of this study are available from the corresponding author upon reasonable request.

\section*{References}
\nocite{*}
\bibliography{Main}

\providecommand{\noopsort}[1]{}\providecommand{\singleletter}[1]{#1}%
\begin{thebibliography}{112}%
\makeatletter
\providecommand \@ifxundefined [1]{%
 \@ifx{#1\undefined}
}%
\providecommand \@ifnum [1]{%
 \ifnum #1\expandafter \@firstoftwo
 \else \expandafter \@secondoftwo
 \fi
}%
\providecommand \@ifx [1]{%
 \ifx #1\expandafter \@firstoftwo
 \else \expandafter \@secondoftwo
 \fi
}%
\providecommand \natexlab [1]{#1}%
\providecommand \enquote  [1]{``#1''}%
\providecommand \bibnamefont  [1]{#1}%
\providecommand \bibfnamefont [1]{#1}%
\providecommand \citenamefont [1]{#1}%
\providecommand \href@noop [0]{\@secondoftwo}%
\providecommand \href [0]{\begingroup \@sanitize@url \@href}%
\providecommand \@href[1]{\@@startlink{#1}\@@href}%
\providecommand \@@href[1]{\endgroup#1\@@endlink}%
\providecommand \@sanitize@url [0]{\catcode `\\12\catcode `\$12\catcode
  `\&12\catcode `\#12\catcode `\^12\catcode `\_12\catcode `\%12\relax}%
\providecommand \@@startlink[1]{}%
\providecommand \@@endlink[0]{}%
\providecommand \url  [0]{\begingroup\@sanitize@url \@url }%
\providecommand \@url [1]{\endgroup\@href {#1}{\urlprefix }}%
\providecommand \urlprefix  [0]{URL }%
\providecommand \Eprint [0]{\href }%
\providecommand \doibase [0]{http://dx.doi.org/}%
\providecommand \selectlanguage [0]{\@gobble}%
\providecommand \bibinfo  [0]{\@secondoftwo}%
\providecommand \bibfield  [0]{\@secondoftwo}%
\providecommand \translation [1]{[#1]}%
\providecommand \BibitemOpen [0]{}%
\providecommand \bibitemStop [0]{}%
\providecommand \bibitemNoStop [0]{.\EOS\space}%
\providecommand \EOS [0]{\spacefactor3000\relax}%
\providecommand \BibitemShut  [1]{\csname bibitem#1\endcsname}%
\let\auto@bib@innerbib\@empty
\bibitem [{\citenamefont {Jie}\ \emph {et~al.}(2015)\citenamefont {Jie},
  \citenamefont {Chen}, \citenamefont {Li}, \citenamefont {Xie}, \citenamefont
  {Hui}, \citenamefont {Lau}, \citenamefont {Cui},\ and\ \citenamefont
  {Hao}}]{doi:10.1002/anie.201409837}%
  \BibitemOpen
  \bibfield  {author} {\bibinfo {author} {\bibfnamefont {W.}~\bibnamefont
  {Jie}}, \bibinfo {author} {\bibfnamefont {X.}~\bibnamefont {Chen}}, \bibinfo
  {author} {\bibfnamefont {D.}~\bibnamefont {Li}}, \bibinfo {author}
  {\bibfnamefont {L.}~\bibnamefont {Xie}}, \bibinfo {author} {\bibfnamefont
  {Y.~Y.}\ \bibnamefont {Hui}}, \bibinfo {author} {\bibfnamefont {S.~P.}\
  \bibnamefont {Lau}}, \bibinfo {author} {\bibfnamefont {X.}~\bibnamefont
  {Cui}}, \ and\ \bibinfo {author} {\bibfnamefont {J.}~\bibnamefont {Hao}},\
  }\bibfield  {title} {\enquote {\bibinfo {title} {Layer-dependent nonlinear
  optical properties and stability of non-centrosymmetric modification in
  few-layer {GaSe} sheets},}\ }\href {\doibase 10.1002/anie.201409837}
  {\bibfield  {journal} {\bibinfo  {journal} {Angewandte Chemie International
  Edition}\ }\textbf {\bibinfo {volume} {54}},\ \bibinfo {pages} {1185--1189}
  (\bibinfo {year} {2015})}\BibitemShut {NoStop}%
\bibitem [{\citenamefont {Karvonen}\ \emph {et~al.}(2015)\citenamefont
  {Karvonen}, \citenamefont {S{\"a}yn{\"a}tjoki}, \citenamefont {Mehravar},
  \citenamefont {Rodriguez}, \citenamefont {Hartmann}, \citenamefont {Zahn},
  \citenamefont {Honkanen}, \citenamefont {Norwood}, \citenamefont
  {Peyghambarian}, \citenamefont {Kieu}, \citenamefont {Lipsanen},\ and\
  \citenamefont {Riikonen}}]{gase-layered}%
  \BibitemOpen
  \bibfield  {author} {\bibinfo {author} {\bibfnamefont {L.}~\bibnamefont
  {Karvonen}}, \bibinfo {author} {\bibfnamefont {A.}~\bibnamefont
  {S{\"a}yn{\"a}tjoki}}, \bibinfo {author} {\bibfnamefont {S.}~\bibnamefont
  {Mehravar}}, \bibinfo {author} {\bibfnamefont {R.~D.}\ \bibnamefont
  {Rodriguez}}, \bibinfo {author} {\bibfnamefont {S.}~\bibnamefont {Hartmann}},
  \bibinfo {author} {\bibfnamefont {D.~R.~T.}\ \bibnamefont {Zahn}}, \bibinfo
  {author} {\bibfnamefont {S.}~\bibnamefont {Honkanen}}, \bibinfo {author}
  {\bibfnamefont {R.~A.}\ \bibnamefont {Norwood}}, \bibinfo {author}
  {\bibfnamefont {N.}~\bibnamefont {Peyghambarian}}, \bibinfo {author}
  {\bibfnamefont {K.}~\bibnamefont {Kieu}}, \bibinfo {author} {\bibfnamefont
  {H.}~\bibnamefont {Lipsanen}}, \ and\ \bibinfo {author} {\bibfnamefont
  {J.}~\bibnamefont {Riikonen}},\ }\bibfield  {title} {\enquote {\bibinfo
  {title} {Investigation of second- and third-harmonic generation in few-layer
  gallium selenide by multiphoton microscopy},}\ }\href {\doibase
  10.1038/srep10334} {\bibfield  {journal} {\bibinfo  {journal} {Scientific
  Reports}\ }\textbf {\bibinfo {volume} {5}},\ \bibinfo {pages} {10334}
  (\bibinfo {year} {2015})}\BibitemShut {NoStop}%
\bibitem [{\citenamefont {Zhou}\ \emph {et~al.}(2015)\citenamefont {Zhou},
  \citenamefont {Cheng}, \citenamefont {Zhou}, \citenamefont {Cao},
  \citenamefont {Hong}, \citenamefont {Liao}, \citenamefont {Wu}, \citenamefont
  {Peng}, \citenamefont {Liu},\ and\ \citenamefont {Yu}}]{gase-jacs}%
  \BibitemOpen
  \bibfield  {author} {\bibinfo {author} {\bibfnamefont {X.}~\bibnamefont
  {Zhou}}, \bibinfo {author} {\bibfnamefont {J.}~\bibnamefont {Cheng}},
  \bibinfo {author} {\bibfnamefont {Y.}~\bibnamefont {Zhou}}, \bibinfo {author}
  {\bibfnamefont {T.}~\bibnamefont {Cao}}, \bibinfo {author} {\bibfnamefont
  {H.}~\bibnamefont {Hong}}, \bibinfo {author} {\bibfnamefont {Z.}~\bibnamefont
  {Liao}}, \bibinfo {author} {\bibfnamefont {S.}~\bibnamefont {Wu}}, \bibinfo
  {author} {\bibfnamefont {H.}~\bibnamefont {Peng}}, \bibinfo {author}
  {\bibfnamefont {K.}~\bibnamefont {Liu}}, \ and\ \bibinfo {author}
  {\bibfnamefont {D.}~\bibnamefont {Yu}},\ }\bibfield  {title} {\enquote
  {\bibinfo {title} {Strong second-harmonic generation in atomic layered
  {GaSe}},}\ }\href {\doibase 10.1021/jacs.5b04305} {\bibfield  {journal}
  {\bibinfo  {journal} {Journal of the American Chemical Society}\ }\textbf
  {\bibinfo {volume} {137}},\ \bibinfo {pages} {7994--7997} (\bibinfo {year}
  {2015})}\BibitemShut {NoStop}%
\bibitem [{\citenamefont {Ni}\ \emph {et~al.}(2013)\citenamefont {Ni},
  \citenamefont {Wu}, \citenamefont {Huang}, \citenamefont {Mao}, \citenamefont
  {Wang},\ and\ \citenamefont {Cheng}}]{NI201310}%
  \BibitemOpen
  \bibfield  {author} {\bibinfo {author} {\bibfnamefont {Y.}~\bibnamefont
  {Ni}}, \bibinfo {author} {\bibfnamefont {H.}~\bibnamefont {Wu}}, \bibinfo
  {author} {\bibfnamefont {C.}~\bibnamefont {Huang}}, \bibinfo {author}
  {\bibfnamefont {M.}~\bibnamefont {Mao}}, \bibinfo {author} {\bibfnamefont
  {Z.}~\bibnamefont {Wang}}, \ and\ \bibinfo {author} {\bibfnamefont
  {X.}~\bibnamefont {Cheng}},\ }\bibfield  {title} {\enquote {\bibinfo {title}
  {{Growth and quality of gallium selenide (GaSe) crystals}},}\ }\href
  {\doibase https://doi.org/10.1016/j.jcrysgro.2013.06.030} {\bibfield
  {journal} {\bibinfo  {journal} {Journal of Crystal Growth}\ }\textbf
  {\bibinfo {volume} {381}},\ \bibinfo {pages} {10 -- 14} (\bibinfo {year}
  {2013})}\BibitemShut {NoStop}%
\bibitem [{\citenamefont {Rybkovskiy}\ \emph {et~al.}(2011)\citenamefont
  {Rybkovskiy}, \citenamefont {Arutyunyan}, \citenamefont {Orekhov},
  \citenamefont {Gromchenko}, \citenamefont {Vorobiev}, \citenamefont
  {Osadchy}, \citenamefont {Salaev}, \citenamefont {Baykara}, \citenamefont
  {Allakhverdiev},\ and\ \citenamefont {Obraztsova}}]{PhysRevB.84.085314}%
  \BibitemOpen
  \bibfield  {author} {\bibinfo {author} {\bibfnamefont {D.~V.}\ \bibnamefont
  {Rybkovskiy}}, \bibinfo {author} {\bibfnamefont {N.~R.}\ \bibnamefont
  {Arutyunyan}}, \bibinfo {author} {\bibfnamefont {A.~S.}\ \bibnamefont
  {Orekhov}}, \bibinfo {author} {\bibfnamefont {I.~A.}\ \bibnamefont
  {Gromchenko}}, \bibinfo {author} {\bibfnamefont {I.~V.}\ \bibnamefont
  {Vorobiev}}, \bibinfo {author} {\bibfnamefont {A.~V.}\ \bibnamefont
  {Osadchy}}, \bibinfo {author} {\bibfnamefont {E.~Y.}\ \bibnamefont {Salaev}},
  \bibinfo {author} {\bibfnamefont {T.~K.}\ \bibnamefont {Baykara}}, \bibinfo
  {author} {\bibfnamefont {K.~R.}\ \bibnamefont {Allakhverdiev}}, \ and\
  \bibinfo {author} {\bibfnamefont {E.~D.}\ \bibnamefont {Obraztsova}},\
  }\bibfield  {title} {\enquote {\bibinfo {title} {{Size-induced effects in
  gallium selenide electronic structure: The influence of interlayer
  interactions}},}\ }\href {\doibase 10.1103/PhysRevB.84.085314} {\bibfield
  {journal} {\bibinfo  {journal} {Phys. Rev. B}\ }\textbf {\bibinfo {volume}
  {84}},\ \bibinfo {pages} {085314} (\bibinfo {year} {2011})}\BibitemShut
  {NoStop}%
\bibitem [{\citenamefont {Z\'olyomi}, \citenamefont {Drummond},\ and\
  \citenamefont {Fal'ko}(2013)}]{PhysRevB.87.195403}%
  \BibitemOpen
  \bibfield  {author} {\bibinfo {author} {\bibfnamefont {V.}~\bibnamefont
  {Z\'olyomi}}, \bibinfo {author} {\bibfnamefont {N.~D.}\ \bibnamefont
  {Drummond}}, \ and\ \bibinfo {author} {\bibfnamefont {V.~I.}\ \bibnamefont
  {Fal'ko}},\ }\bibfield  {title} {\enquote {\bibinfo {title} {{Band structure
  and optical transitions in atomic layers of hexagonal gallium
  chalcogenides}},}\ }\href {\doibase 10.1103/PhysRevB.87.195403} {\bibfield
  {journal} {\bibinfo  {journal} {Phys. Rev. B}\ }\textbf {\bibinfo {volume}
  {87}},\ \bibinfo {pages} {195403} (\bibinfo {year} {2013})}\BibitemShut
  {NoStop}%
\bibitem [{\citenamefont {Pozo-Zamudio}\ \emph {et~al.}(2015)\citenamefont
  {Pozo-Zamudio}, \citenamefont {Schwarz}, \citenamefont {Sich}, \citenamefont
  {Akimov}, \citenamefont {Bayer}, \citenamefont {Schofield}, \citenamefont
  {Chekhovich}, \citenamefont {Robinson}, \citenamefont {Kay}, \citenamefont
  {Kolosov}, \citenamefont {Dmitriev}, \citenamefont {Lashkarev}, \citenamefont
  {Borisenko}, \citenamefont {Kolesnikov},\ and\ \citenamefont
  {Tartakovskii}}]{Pozo_Zamudio_2015}%
  \BibitemOpen
  \bibfield  {author} {\bibinfo {author} {\bibfnamefont {O.~D.}\ \bibnamefont
  {Pozo-Zamudio}}, \bibinfo {author} {\bibfnamefont {S.}~\bibnamefont
  {Schwarz}}, \bibinfo {author} {\bibfnamefont {M.}~\bibnamefont {Sich}},
  \bibinfo {author} {\bibfnamefont {I.~A.}\ \bibnamefont {Akimov}}, \bibinfo
  {author} {\bibfnamefont {M.}~\bibnamefont {Bayer}}, \bibinfo {author}
  {\bibfnamefont {R.~C.}\ \bibnamefont {Schofield}}, \bibinfo {author}
  {\bibfnamefont {E.~A.}\ \bibnamefont {Chekhovich}}, \bibinfo {author}
  {\bibfnamefont {B.~J.}\ \bibnamefont {Robinson}}, \bibinfo {author}
  {\bibfnamefont {N.~D.}\ \bibnamefont {Kay}}, \bibinfo {author} {\bibfnamefont
  {O.~V.}\ \bibnamefont {Kolosov}}, \bibinfo {author} {\bibfnamefont {A.~I.}\
  \bibnamefont {Dmitriev}}, \bibinfo {author} {\bibfnamefont {G.~V.}\
  \bibnamefont {Lashkarev}}, \bibinfo {author} {\bibfnamefont {D.~N.}\
  \bibnamefont {Borisenko}}, \bibinfo {author} {\bibfnamefont {N.~N.}\
  \bibnamefont {Kolesnikov}}, \ and\ \bibinfo {author} {\bibfnamefont {A.~I.}\
  \bibnamefont {Tartakovskii}},\ }\bibfield  {title} {\enquote {\bibinfo
  {title} {{Photoluminescence of two-dimensional {GaTe} and {GaSe} films}},}\
  }\href {\doibase 10.1088/2053-1583/2/3/035010} {\bibfield  {journal}
  {\bibinfo  {journal} {2D Materials}\ }\textbf {\bibinfo {volume} {2}},\
  \bibinfo {pages} {035010} (\bibinfo {year} {2015})}\BibitemShut {NoStop}%
\bibitem [{\citenamefont {Zhuang}\ and\ \citenamefont
  {Hennig}(2013)}]{water-splitting-mx}%
  \BibitemOpen
  \bibfield  {author} {\bibinfo {author} {\bibfnamefont {H.~L.}\ \bibnamefont
  {Zhuang}}\ and\ \bibinfo {author} {\bibfnamefont {R.~G.}\ \bibnamefont
  {Hennig}},\ }\bibfield  {title} {\enquote {\bibinfo {title} {{Single-Layer
  Group-III Monochalcogenide Photocatalysts for Water Splitting}},}\ }\href
  {\doibase 10.1021/cm401661x} {\bibfield  {journal} {\bibinfo  {journal}
  {Chemistry of Materials}\ }\textbf {\bibinfo {volume} {25}},\ \bibinfo
  {pages} {3232--3238} (\bibinfo {year} {2013})}\BibitemShut {NoStop}%
\bibitem [{\citenamefont {Novoselov}\ \emph {et~al.}(2004)\citenamefont
  {Novoselov}, \citenamefont {Geim}, \citenamefont {Morozov}, \citenamefont
  {Jiang}, \citenamefont {Zhang}, \citenamefont {Dubonos}, \citenamefont
  {Grigorieva},\ and\ \citenamefont {Firsov}}]{Novoselov666}%
  \BibitemOpen
  \bibfield  {author} {\bibinfo {author} {\bibfnamefont {K.~S.}\ \bibnamefont
  {Novoselov}}, \bibinfo {author} {\bibfnamefont {A.~K.}\ \bibnamefont {Geim}},
  \bibinfo {author} {\bibfnamefont {S.~V.}\ \bibnamefont {Morozov}}, \bibinfo
  {author} {\bibfnamefont {D.}~\bibnamefont {Jiang}}, \bibinfo {author}
  {\bibfnamefont {Y.}~\bibnamefont {Zhang}}, \bibinfo {author} {\bibfnamefont
  {S.~V.}\ \bibnamefont {Dubonos}}, \bibinfo {author} {\bibfnamefont {I.~V.}\
  \bibnamefont {Grigorieva}}, \ and\ \bibinfo {author} {\bibfnamefont {A.~A.}\
  \bibnamefont {Firsov}},\ }\bibfield  {title} {\enquote {\bibinfo {title}
  {Electric field effect in atomically thin carbon films},}\ }\href {\doibase
  10.1126/science.1102896} {\bibfield  {journal} {\bibinfo  {journal}
  {Science}\ }\textbf {\bibinfo {volume} {306}},\ \bibinfo {pages} {666--669}
  (\bibinfo {year} {2004})}\BibitemShut {NoStop}%
\bibitem [{\citenamefont {Castro~Neto}\ \emph {et~al.}(2009)\citenamefont
  {Castro~Neto}, \citenamefont {Guinea}, \citenamefont {Peres}, \citenamefont
  {Novoselov},\ and\ \citenamefont {Geim}}]{RevModPhys.81.109}%
  \BibitemOpen
  \bibfield  {author} {\bibinfo {author} {\bibfnamefont {A.~H.}\ \bibnamefont
  {Castro~Neto}}, \bibinfo {author} {\bibfnamefont {F.}~\bibnamefont {Guinea}},
  \bibinfo {author} {\bibfnamefont {N.~M.~R.}\ \bibnamefont {Peres}}, \bibinfo
  {author} {\bibfnamefont {K.~S.}\ \bibnamefont {Novoselov}}, \ and\ \bibinfo
  {author} {\bibfnamefont {A.~K.}\ \bibnamefont {Geim}},\ }\bibfield  {title}
  {\enquote {\bibinfo {title} {{The electronic properties of graphene}},}\
  }\href {\doibase 10.1103/RevModPhys.81.109} {\bibfield  {journal} {\bibinfo
  {journal} {Rev. Mod. Phys.}\ }\textbf {\bibinfo {volume} {81}},\ \bibinfo
  {pages} {109--162} (\bibinfo {year} {2009})}\BibitemShut {NoStop}%
\bibitem [{\citenamefont {Splendiani}\ \emph {et~al.}(2010)\citenamefont
  {Splendiani}, \citenamefont {Sun}, \citenamefont {Zhang}, \citenamefont {Li},
  \citenamefont {Kim}, \citenamefont {Chim}, \citenamefont {Galli},\ and\
  \citenamefont {Wang}}]{mos2-photo}%
  \BibitemOpen
  \bibfield  {author} {\bibinfo {author} {\bibfnamefont {A.}~\bibnamefont
  {Splendiani}}, \bibinfo {author} {\bibfnamefont {L.}~\bibnamefont {Sun}},
  \bibinfo {author} {\bibfnamefont {Y.}~\bibnamefont {Zhang}}, \bibinfo
  {author} {\bibfnamefont {T.}~\bibnamefont {Li}}, \bibinfo {author}
  {\bibfnamefont {J.}~\bibnamefont {Kim}}, \bibinfo {author} {\bibfnamefont
  {C.-Y.}\ \bibnamefont {Chim}}, \bibinfo {author} {\bibfnamefont
  {G.}~\bibnamefont {Galli}}, \ and\ \bibinfo {author} {\bibfnamefont
  {F.}~\bibnamefont {Wang}},\ }\bibfield  {title} {\enquote {\bibinfo {title}
  {Emerging photoluminescence in monolayer {MoS$_2$}},}\ }\href {\doibase
  10.1021/nl903868w} {\bibfield  {journal} {\bibinfo  {journal} {Nano Letters}\
  }\textbf {\bibinfo {volume} {10}},\ \bibinfo {pages} {1271--1275} (\bibinfo
  {year} {2010})}\BibitemShut {NoStop}%
\bibitem [{\citenamefont {Mak}\ \emph {et~al.}(2010)\citenamefont {Mak},
  \citenamefont {Lee}, \citenamefont {Hone}, \citenamefont {Shan},\ and\
  \citenamefont {Heinz}}]{PhysRevLett.105.136805}%
  \BibitemOpen
  \bibfield  {author} {\bibinfo {author} {\bibfnamefont {K.~F.}\ \bibnamefont
  {Mak}}, \bibinfo {author} {\bibfnamefont {C.}~\bibnamefont {Lee}}, \bibinfo
  {author} {\bibfnamefont {J.}~\bibnamefont {Hone}}, \bibinfo {author}
  {\bibfnamefont {J.}~\bibnamefont {Shan}}, \ and\ \bibinfo {author}
  {\bibfnamefont {T.~F.}\ \bibnamefont {Heinz}},\ }\bibfield  {title} {\enquote
  {\bibinfo {title} {Atomically thin {${\mathrm{MoS}}_{2}$}: A new direct-gap
  semiconductor},}\ }\href {\doibase 10.1103/PhysRevLett.105.136805} {\bibfield
   {journal} {\bibinfo  {journal} {Phys. Rev. Lett.}\ }\textbf {\bibinfo
  {volume} {105}},\ \bibinfo {pages} {136805} (\bibinfo {year}
  {2010})}\BibitemShut {NoStop}%
\bibitem [{\citenamefont {Radisavljevic}\ \emph {et~al.}(2011)\citenamefont
  {Radisavljevic}, \citenamefont {Radenovic}, \citenamefont {Brivio},
  \citenamefont {Giacometti},\ and\ \citenamefont {Kis}}]{mos2-nature}%
  \BibitemOpen
  \bibfield  {author} {\bibinfo {author} {\bibfnamefont {B.}~\bibnamefont
  {Radisavljevic}}, \bibinfo {author} {\bibfnamefont {A.}~\bibnamefont
  {Radenovic}}, \bibinfo {author} {\bibfnamefont {J.}~\bibnamefont {Brivio}},
  \bibinfo {author} {\bibfnamefont {V.}~\bibnamefont {Giacometti}}, \ and\
  \bibinfo {author} {\bibfnamefont {A.}~\bibnamefont {Kis}},\ }\bibfield
  {title} {\enquote {\bibinfo {title} {{Single-layer MoS$_2$ transistors}},}\
  }\href {\doibase 10.1038/nnano.2010.279} {\bibfield  {journal} {\bibinfo
  {journal} {Nature Nanotechnology}\ }\textbf {\bibinfo {volume} {6}},\
  \bibinfo {pages} {147--150} (\bibinfo {year} {2011})}\BibitemShut {NoStop}%
\bibitem [{\citenamefont {Zeng}\ \emph {et~al.}(2012)\citenamefont {Zeng},
  \citenamefont {Dai}, \citenamefont {Yao}, \citenamefont {Xiao},\ and\
  \citenamefont {Cui}}]{valley}%
  \BibitemOpen
  \bibfield  {author} {\bibinfo {author} {\bibfnamefont {H.}~\bibnamefont
  {Zeng}}, \bibinfo {author} {\bibfnamefont {J.}~\bibnamefont {Dai}}, \bibinfo
  {author} {\bibfnamefont {W.}~\bibnamefont {Yao}}, \bibinfo {author}
  {\bibfnamefont {D.}~\bibnamefont {Xiao}}, \ and\ \bibinfo {author}
  {\bibfnamefont {X.}~\bibnamefont {Cui}},\ }\bibfield  {title} {\enquote
  {\bibinfo {title} {{Valley polarization in MoS$_2$ monolayers by optical
  pumping}},}\ }\href {\doibase 10.1038/nnano.2012.95} {\bibfield  {journal}
  {\bibinfo  {journal} {Nature Nanotechnology}\ }\textbf {\bibinfo {volume}
  {7}},\ \bibinfo {pages} {490--493} (\bibinfo {year} {2012})}\BibitemShut
  {NoStop}%
\bibitem [{\citenamefont {Wang}\ \emph {et~al.}(2012)\citenamefont {Wang},
  \citenamefont {Kalantar-Zadeh}, \citenamefont {Kis}, \citenamefont
  {Coleman},\ and\ \citenamefont {Strano}}]{tmddesign}%
  \BibitemOpen
  \bibfield  {author} {\bibinfo {author} {\bibfnamefont {Q.~H.}\ \bibnamefont
  {Wang}}, \bibinfo {author} {\bibfnamefont {K.}~\bibnamefont
  {Kalantar-Zadeh}}, \bibinfo {author} {\bibfnamefont {A.}~\bibnamefont {Kis}},
  \bibinfo {author} {\bibfnamefont {J.~N.}\ \bibnamefont {Coleman}}, \ and\
  \bibinfo {author} {\bibfnamefont {M.~S.}\ \bibnamefont {Strano}},\ }\bibfield
   {title} {\enquote {\bibinfo {title} {{Electronics and optoelectronics of
  two-dimensional transition metal dichalcogenides}},}\ }\href {\doibase
  10.1038/nnano.2012.193} {\bibfield  {journal} {\bibinfo  {journal} {Nature
  Nanotechnology}\ }\textbf {\bibinfo {volume} {7}},\ \bibinfo {pages}
  {699--712} (\bibinfo {year} {2012})}\BibitemShut {NoStop}%
\bibitem [{\citenamefont {Zhang}\ \emph {et~al.}(2014)\citenamefont {Zhang},
  \citenamefont {Chang}, \citenamefont {Zhou}, \citenamefont {Cui},
  \citenamefont {Yan}, \citenamefont {Liu}, \citenamefont {Schmitt},
  \citenamefont {Lee}, \citenamefont {Moore}, \citenamefont {Chen},
  \citenamefont {Lin}, \citenamefont {Jeng}, \citenamefont {Mo}, \citenamefont
  {Hussain}, \citenamefont {Bansil},\ and\ \citenamefont
  {Shen}}]{nature-mose2}%
  \BibitemOpen
  \bibfield  {author} {\bibinfo {author} {\bibfnamefont {Y.}~\bibnamefont
  {Zhang}}, \bibinfo {author} {\bibfnamefont {T.-R.}\ \bibnamefont {Chang}},
  \bibinfo {author} {\bibfnamefont {B.}~\bibnamefont {Zhou}}, \bibinfo {author}
  {\bibfnamefont {Y.-T.}\ \bibnamefont {Cui}}, \bibinfo {author} {\bibfnamefont
  {H.}~\bibnamefont {Yan}}, \bibinfo {author} {\bibfnamefont {Z.}~\bibnamefont
  {Liu}}, \bibinfo {author} {\bibfnamefont {F.}~\bibnamefont {Schmitt}},
  \bibinfo {author} {\bibfnamefont {J.}~\bibnamefont {Lee}}, \bibinfo {author}
  {\bibfnamefont {R.}~\bibnamefont {Moore}}, \bibinfo {author} {\bibfnamefont
  {Y.}~\bibnamefont {Chen}}, \bibinfo {author} {\bibfnamefont {H.}~\bibnamefont
  {Lin}}, \bibinfo {author} {\bibfnamefont {H.-T.}\ \bibnamefont {Jeng}},
  \bibinfo {author} {\bibfnamefont {S.-K.}\ \bibnamefont {Mo}}, \bibinfo
  {author} {\bibfnamefont {Z.}~\bibnamefont {Hussain}}, \bibinfo {author}
  {\bibfnamefont {A.}~\bibnamefont {Bansil}}, \ and\ \bibinfo {author}
  {\bibfnamefont {Z.-X.}\ \bibnamefont {Shen}},\ }\bibfield  {title} {\enquote
  {\bibinfo {title} {{Direct observation of the transition from indirect to
  direct bandgap in atomically thin epitaxial MoSe$_2$}},}\ }\href {\doibase
  10.1038/nnano.2013.277} {\bibfield  {journal} {\bibinfo  {journal} {Nature
  Nanotechnology}\ }\textbf {\bibinfo {volume} {9}},\ \bibinfo {pages}
  {111--115} (\bibinfo {year} {2014})}\BibitemShut {NoStop}%
\bibitem [{\citenamefont {Lei}\ \emph {et~al.}(2013)\citenamefont {Lei},
  \citenamefont {Ge}, \citenamefont {Liu}, \citenamefont {Najmaei},
  \citenamefont {Shi}, \citenamefont {You}, \citenamefont {Lou}, \citenamefont
  {Vajtai},\ and\ \citenamefont {Ajayan}}]{gase-atomic-layers}%
  \BibitemOpen
  \bibfield  {author} {\bibinfo {author} {\bibfnamefont {S.}~\bibnamefont
  {Lei}}, \bibinfo {author} {\bibfnamefont {L.}~\bibnamefont {Ge}}, \bibinfo
  {author} {\bibfnamefont {Z.}~\bibnamefont {Liu}}, \bibinfo {author}
  {\bibfnamefont {S.}~\bibnamefont {Najmaei}}, \bibinfo {author} {\bibfnamefont
  {G.}~\bibnamefont {Shi}}, \bibinfo {author} {\bibfnamefont {G.}~\bibnamefont
  {You}}, \bibinfo {author} {\bibfnamefont {J.}~\bibnamefont {Lou}}, \bibinfo
  {author} {\bibfnamefont {R.}~\bibnamefont {Vajtai}}, \ and\ \bibinfo {author}
  {\bibfnamefont {P.~M.}\ \bibnamefont {Ajayan}},\ }\bibfield  {title}
  {\enquote {\bibinfo {title} {Synthesis and photoresponse of large {GaSe}
  atomic layers},}\ }\href {\doibase 10.1021/nl4010089} {\bibfield  {journal}
  {\bibinfo  {journal} {Nano Letters}\ }\textbf {\bibinfo {volume} {13}},\
  \bibinfo {pages} {2777--2781} (\bibinfo {year} {2013})}\BibitemShut {NoStop}%
\bibitem [{\citenamefont {Ma}\ \emph {et~al.}(2013)\citenamefont {Ma},
  \citenamefont {Dai}, \citenamefont {Guo}, \citenamefont {Yu},\ and\
  \citenamefont {Huang}}]{C3CP50233C}%
  \BibitemOpen
  \bibfield  {author} {\bibinfo {author} {\bibfnamefont {Y.}~\bibnamefont
  {Ma}}, \bibinfo {author} {\bibfnamefont {Y.}~\bibnamefont {Dai}}, \bibinfo
  {author} {\bibfnamefont {M.}~\bibnamefont {Guo}}, \bibinfo {author}
  {\bibfnamefont {L.}~\bibnamefont {Yu}}, \ and\ \bibinfo {author}
  {\bibfnamefont {B.}~\bibnamefont {Huang}},\ }\bibfield  {title} {\enquote
  {\bibinfo {title} {Tunable electronic and dielectric behavior of {GaS} and
  {GaSe} monolayers},}\ }\href {\doibase 10.1039/C3CP50233C} {\bibfield
  {journal} {\bibinfo  {journal} {Phys. Chem. Chem. Phys.}\ }\textbf {\bibinfo
  {volume} {15}},\ \bibinfo {pages} {7098--7105} (\bibinfo {year}
  {2013})}\BibitemShut {NoStop}%
\bibitem [{\citenamefont {Late}\ \emph {et~al.}(2012)\citenamefont {Late},
  \citenamefont {Liu}, \citenamefont {Luo}, \citenamefont {Yan}, \citenamefont
  {Matte}, \citenamefont {Grayson}, \citenamefont {Rao},\ and\ \citenamefont
  {Dravid}}]{doi:10.1002/adma.201201361}%
  \BibitemOpen
  \bibfield  {author} {\bibinfo {author} {\bibfnamefont {D.~J.}\ \bibnamefont
  {Late}}, \bibinfo {author} {\bibfnamefont {B.}~\bibnamefont {Liu}}, \bibinfo
  {author} {\bibfnamefont {J.}~\bibnamefont {Luo}}, \bibinfo {author}
  {\bibfnamefont {A.}~\bibnamefont {Yan}}, \bibinfo {author} {\bibfnamefont
  {H.~S. S.~R.}\ \bibnamefont {Matte}}, \bibinfo {author} {\bibfnamefont
  {M.}~\bibnamefont {Grayson}}, \bibinfo {author} {\bibfnamefont {C.~N.~R.}\
  \bibnamefont {Rao}}, \ and\ \bibinfo {author} {\bibfnamefont {V.~P.}\
  \bibnamefont {Dravid}},\ }\bibfield  {title} {\enquote {\bibinfo {title}
  {{GaS} and {GaSe} ultrathin layer transistors},}\ }\href {\doibase
  10.1002/adma.201201361} {\bibfield  {journal} {\bibinfo  {journal} {Advanced
  Materials}\ }\textbf {\bibinfo {volume} {24}},\ \bibinfo {pages} {3549--3554}
  (\bibinfo {year} {2012})},\ \Eprint
  {http://arxiv.org/abs/https://onlinelibrary.wiley.com/doi/pdf/10.1002/adma.201201361}
  {https://onlinelibrary.wiley.com/doi/pdf/10.1002/adma.201201361} \BibitemShut
  {NoStop}%
\bibitem [{\citenamefont {Depeursinge}(1977)}]{bulk-gase}%
  \BibitemOpen
  \bibfield  {author} {\bibinfo {author} {\bibfnamefont {Y.}~\bibnamefont
  {Depeursinge}},\ }\bibfield  {title} {\enquote {\bibinfo {title} {Electronic
  band structure for the polytypes of {GaSe}},}\ }\href {\doibase
  10.1007/BF02723482} {\bibfield  {journal} {\bibinfo  {journal} {Il Nuovo
  Cimento B (1971-1996)}\ }\textbf {\bibinfo {volume} {38}},\ \bibinfo {pages}
  {153--158} (\bibinfo {year} {1977})}\BibitemShut {NoStop}%
\bibitem [{\citenamefont {Bahuguna}\ \emph {et~al.}(2018)\citenamefont
  {Bahuguna}, \citenamefont {Saini}, \citenamefont {Sharma},\ and\
  \citenamefont {Tiwari}}]{C8CP04723E}%
  \BibitemOpen
  \bibfield  {author} {\bibinfo {author} {\bibfnamefont {B.~P.}\ \bibnamefont
  {Bahuguna}}, \bibinfo {author} {\bibfnamefont {L.~K.}\ \bibnamefont {Saini}},
  \bibinfo {author} {\bibfnamefont {R.~O.}\ \bibnamefont {Sharma}}, \ and\
  \bibinfo {author} {\bibfnamefont {B.}~\bibnamefont {Tiwari}},\ }\bibfield
  {title} {\enquote {\bibinfo {title} {Hybrid functional calculations of
  electronic and thermoelectric properties of {GaS}{,} {GaSe}{,} and {GaTe}
  monolayers},}\ }\href {\doibase 10.1039/C8CP04723E} {\bibfield  {journal}
  {\bibinfo  {journal} {Phys. Chem. Chem. Phys.}\ }\textbf {\bibinfo {volume}
  {20}},\ \bibinfo {pages} {28575--28582} (\bibinfo {year} {2018})}\BibitemShut
  {NoStop}%
\bibitem [{\citenamefont {Jappor}\ and\ \citenamefont
  {Habeeb}(2018)}]{JAPPOR2018251}%
  \BibitemOpen
  \bibfield  {author} {\bibinfo {author} {\bibfnamefont {H.~R.}\ \bibnamefont
  {Jappor}}\ and\ \bibinfo {author} {\bibfnamefont {M.~A.}\ \bibnamefont
  {Habeeb}},\ }\bibfield  {title} {\enquote {\bibinfo {title} {Optical
  properties of two-dimensional {GaS} and {GaSe} monolayers},}\ }\href
  {\doibase https://doi.org/10.1016/j.physe.2018.04.019} {\bibfield  {journal}
  {\bibinfo  {journal} {Physica E: Low-dimensional Systems and Nanostructures}\
  }\textbf {\bibinfo {volume} {101}},\ \bibinfo {pages} {251 -- 255} (\bibinfo
  {year} {2018})}\BibitemShut {NoStop}%
\bibitem [{\citenamefont {Wang}\ \emph
  {et~al.}(2019{\natexlab{a}})\citenamefont {Wang}, \citenamefont {Qin},
  \citenamefont {Yang}, \citenamefont {Qin}, \citenamefont {Yao}, \citenamefont
  {Wang},\ and\ \citenamefont {Hu}}]{doi:10.1063/1.5094663}%
  \BibitemOpen
  \bibfield  {author} {\bibinfo {author} {\bibfnamefont {H.}~\bibnamefont
  {Wang}}, \bibinfo {author} {\bibfnamefont {G.}~\bibnamefont {Qin}}, \bibinfo
  {author} {\bibfnamefont {J.}~\bibnamefont {Yang}}, \bibinfo {author}
  {\bibfnamefont {Z.}~\bibnamefont {Qin}}, \bibinfo {author} {\bibfnamefont
  {Y.}~\bibnamefont {Yao}}, \bibinfo {author} {\bibfnamefont {Q.}~\bibnamefont
  {Wang}}, \ and\ \bibinfo {author} {\bibfnamefont {M.}~\bibnamefont {Hu}},\
  }\bibfield  {title} {\enquote {\bibinfo {title} {First-principles study of
  electronic, optical and thermal transport properties of group {III–VI}
  monolayer {MX} {(M=Ga, In; X=S, Se)}},}\ }\href {\doibase 10.1063/1.5094663}
  {\bibfield  {journal} {\bibinfo  {journal} {Journal of Applied Physics}\
  }\textbf {\bibinfo {volume} {125}},\ \bibinfo {pages} {245104} (\bibinfo
  {year} {2019}{\natexlab{a}})}\BibitemShut {NoStop}%
\bibitem [{\citenamefont {Longuinhos}\ and\ \citenamefont
  {Ribeiro-Soares}(2019)}]{PhysRevApplied.11.024012}%
  \BibitemOpen
  \bibfield  {author} {\bibinfo {author} {\bibfnamefont {R.}~\bibnamefont
  {Longuinhos}}\ and\ \bibinfo {author} {\bibfnamefont {J.}~\bibnamefont
  {Ribeiro-Soares}},\ }\bibfield  {title} {\enquote {\bibinfo {title}
  {Monitoring the applied strain in monolayer gallium selenide through
  vibrational spectroscopies: A first-principles investigation},}\ }\href
  {\doibase 10.1103/PhysRevApplied.11.024012} {\bibfield  {journal} {\bibinfo
  {journal} {Phys. Rev. Applied}\ }\textbf {\bibinfo {volume} {11}},\ \bibinfo
  {pages} {024012} (\bibinfo {year} {2019})}\BibitemShut {NoStop}%
\bibitem [{\citenamefont {Yagmurcukardes}\ \emph {et~al.}(2016)\citenamefont
  {Yagmurcukardes}, \citenamefont {Senger}, \citenamefont {Peeters},\ and\
  \citenamefont {Sahin}}]{PhysRevB.94.245407}%
  \BibitemOpen
  \bibfield  {author} {\bibinfo {author} {\bibfnamefont {M.}~\bibnamefont
  {Yagmurcukardes}}, \bibinfo {author} {\bibfnamefont {R.~T.}\ \bibnamefont
  {Senger}}, \bibinfo {author} {\bibfnamefont {F.~M.}\ \bibnamefont {Peeters}},
  \ and\ \bibinfo {author} {\bibfnamefont {H.}~\bibnamefont {Sahin}},\
  }\bibfield  {title} {\enquote {\bibinfo {title} {Mechanical properties of
  monolayer {GaS} and {GaSe} crystals},}\ }\href {\doibase
  10.1103/PhysRevB.94.245407} {\bibfield  {journal} {\bibinfo  {journal} {Phys.
  Rev. B}\ }\textbf {\bibinfo {volume} {94}},\ \bibinfo {pages} {245407}
  (\bibinfo {year} {2016})}\BibitemShut {NoStop}%
\bibitem [{\citenamefont {Liu}\ \emph {et~al.}(2015)\citenamefont {Liu},
  \citenamefont {Xia}, \citenamefont {Hou}, \citenamefont {Gao},\ and\
  \citenamefont {Zhang}}]{doi:10.1142/S0217984915500499}%
  \BibitemOpen
  \bibfield  {author} {\bibinfo {author} {\bibfnamefont {G.}~\bibnamefont
  {Liu}}, \bibinfo {author} {\bibfnamefont {S.}~\bibnamefont {Xia}}, \bibinfo
  {author} {\bibfnamefont {B.}~\bibnamefont {Hou}}, \bibinfo {author}
  {\bibfnamefont {T.}~\bibnamefont {Gao}}, \ and\ \bibinfo {author}
  {\bibfnamefont {R.}~\bibnamefont {Zhang}},\ }\bibfield  {title} {\enquote
  {\bibinfo {title} {Mechanical stabilities and nonlinear properties of
  monolayer gallium selenide under tension},}\ }\href {\doibase
  10.1142/S0217984915500499} {\bibfield  {journal} {\bibinfo  {journal} {Modern
  Physics Letters B}\ }\textbf {\bibinfo {volume} {29}},\ \bibinfo {pages}
  {1550049} (\bibinfo {year} {2015})}\BibitemShut {NoStop}%
\bibitem [{\citenamefont {Wang}\ \emph {et~al.}(2016)\citenamefont {Wang},
  \citenamefont {Yang}, \citenamefont {Cai}, \citenamefont {Ataca},
  \citenamefont {Chen}, \citenamefont {Zhang}, \citenamefont {Xu},
  \citenamefont {Chen}, \citenamefont {Wu}, \citenamefont {Zhang},
  \citenamefont {Liu}, \citenamefont {Li}, \citenamefont {Grossman},
  \citenamefont {Tongay},\ and\ \citenamefont {Liu}}]{C5NR08692B}%
  \BibitemOpen
  \bibfield  {author} {\bibinfo {author} {\bibfnamefont {C.}~\bibnamefont
  {Wang}}, \bibinfo {author} {\bibfnamefont {S.}~\bibnamefont {Yang}}, \bibinfo
  {author} {\bibfnamefont {H.}~\bibnamefont {Cai}}, \bibinfo {author}
  {\bibfnamefont {C.}~\bibnamefont {Ataca}}, \bibinfo {author} {\bibfnamefont
  {H.}~\bibnamefont {Chen}}, \bibinfo {author} {\bibfnamefont {X.}~\bibnamefont
  {Zhang}}, \bibinfo {author} {\bibfnamefont {J.}~\bibnamefont {Xu}}, \bibinfo
  {author} {\bibfnamefont {B.}~\bibnamefont {Chen}}, \bibinfo {author}
  {\bibfnamefont {K.}~\bibnamefont {Wu}}, \bibinfo {author} {\bibfnamefont
  {H.}~\bibnamefont {Zhang}}, \bibinfo {author} {\bibfnamefont
  {L.}~\bibnamefont {Liu}}, \bibinfo {author} {\bibfnamefont {J.}~\bibnamefont
  {Li}}, \bibinfo {author} {\bibfnamefont {J.~C.}\ \bibnamefont {Grossman}},
  \bibinfo {author} {\bibfnamefont {S.}~\bibnamefont {Tongay}}, \ and\ \bibinfo
  {author} {\bibfnamefont {Q.}~\bibnamefont {Liu}},\ }\bibfield  {title}
  {\enquote {\bibinfo {title} {Enhancing light emission efficiency without
  color change in post-transition metal chalcogenides},}\ }\href {\doibase
  10.1039/C5NR08692B} {\bibfield  {journal} {\bibinfo  {journal} {Nanoscale}\
  }\textbf {\bibinfo {volume} {8}},\ \bibinfo {pages} {5820--5825} (\bibinfo
  {year} {2016})}\BibitemShut {NoStop}%
\bibitem [{\citenamefont {Li}\ \emph {et~al.}(2016)\citenamefont {Li},
  \citenamefont {Lin}, \citenamefont {Lin}, \citenamefont {Huang},
  \citenamefont {Puretzky}, \citenamefont {Ma}, \citenamefont {Wang},
  \citenamefont {Zhou}, \citenamefont {Pantelides}, \citenamefont {Chi},
  \citenamefont {Kravchenko}, \citenamefont {Fowlkes}, \citenamefont {Rouleau},
  \citenamefont {Geohegan},\ and\ \citenamefont {Xiao}}]{Lie1501882}%
  \BibitemOpen
  \bibfield  {author} {\bibinfo {author} {\bibfnamefont {X.}~\bibnamefont
  {Li}}, \bibinfo {author} {\bibfnamefont {M.-W.}\ \bibnamefont {Lin}},
  \bibinfo {author} {\bibfnamefont {J.}~\bibnamefont {Lin}}, \bibinfo {author}
  {\bibfnamefont {B.}~\bibnamefont {Huang}}, \bibinfo {author} {\bibfnamefont
  {A.~A.}\ \bibnamefont {Puretzky}}, \bibinfo {author} {\bibfnamefont
  {C.}~\bibnamefont {Ma}}, \bibinfo {author} {\bibfnamefont {K.}~\bibnamefont
  {Wang}}, \bibinfo {author} {\bibfnamefont {W.}~\bibnamefont {Zhou}}, \bibinfo
  {author} {\bibfnamefont {S.~T.}\ \bibnamefont {Pantelides}}, \bibinfo
  {author} {\bibfnamefont {M.}~\bibnamefont {Chi}}, \bibinfo {author}
  {\bibfnamefont {I.}~\bibnamefont {Kravchenko}}, \bibinfo {author}
  {\bibfnamefont {J.}~\bibnamefont {Fowlkes}}, \bibinfo {author} {\bibfnamefont
  {C.~M.}\ \bibnamefont {Rouleau}}, \bibinfo {author} {\bibfnamefont {D.~B.}\
  \bibnamefont {Geohegan}}, \ and\ \bibinfo {author} {\bibfnamefont
  {K.}~\bibnamefont {Xiao}},\ }\bibfield  {title} {\enquote {\bibinfo {title}
  {Two-dimensional {GaSe/MoSe$_2$} misfit bilayer heterojunctions by van der
  waals epitaxy},}\ }\href {\doibase 10.1126/sciadv.1501882} {\bibfield
  {journal} {\bibinfo  {journal} {Science Advances}\ }\textbf {\bibinfo
  {volume} {2}} (\bibinfo {year} {2016}),\ 10.1126/sciadv.1501882}\BibitemShut
  {NoStop}%
\bibitem [{\citenamefont {Liu}\ \emph {et~al.}(2018)\citenamefont {Liu},
  \citenamefont {Zhou}, \citenamefont {Gao},\ and\ \citenamefont
  {Zhao}}]{C8CP03740J}%
  \BibitemOpen
  \bibfield  {author} {\bibinfo {author} {\bibfnamefont {N.}~\bibnamefont
  {Liu}}, \bibinfo {author} {\bibfnamefont {S.}~\bibnamefont {Zhou}}, \bibinfo
  {author} {\bibfnamefont {N.}~\bibnamefont {Gao}}, \ and\ \bibinfo {author}
  {\bibfnamefont {J.}~\bibnamefont {Zhao}},\ }\bibfield  {title} {\enquote
  {\bibinfo {title} {{Tuning Schottky barriers for monolayer GaSe FETs by
  exploiting a weak Fermi level pinning effect}},}\ }\href {\doibase
  10.1039/C8CP03740J} {\bibfield  {journal} {\bibinfo  {journal} {Phys. Chem.
  Chem. Phys.}\ }\textbf {\bibinfo {volume} {20}},\ \bibinfo {pages}
  {21732--21738} (\bibinfo {year} {2018})}\BibitemShut {NoStop}%
\bibitem [{\citenamefont {Yang}\ \emph {et~al.}(2016)\citenamefont {Yang},
  \citenamefont {Wang}, \citenamefont {Ataca}, \citenamefont {Li},
  \citenamefont {Chen}, \citenamefont {Cai}, \citenamefont {Suslu},
  \citenamefont {Grossman}, \citenamefont {Jiang}, \citenamefont {Liu},\ and\
  \citenamefont {Tongay}}]{gate-mos2}%
  \BibitemOpen
  \bibfield  {author} {\bibinfo {author} {\bibfnamefont {S.}~\bibnamefont
  {Yang}}, \bibinfo {author} {\bibfnamefont {C.}~\bibnamefont {Wang}}, \bibinfo
  {author} {\bibfnamefont {C.}~\bibnamefont {Ataca}}, \bibinfo {author}
  {\bibfnamefont {Y.}~\bibnamefont {Li}}, \bibinfo {author} {\bibfnamefont
  {H.}~\bibnamefont {Chen}}, \bibinfo {author} {\bibfnamefont {H.}~\bibnamefont
  {Cai}}, \bibinfo {author} {\bibfnamefont {A.}~\bibnamefont {Suslu}}, \bibinfo
  {author} {\bibfnamefont {J.~C.}\ \bibnamefont {Grossman}}, \bibinfo {author}
  {\bibfnamefont {C.}~\bibnamefont {Jiang}}, \bibinfo {author} {\bibfnamefont
  {Q.}~\bibnamefont {Liu}}, \ and\ \bibinfo {author} {\bibfnamefont
  {S.}~\bibnamefont {Tongay}},\ }\bibfield  {title} {\enquote {\bibinfo {title}
  {Self-driven photodetector and ambipolar transistor in atomically thin
  {GaTe-MoS$_2$} p--n vdw heterostructure},}\ }\href {\doibase
  10.1021/acsami.5b10001} {\bibfield  {journal} {\bibinfo  {journal} {ACS
  Applied Materials \& Interfaces}\ }\textbf {\bibinfo {volume} {8}},\ \bibinfo
  {pages} {2533--2539} (\bibinfo {year} {2016})}\BibitemShut {NoStop}%
\bibitem [{\citenamefont {Jappor}(2017)}]{JAPPOR2017109}%
  \BibitemOpen
  \bibfield  {author} {\bibinfo {author} {\bibfnamefont {H.~R.}\ \bibnamefont
  {Jappor}},\ }\bibfield  {title} {\enquote {\bibinfo {title} {Electronic
  structure of novel {GaS/GaSe} heterostructures based on {GaS} and {GaSe}
  monolayers},}\ }\href {\doibase https://doi.org/10.1016/j.physb.2017.08.054}
  {\bibfield  {journal} {\bibinfo  {journal} {Physica B: Condensed Matter}\
  }\textbf {\bibinfo {volume} {524}},\ \bibinfo {pages} {109 -- 117} (\bibinfo
  {year} {2017})}\BibitemShut {NoStop}%
\bibitem [{\citenamefont {Zhou}\ \emph {et~al.}(2018)\citenamefont {Zhou},
  \citenamefont {Liu}, \citenamefont {Zhao},\ and\ \citenamefont
  {Yao}}]{o-func}%
  \BibitemOpen
  \bibfield  {author} {\bibinfo {author} {\bibfnamefont {S.}~\bibnamefont
  {Zhou}}, \bibinfo {author} {\bibfnamefont {C.-C.}\ \bibnamefont {Liu}},
  \bibinfo {author} {\bibfnamefont {J.}~\bibnamefont {Zhao}}, \ and\ \bibinfo
  {author} {\bibfnamefont {Y.}~\bibnamefont {Yao}},\ }\bibfield  {title}
  {\enquote {\bibinfo {title} {Monolayer group-{III} monochalcogenides by
  oxygen functionalization: a promising class of two-dimensional topological
  insulators},}\ }\href {\doibase 10.1038/s41535-018-0089-0} {\bibfield
  {journal} {\bibinfo  {journal} {npj Quantum Materials}\ }\textbf {\bibinfo
  {volume} {3}},\ \bibinfo {pages} {16} (\bibinfo {year} {2018})}\BibitemShut
  {NoStop}%
\bibitem [{\citenamefont {Ao}\ \emph {et~al.}(2015)\citenamefont {Ao},
  \citenamefont {Xiao}, \citenamefont {Xiang}, \citenamefont {Li},
  \citenamefont {Liu}, \citenamefont {Huang},\ and\ \citenamefont
  {Zu}}]{C5CP00397K}%
  \BibitemOpen
  \bibfield  {author} {\bibinfo {author} {\bibfnamefont {L.}~\bibnamefont
  {Ao}}, \bibinfo {author} {\bibfnamefont {H.~Y.}\ \bibnamefont {Xiao}},
  \bibinfo {author} {\bibfnamefont {X.}~\bibnamefont {Xiang}}, \bibinfo
  {author} {\bibfnamefont {S.}~\bibnamefont {Li}}, \bibinfo {author}
  {\bibfnamefont {K.~Z.}\ \bibnamefont {Liu}}, \bibinfo {author} {\bibfnamefont
  {H.}~\bibnamefont {Huang}}, \ and\ \bibinfo {author} {\bibfnamefont {X.~T.}\
  \bibnamefont {Zu}},\ }\bibfield  {title} {\enquote {\bibinfo {title}
  {Functionalization of a {GaSe} monolayer by vacancy and chemical element
  doping},}\ }\href {\doibase 10.1039/C5CP00397K} {\bibfield  {journal}
  {\bibinfo  {journal} {Phys. Chem. Chem. Phys.}\ }\textbf {\bibinfo {volume}
  {17}},\ \bibinfo {pages} {10737--10748} (\bibinfo {year} {2015})}\BibitemShut
  {NoStop}%
\bibitem [{\citenamefont {Meng}\ \emph {et~al.}(2015)\citenamefont {Meng},
  \citenamefont {Pant}, \citenamefont {Cai}, \citenamefont {Kang},
  \citenamefont {Sahin}, \citenamefont {Chen}, \citenamefont {Wu},
  \citenamefont {Yang}, \citenamefont {Suslu}, \citenamefont {Peeters},\ and\
  \citenamefont {Tongay}}]{pyradine}%
  \BibitemOpen
  \bibfield  {author} {\bibinfo {author} {\bibfnamefont {X.}~\bibnamefont
  {Meng}}, \bibinfo {author} {\bibfnamefont {A.}~\bibnamefont {Pant}}, \bibinfo
  {author} {\bibfnamefont {H.}~\bibnamefont {Cai}}, \bibinfo {author}
  {\bibfnamefont {J.}~\bibnamefont {Kang}}, \bibinfo {author} {\bibfnamefont
  {H.}~\bibnamefont {Sahin}}, \bibinfo {author} {\bibfnamefont
  {B.}~\bibnamefont {Chen}}, \bibinfo {author} {\bibfnamefont {K.}~\bibnamefont
  {Wu}}, \bibinfo {author} {\bibfnamefont {S.}~\bibnamefont {Yang}}, \bibinfo
  {author} {\bibfnamefont {A.}~\bibnamefont {Suslu}}, \bibinfo {author}
  {\bibfnamefont {F.~M.}\ \bibnamefont {Peeters}}, \ and\ \bibinfo {author}
  {\bibfnamefont {S.}~\bibnamefont {Tongay}},\ }\bibfield  {title} {\enquote
  {\bibinfo {title} {Engineering excitonic dynamics and environmental stability
  of post-transition metal chalcogenides by pyridine functionalization
  technique},}\ }\href {\doibase 10.1039/C5NR04879F} {\bibfield  {journal}
  {\bibinfo  {journal} {Nanoscale}\ }\textbf {\bibinfo {volume} {7}},\ \bibinfo
  {pages} {17109--17115} (\bibinfo {year} {2015})}\BibitemShut {NoStop}%
\bibitem [{\citenamefont {Wines}\ \emph {et~al.}(2020)\citenamefont {Wines},
  \citenamefont {Kropp}, \citenamefont {Chaney}, \citenamefont {Ersan},\ and\
  \citenamefont {Ataca}}]{D0CP00357C}%
  \BibitemOpen
  \bibfield  {author} {\bibinfo {author} {\bibfnamefont {D.}~\bibnamefont
  {Wines}}, \bibinfo {author} {\bibfnamefont {J.~A.}\ \bibnamefont {Kropp}},
  \bibinfo {author} {\bibfnamefont {G.}~\bibnamefont {Chaney}}, \bibinfo
  {author} {\bibfnamefont {F.}~\bibnamefont {Ersan}}, \ and\ \bibinfo {author}
  {\bibfnamefont {C.}~\bibnamefont {Ataca}},\ }\bibfield  {title} {\enquote
  {\bibinfo {title} {Electronic properties of bare and functionalized
  two-dimensional {(2D)} tellurene structures},}\ }\href {\doibase
  10.1039/D0CP00357C} {\bibfield  {journal} {\bibinfo  {journal} {Phys. Chem.
  Chem. Phys.}\ }\textbf {\bibinfo {volume} {22}},\ \bibinfo {pages}
  {6727--6737} (\bibinfo {year} {2020})}\BibitemShut {NoStop}%
\bibitem [{\citenamefont {Yang}\ \emph {et~al.}(2018)\citenamefont {Yang},
  \citenamefont {Chen}, \citenamefont {Qin}, \citenamefont {Zhou},
  \citenamefont {Liu}, \citenamefont {Durso}, \citenamefont {Zhuang},
  \citenamefont {Shen},\ and\ \citenamefont
  {Tongay}}]{PhysRevMaterials.2.104002}%
  \BibitemOpen
  \bibfield  {author} {\bibinfo {author} {\bibfnamefont {S.}~\bibnamefont
  {Yang}}, \bibinfo {author} {\bibfnamefont {B.}~\bibnamefont {Chen}}, \bibinfo
  {author} {\bibfnamefont {Y.}~\bibnamefont {Qin}}, \bibinfo {author}
  {\bibfnamefont {Y.}~\bibnamefont {Zhou}}, \bibinfo {author} {\bibfnamefont
  {L.}~\bibnamefont {Liu}}, \bibinfo {author} {\bibfnamefont {M.}~\bibnamefont
  {Durso}}, \bibinfo {author} {\bibfnamefont {H.}~\bibnamefont {Zhuang}},
  \bibinfo {author} {\bibfnamefont {Y.}~\bibnamefont {Shen}}, \ and\ \bibinfo
  {author} {\bibfnamefont {S.}~\bibnamefont {Tongay}},\ }\bibfield  {title}
  {\enquote {\bibinfo {title} {Highly crystalline synthesis of tellurene sheets
  on two-dimensional surfaces: Control over helical chain direction of
  tellurene},}\ }\href {\doibase 10.1103/PhysRevMaterials.2.104002} {\bibfield
  {journal} {\bibinfo  {journal} {Phys. Rev. Materials}\ }\textbf {\bibinfo
  {volume} {2}},\ \bibinfo {pages} {104002} (\bibinfo {year}
  {2018})}\BibitemShut {NoStop}%
\bibitem [{\citenamefont {Ben~Aziza}\ \emph {et~al.}(2017)\citenamefont
  {Ben~Aziza}, \citenamefont {Pierucci}, \citenamefont {Henck}, \citenamefont
  {Silly}, \citenamefont {David}, \citenamefont {Yoon}, \citenamefont
  {Sirotti}, \citenamefont {Xiao}, \citenamefont {Eddrief}, \citenamefont
  {Girard},\ and\ \citenamefont {Ouerghi}}]{PhysRevB.96.035407}%
  \BibitemOpen
  \bibfield  {author} {\bibinfo {author} {\bibfnamefont {Z.}~\bibnamefont
  {Ben~Aziza}}, \bibinfo {author} {\bibfnamefont {D.}~\bibnamefont {Pierucci}},
  \bibinfo {author} {\bibfnamefont {H.}~\bibnamefont {Henck}}, \bibinfo
  {author} {\bibfnamefont {M.~G.}\ \bibnamefont {Silly}}, \bibinfo {author}
  {\bibfnamefont {C.}~\bibnamefont {David}}, \bibinfo {author} {\bibfnamefont
  {M.}~\bibnamefont {Yoon}}, \bibinfo {author} {\bibfnamefont {F.}~\bibnamefont
  {Sirotti}}, \bibinfo {author} {\bibfnamefont {K.}~\bibnamefont {Xiao}},
  \bibinfo {author} {\bibfnamefont {M.}~\bibnamefont {Eddrief}}, \bibinfo
  {author} {\bibfnamefont {J.-C.}\ \bibnamefont {Girard}}, \ and\ \bibinfo
  {author} {\bibfnamefont {A.}~\bibnamefont {Ouerghi}},\ }\bibfield  {title}
  {\enquote {\bibinfo {title} {Tunable quasiparticle band gap in few-layer
  {GaSe}/graphene van der waals heterostructures},}\ }\href {\doibase
  10.1103/PhysRevB.96.035407} {\bibfield  {journal} {\bibinfo  {journal} {Phys.
  Rev. B}\ }\textbf {\bibinfo {volume} {96}},\ \bibinfo {pages} {035407}
  (\bibinfo {year} {2017})}\BibitemShut {NoStop}%
\bibitem [{\citenamefont {Susoma}\ \emph {et~al.}(2017)\citenamefont {Susoma},
  \citenamefont {Lahtinen}, \citenamefont {Kim}, \citenamefont {Riikonen},\
  and\ \citenamefont {Lipsanen}}]{doi:10.1063/1.4973918}%
  \BibitemOpen
  \bibfield  {author} {\bibinfo {author} {\bibfnamefont {J.}~\bibnamefont
  {Susoma}}, \bibinfo {author} {\bibfnamefont {J.}~\bibnamefont {Lahtinen}},
  \bibinfo {author} {\bibfnamefont {M.}~\bibnamefont {Kim}}, \bibinfo {author}
  {\bibfnamefont {J.}~\bibnamefont {Riikonen}}, \ and\ \bibinfo {author}
  {\bibfnamefont {H.}~\bibnamefont {Lipsanen}},\ }\bibfield  {title} {\enquote
  {\bibinfo {title} {{Crystal quality of two-dimensional gallium telluride and
  gallium selenide using Raman fingerprint}},}\ }\href {\doibase
  10.1063/1.4973918} {\bibfield  {journal} {\bibinfo  {journal} {AIP Advances}\
  }\textbf {\bibinfo {volume} {7}},\ \bibinfo {pages} {015014} (\bibinfo {year}
  {2017})}\BibitemShut {NoStop}%
\bibitem [{\citenamefont {Jung}\ \emph {et~al.}(2015)\citenamefont {Jung},
  \citenamefont {Shojaei}, \citenamefont {Park}, \citenamefont {Oh},
  \citenamefont {Im}, \citenamefont {Jang}, \citenamefont {Park},\ and\
  \citenamefont {Kang}}]{GaSe-optical}%
  \BibitemOpen
  \bibfield  {author} {\bibinfo {author} {\bibfnamefont {C.~S.}\ \bibnamefont
  {Jung}}, \bibinfo {author} {\bibfnamefont {F.}~\bibnamefont {Shojaei}},
  \bibinfo {author} {\bibfnamefont {K.}~\bibnamefont {Park}}, \bibinfo {author}
  {\bibfnamefont {J.~Y.}\ \bibnamefont {Oh}}, \bibinfo {author} {\bibfnamefont
  {H.~S.}\ \bibnamefont {Im}}, \bibinfo {author} {\bibfnamefont {D.~M.}\
  \bibnamefont {Jang}}, \bibinfo {author} {\bibfnamefont {J.}~\bibnamefont
  {Park}}, \ and\ \bibinfo {author} {\bibfnamefont {H.~S.}\ \bibnamefont
  {Kang}},\ }\bibfield  {title} {\enquote {\bibinfo {title} {Red-to-ultraviolet
  emission tuning of two-dimensional gallium sulfide/selenide},}\ }\href
  {\doibase 10.1021/acsnano.5b04876} {\bibfield  {journal} {\bibinfo  {journal}
  {ACS Nano}\ }\textbf {\bibinfo {volume} {9}},\ \bibinfo {pages} {9585--9593}
  (\bibinfo {year} {2015})}\BibitemShut {NoStop}%
\bibitem [{\citenamefont {Chitara}\ and\ \citenamefont
  {Ya{'}akobovitz}(2018)}]{C8NR01065J}%
  \BibitemOpen
  \bibfield  {author} {\bibinfo {author} {\bibfnamefont {B.}~\bibnamefont
  {Chitara}}\ and\ \bibinfo {author} {\bibfnamefont {A.}~\bibnamefont
  {Ya{'}akobovitz}},\ }\bibfield  {title} {\enquote {\bibinfo {title} {{Elastic
  properties and breaking strengths of GaS{,} GaSe and GaTe nanosheets}},}\
  }\href {\doibase 10.1039/C8NR01065J} {\bibfield  {journal} {\bibinfo
  {journal} {Nanoscale}\ }\textbf {\bibinfo {volume} {10}},\ \bibinfo {pages}
  {13022--13027} (\bibinfo {year} {2018})}\BibitemShut {NoStop}%
\bibitem [{\citenamefont {Li}\ \emph {et~al.}(2014)\citenamefont {Li},
  \citenamefont {Lin}, \citenamefont {Puretzky}, \citenamefont {Idrobo},
  \citenamefont {Ma}, \citenamefont {Chi}, \citenamefont {Yoon}, \citenamefont
  {Rouleau}, \citenamefont {Kravchenko}, \citenamefont {Geohegan},\ and\
  \citenamefont {Xiao}}]{GaSe-nature-synthesis}%
  \BibitemOpen
  \bibfield  {author} {\bibinfo {author} {\bibfnamefont {X.}~\bibnamefont
  {Li}}, \bibinfo {author} {\bibfnamefont {M.-W.}\ \bibnamefont {Lin}},
  \bibinfo {author} {\bibfnamefont {A.~A.}\ \bibnamefont {Puretzky}}, \bibinfo
  {author} {\bibfnamefont {J.~C.}\ \bibnamefont {Idrobo}}, \bibinfo {author}
  {\bibfnamefont {C.}~\bibnamefont {Ma}}, \bibinfo {author} {\bibfnamefont
  {M.}~\bibnamefont {Chi}}, \bibinfo {author} {\bibfnamefont {M.}~\bibnamefont
  {Yoon}}, \bibinfo {author} {\bibfnamefont {C.~M.}\ \bibnamefont {Rouleau}},
  \bibinfo {author} {\bibfnamefont {I.~I.}\ \bibnamefont {Kravchenko}},
  \bibinfo {author} {\bibfnamefont {D.~B.}\ \bibnamefont {Geohegan}}, \ and\
  \bibinfo {author} {\bibfnamefont {K.}~\bibnamefont {Xiao}},\ }\bibfield
  {title} {\enquote {\bibinfo {title} {Controlled vapor phase growth of single
  crystalline, two-dimensional {GaSe} crystals with high photoresponse},}\
  }\href {\doibase 10.1038/srep05497} {\bibfield  {journal} {\bibinfo
  {journal} {Scientific Reports}\ }\textbf {\bibinfo {volume} {4}},\ \bibinfo
  {pages} {5497} (\bibinfo {year} {2014})}\BibitemShut {NoStop}%
\bibitem [{\citenamefont {Rahaman}\ \emph {et~al.}(2018)\citenamefont
  {Rahaman}, \citenamefont {Bejani}, \citenamefont {Salvan}, \citenamefont
  {Lopez-Rivera}, \citenamefont {Pulci}, \citenamefont {Bechstedt},\ and\
  \citenamefont {Zahn}}]{Rahaman_2018}%
  \BibitemOpen
  \bibfield  {author} {\bibinfo {author} {\bibfnamefont {M.}~\bibnamefont
  {Rahaman}}, \bibinfo {author} {\bibfnamefont {M.}~\bibnamefont {Bejani}},
  \bibinfo {author} {\bibfnamefont {G.}~\bibnamefont {Salvan}}, \bibinfo
  {author} {\bibfnamefont {S.~A.}\ \bibnamefont {Lopez-Rivera}}, \bibinfo
  {author} {\bibfnamefont {O.}~\bibnamefont {Pulci}}, \bibinfo {author}
  {\bibfnamefont {F.}~\bibnamefont {Bechstedt}}, \ and\ \bibinfo {author}
  {\bibfnamefont {D.~R.~T.}\ \bibnamefont {Zahn}},\ }\bibfield  {title}
  {\enquote {\bibinfo {title} {{Vibrational properties of {GaSe}: a layer
  dependent study from experiments to theory}},}\ }\href {\doibase
  10.1088/1361-6641/aae4c7} {\bibfield  {journal} {\bibinfo  {journal}
  {Semiconductor Science and Technology}\ }\textbf {\bibinfo {volume} {33}},\
  \bibinfo {pages} {125008} (\bibinfo {year} {2018})}\BibitemShut {NoStop}%
\bibitem [{\citenamefont {Rahaman}\ \emph {et~al.}(2017)\citenamefont
  {Rahaman}, \citenamefont {Rodriguez}, \citenamefont {Monecke}, \citenamefont
  {Lopez-Rivera},\ and\ \citenamefont {Zahn}}]{Rahaman_2017}%
  \BibitemOpen
  \bibfield  {author} {\bibinfo {author} {\bibfnamefont {M.}~\bibnamefont
  {Rahaman}}, \bibinfo {author} {\bibfnamefont {R.~D.}\ \bibnamefont
  {Rodriguez}}, \bibinfo {author} {\bibfnamefont {M.}~\bibnamefont {Monecke}},
  \bibinfo {author} {\bibfnamefont {S.~A.}\ \bibnamefont {Lopez-Rivera}}, \
  and\ \bibinfo {author} {\bibfnamefont {D.~R.~T.}\ \bibnamefont {Zahn}},\
  }\bibfield  {title} {\enquote {\bibinfo {title} {{{GaSe} oxidation in air:
  from bulk to monolayers}},}\ }\href {\doibase 10.1088/1361-6641/aa8441}
  {\bibfield  {journal} {\bibinfo  {journal} {Semiconductor Science and
  Technology}\ }\textbf {\bibinfo {volume} {32}},\ \bibinfo {pages} {105004}
  (\bibinfo {year} {2017})}\BibitemShut {NoStop}%
\bibitem [{\citenamefont {Cai}\ \emph {et~al.}(2016)\citenamefont {Cai},
  \citenamefont {Soignard}, \citenamefont {Ataca}, \citenamefont {Chen},
  \citenamefont {Ko}, \citenamefont {Aoki}, \citenamefont {Pant}, \citenamefont
  {Meng}, \citenamefont {Yang}, \citenamefont {Grossman}, \citenamefont
  {Ogletree},\ and\ \citenamefont {Tongay}}]{doi:10.1002/adma.201601184}%
  \BibitemOpen
  \bibfield  {author} {\bibinfo {author} {\bibfnamefont {H.}~\bibnamefont
  {Cai}}, \bibinfo {author} {\bibfnamefont {E.}~\bibnamefont {Soignard}},
  \bibinfo {author} {\bibfnamefont {C.}~\bibnamefont {Ataca}}, \bibinfo
  {author} {\bibfnamefont {B.}~\bibnamefont {Chen}}, \bibinfo {author}
  {\bibfnamefont {C.}~\bibnamefont {Ko}}, \bibinfo {author} {\bibfnamefont
  {T.}~\bibnamefont {Aoki}}, \bibinfo {author} {\bibfnamefont {A.}~\bibnamefont
  {Pant}}, \bibinfo {author} {\bibfnamefont {X.}~\bibnamefont {Meng}}, \bibinfo
  {author} {\bibfnamefont {S.}~\bibnamefont {Yang}}, \bibinfo {author}
  {\bibfnamefont {J.}~\bibnamefont {Grossman}}, \bibinfo {author}
  {\bibfnamefont {F.~D.}\ \bibnamefont {Ogletree}}, \ and\ \bibinfo {author}
  {\bibfnamefont {S.}~\bibnamefont {Tongay}},\ }\bibfield  {title} {\enquote
  {\bibinfo {title} {Band engineering by controlling vdw epitaxy growth mode in
  {2D} gallium chalcogenides},}\ }\href {\doibase 10.1002/adma.201601184}
  {\bibfield  {journal} {\bibinfo  {journal} {Advanced Materials}\ }\textbf
  {\bibinfo {volume} {28}},\ \bibinfo {pages} {7375--7382} (\bibinfo {year}
  {2016})}\BibitemShut {NoStop}%
\bibitem [{\citenamefont {Mueller}\ and\ \citenamefont
  {Malic}(2018)}]{2dexciton}%
  \BibitemOpen
  \bibfield  {author} {\bibinfo {author} {\bibfnamefont {T.}~\bibnamefont
  {Mueller}}\ and\ \bibinfo {author} {\bibfnamefont {E.}~\bibnamefont
  {Malic}},\ }\bibfield  {title} {\enquote {\bibinfo {title} {Exciton physics
  and device application of two-dimensional transition metal dichalcogenide
  semiconductors},}\ }\href {\doibase 10.1038/s41699-018-0074-2} {\bibfield
  {journal} {\bibinfo  {journal} {npj 2D Materials and Applications}\ }\textbf
  {\bibinfo {volume} {2}},\ \bibinfo {pages} {29} (\bibinfo {year}
  {2018})}\BibitemShut {NoStop}%
\bibitem [{\citenamefont {Hohenberg}\ and\ \citenamefont
  {Kohn}(1964)}]{PhysRev.136.B864}%
  \BibitemOpen
  \bibfield  {author} {\bibinfo {author} {\bibfnamefont {P.}~\bibnamefont
  {Hohenberg}}\ and\ \bibinfo {author} {\bibfnamefont {W.}~\bibnamefont
  {Kohn}},\ }\bibfield  {title} {\enquote {\bibinfo {title} {Inhomogeneous
  electron gas},}\ }\href {\doibase 10.1103/PhysRev.136.B864} {\bibfield
  {journal} {\bibinfo  {journal} {Phys. Rev.}\ }\textbf {\bibinfo {volume}
  {136}},\ \bibinfo {pages} {B864--B871} (\bibinfo {year} {1964})}\BibitemShut
  {NoStop}%
\bibitem [{\citenamefont {Perdew}, \citenamefont {Burke},\ and\ \citenamefont
  {Ernzerhof}(1996)}]{PhysRevLett.77.3865}%
  \BibitemOpen
  \bibfield  {author} {\bibinfo {author} {\bibfnamefont {J.~P.}\ \bibnamefont
  {Perdew}}, \bibinfo {author} {\bibfnamefont {K.}~\bibnamefont {Burke}}, \
  and\ \bibinfo {author} {\bibfnamefont {M.}~\bibnamefont {Ernzerhof}},\
  }\bibfield  {title} {\enquote {\bibinfo {title} {Generalized gradient
  approximation made simple},}\ }\href {\doibase 10.1103/PhysRevLett.77.3865}
  {\bibfield  {journal} {\bibinfo  {journal} {Phys. Rev. Lett.}\ }\textbf
  {\bibinfo {volume} {77}},\ \bibinfo {pages} {3865--3868} (\bibinfo {year}
  {1996})}\BibitemShut {NoStop}%
\bibitem [{\citenamefont {Sun}, \citenamefont {Ruzsinszky},\ and\ \citenamefont
  {Perdew}(2015)}]{PhysRevLett.115.036402}%
  \BibitemOpen
  \bibfield  {author} {\bibinfo {author} {\bibfnamefont {J.}~\bibnamefont
  {Sun}}, \bibinfo {author} {\bibfnamefont {A.}~\bibnamefont {Ruzsinszky}}, \
  and\ \bibinfo {author} {\bibfnamefont {J.~P.}\ \bibnamefont {Perdew}},\
  }\bibfield  {title} {\enquote {\bibinfo {title} {Strongly constrained and
  appropriately normed semilocal density functional},}\ }\href {\doibase
  10.1103/PhysRevLett.115.036402} {\bibfield  {journal} {\bibinfo  {journal}
  {Phys. Rev. Lett.}\ }\textbf {\bibinfo {volume} {115}},\ \bibinfo {pages}
  {036402} (\bibinfo {year} {2015})}\BibitemShut {NoStop}%
\bibitem [{\citenamefont {Heyd}, \citenamefont {Scuseria},\ and\ \citenamefont
  {Ernzerhof}(2003)}]{doi:10.1063/1.1564060}%
  \BibitemOpen
  \bibfield  {author} {\bibinfo {author} {\bibfnamefont {J.}~\bibnamefont
  {Heyd}}, \bibinfo {author} {\bibfnamefont {G.~E.}\ \bibnamefont {Scuseria}},
  \ and\ \bibinfo {author} {\bibfnamefont {M.}~\bibnamefont {Ernzerhof}},\
  }\bibfield  {title} {\enquote {\bibinfo {title} {Hybrid functionals based on
  a screened coulomb potential},}\ }\href {\doibase 10.1063/1.1564060}
  {\bibfield  {journal} {\bibinfo  {journal} {The Journal of Chemical Physics}\
  }\textbf {\bibinfo {volume} {118}},\ \bibinfo {pages} {8207--8215} (\bibinfo
  {year} {2003})}\BibitemShut {NoStop}%
\bibitem [{\citenamefont {Krukau}\ \emph {et~al.}(2006)\citenamefont {Krukau},
  \citenamefont {Vydrov}, \citenamefont {Izmaylov},\ and\ \citenamefont
  {Scuseria}}]{doi:10.1063/1.2404663}%
  \BibitemOpen
  \bibfield  {author} {\bibinfo {author} {\bibfnamefont {A.~V.}\ \bibnamefont
  {Krukau}}, \bibinfo {author} {\bibfnamefont {O.~A.}\ \bibnamefont {Vydrov}},
  \bibinfo {author} {\bibfnamefont {A.~F.}\ \bibnamefont {Izmaylov}}, \ and\
  \bibinfo {author} {\bibfnamefont {G.~E.}\ \bibnamefont {Scuseria}},\
  }\bibfield  {title} {\enquote {\bibinfo {title} {{Influence of the exchange
  screening parameter on the performance of screened hybrid functionals}},}\
  }\href {\doibase 10.1063/1.2404663} {\bibfield  {journal} {\bibinfo
  {journal} {The Journal of Chemical Physics}\ }\textbf {\bibinfo {volume}
  {125}},\ \bibinfo {pages} {224106} (\bibinfo {year} {2006})}\BibitemShut
  {NoStop}%
\bibitem [{\citenamefont {Hybertsen}\ and\ \citenamefont
  {Louie}(1986)}]{PhysRevB.34.5390}%
  \BibitemOpen
  \bibfield  {author} {\bibinfo {author} {\bibfnamefont {M.~S.}\ \bibnamefont
  {Hybertsen}}\ and\ \bibinfo {author} {\bibfnamefont {S.~G.}\ \bibnamefont
  {Louie}},\ }\bibfield  {title} {\enquote {\bibinfo {title} {{Electron
  correlation in semiconductors and insulators: Band gaps and quasiparticle
  energies}},}\ }\href {\doibase 10.1103/PhysRevB.34.5390} {\bibfield
  {journal} {\bibinfo  {journal} {Phys. Rev. B}\ }\textbf {\bibinfo {volume}
  {34}},\ \bibinfo {pages} {5390--5413} (\bibinfo {year} {1986})}\BibitemShut
  {NoStop}%
\bibitem [{\citenamefont {Hedin}(1965)}]{PhysRev.139.A796}%
  \BibitemOpen
  \bibfield  {author} {\bibinfo {author} {\bibfnamefont {L.}~\bibnamefont
  {Hedin}},\ }\bibfield  {title} {\enquote {\bibinfo {title} {New method for
  calculating the one-particle {Green's} function with application to the
  electron-gas problem},}\ }\href {\doibase 10.1103/PhysRev.139.A796}
  {\bibfield  {journal} {\bibinfo  {journal} {Phys. Rev.}\ }\textbf {\bibinfo
  {volume} {139}},\ \bibinfo {pages} {A796--A823} (\bibinfo {year}
  {1965})}\BibitemShut {NoStop}%
\bibitem [{\citenamefont {Onida}, \citenamefont {Reining},\ and\ \citenamefont
  {Rubio}(2002)}]{RevModPhys.74.601}%
  \BibitemOpen
  \bibfield  {author} {\bibinfo {author} {\bibfnamefont {G.}~\bibnamefont
  {Onida}}, \bibinfo {author} {\bibfnamefont {L.}~\bibnamefont {Reining}}, \
  and\ \bibinfo {author} {\bibfnamefont {A.}~\bibnamefont {Rubio}},\ }\bibfield
   {title} {\enquote {\bibinfo {title} {Electronic excitations:
  density-functional versus many-body {Green}'s-function approaches},}\ }\href
  {\doibase 10.1103/RevModPhys.74.601} {\bibfield  {journal} {\bibinfo
  {journal} {Rev. Mod. Phys.}\ }\textbf {\bibinfo {volume} {74}},\ \bibinfo
  {pages} {601--659} (\bibinfo {year} {2002})}\BibitemShut {NoStop}%
\bibitem [{\citenamefont {Hao}\ \emph {et~al.}(2018)\citenamefont {Hao},
  \citenamefont {Shee}, \citenamefont {Upadhyay}, \citenamefont {Ataca},
  \citenamefont {Jordan},\ and\ \citenamefont {Rubenstein}}]{ataca_qmc}%
  \BibitemOpen
  \bibfield  {author} {\bibinfo {author} {\bibfnamefont {H.}~\bibnamefont
  {Hao}}, \bibinfo {author} {\bibfnamefont {J.}~\bibnamefont {Shee}}, \bibinfo
  {author} {\bibfnamefont {S.}~\bibnamefont {Upadhyay}}, \bibinfo {author}
  {\bibfnamefont {C.}~\bibnamefont {Ataca}}, \bibinfo {author} {\bibfnamefont
  {K.~D.}\ \bibnamefont {Jordan}}, \ and\ \bibinfo {author} {\bibfnamefont
  {B.~M.}\ \bibnamefont {Rubenstein}},\ }\bibfield  {title} {\enquote {\bibinfo
  {title} {Accurate predictions of electron binding energies of dipole-bound
  anions via quantum monte carlo methods},}\ }\href {\doibase
  10.1021/acs.jpclett.8b02733} {\bibfield  {journal} {\bibinfo  {journal} {The
  Journal of Physical Chemistry Letters}\ }\textbf {\bibinfo {volume} {9}},\
  \bibinfo {pages} {6185--6190} (\bibinfo {year} {2018})}\BibitemShut {NoStop}%
\bibitem [{\citenamefont {Mostaani}, \citenamefont {Drummond},\ and\
  \citenamefont {Fal'ko}(2015)}]{PhysRevLett.115.115501}%
  \BibitemOpen
  \bibfield  {author} {\bibinfo {author} {\bibfnamefont {E.}~\bibnamefont
  {Mostaani}}, \bibinfo {author} {\bibfnamefont {N.~D.}\ \bibnamefont
  {Drummond}}, \ and\ \bibinfo {author} {\bibfnamefont {V.~I.}\ \bibnamefont
  {Fal'ko}},\ }\bibfield  {title} {\enquote {\bibinfo {title} {{Quantum Monte
  Carlo} calculation of the binding energy of bilayer graphene},}\ }\href
  {\doibase 10.1103/PhysRevLett.115.115501} {\bibfield  {journal} {\bibinfo
  {journal} {Phys. Rev. Lett.}\ }\textbf {\bibinfo {volume} {115}},\ \bibinfo
  {pages} {115501} (\bibinfo {year} {2015})}\BibitemShut {NoStop}%
\bibitem [{\citenamefont {Shulenburger}\ \emph {et~al.}(2015)\citenamefont
  {Shulenburger}, \citenamefont {Baczewski}, \citenamefont {Zhu}, \citenamefont
  {Guan},\ and\ \citenamefont {Tom{\'a}nek}}]{bilayer-phos}%
  \BibitemOpen
  \bibfield  {author} {\bibinfo {author} {\bibfnamefont {L.}~\bibnamefont
  {Shulenburger}}, \bibinfo {author} {\bibfnamefont {A.~D.}\ \bibnamefont
  {Baczewski}}, \bibinfo {author} {\bibfnamefont {Z.}~\bibnamefont {Zhu}},
  \bibinfo {author} {\bibfnamefont {J.}~\bibnamefont {Guan}}, \ and\ \bibinfo
  {author} {\bibfnamefont {D.}~\bibnamefont {Tom{\'a}nek}},\ }\bibfield
  {title} {\enquote {\bibinfo {title} {The nature of the interlayer interaction
  in bulk and few-layer phosphorus},}\ }\href {\doibase
  10.1021/acs.nanolett.5b03615} {\bibfield  {journal} {\bibinfo  {journal}
  {Nano Letters}\ }\textbf {\bibinfo {volume} {15}},\ \bibinfo {pages}
  {8170--8175} (\bibinfo {year} {2015})}\BibitemShut {NoStop}%
\bibitem [{\citenamefont {Kadioglu}\ \emph {et~al.}(2018)\citenamefont
  {Kadioglu}, \citenamefont {Santana}, \citenamefont {Ozaydin}, \citenamefont
  {Ersan}, \citenamefont {Akturk}, \citenamefont {Akturk},\ and\ \citenamefont
  {Reboredo}}]{doi:10.1063/1.5026120}%
  \BibitemOpen
  \bibfield  {author} {\bibinfo {author} {\bibfnamefont {Y.}~\bibnamefont
  {Kadioglu}}, \bibinfo {author} {\bibfnamefont {J.~A.}\ \bibnamefont
  {Santana}}, \bibinfo {author} {\bibfnamefont {H.~D.}\ \bibnamefont
  {Ozaydin}}, \bibinfo {author} {\bibfnamefont {F.}~\bibnamefont {Ersan}},
  \bibinfo {author} {\bibfnamefont {O.~U.}\ \bibnamefont {Akturk}}, \bibinfo
  {author} {\bibfnamefont {E.}~\bibnamefont {Akturk}}, \ and\ \bibinfo {author}
  {\bibfnamefont {F.~A.}\ \bibnamefont {Reboredo}},\ }\bibfield  {title}
  {\enquote {\bibinfo {title} {{Diffusion Quantum Monte Carlo and density
  functional calculations of the structural stability of bilayer arsenene}},}\
  }\href {\doibase 10.1063/1.5026120} {\bibfield  {journal} {\bibinfo
  {journal} {The Journal of Chemical Physics}\ }\textbf {\bibinfo {volume}
  {148}},\ \bibinfo {pages} {214706} (\bibinfo {year} {2018})}\BibitemShut
  {NoStop}%
\bibitem [{\citenamefont {Szyniszewski}\ \emph
  {et~al.}(2017{\natexlab{a}})\citenamefont {Szyniszewski}, \citenamefont
  {Mostaani}, \citenamefont {Drummond},\ and\ \citenamefont
  {Fal'ko}}]{PhysRevB.95.081301}%
  \BibitemOpen
  \bibfield  {author} {\bibinfo {author} {\bibfnamefont {M.}~\bibnamefont
  {Szyniszewski}}, \bibinfo {author} {\bibfnamefont {E.}~\bibnamefont
  {Mostaani}}, \bibinfo {author} {\bibfnamefont {N.~D.}\ \bibnamefont
  {Drummond}}, \ and\ \bibinfo {author} {\bibfnamefont {V.~I.}\ \bibnamefont
  {Fal'ko}},\ }\bibfield  {title} {\enquote {\bibinfo {title} {{Binding
  energies of trions and biexcitons in two-dimensional semiconductors from
  Diffusion Quantum Monte Carlo calculations}},}\ }\href {\doibase
  10.1103/PhysRevB.95.081301} {\bibfield  {journal} {\bibinfo  {journal} {Phys.
  Rev. B}\ }\textbf {\bibinfo {volume} {95}},\ \bibinfo {pages} {081301}
  (\bibinfo {year} {2017}{\natexlab{a}})}\BibitemShut {NoStop}%
\bibitem [{\citenamefont {Szyniszewski}\ \emph
  {et~al.}(2017{\natexlab{b}})\citenamefont {Szyniszewski}, \citenamefont
  {Mostaani}, \citenamefont {Drummond},\ and\ \citenamefont
  {Fal'ko}}]{PhysRevB.96.119902}%
  \BibitemOpen
  \bibfield  {author} {\bibinfo {author} {\bibfnamefont {M.}~\bibnamefont
  {Szyniszewski}}, \bibinfo {author} {\bibfnamefont {E.}~\bibnamefont
  {Mostaani}}, \bibinfo {author} {\bibfnamefont {N.~D.}\ \bibnamefont
  {Drummond}}, \ and\ \bibinfo {author} {\bibfnamefont {V.~I.}\ \bibnamefont
  {Fal'ko}},\ }\bibfield  {title} {\enquote {\bibinfo {title} {{Erratum:
  Binding energies of trions and biexcitons in two-dimensional semiconductors
  from Diffusion Quantum Monte Carlo calculations [Phys. Rev. B 95, 081301(R)
  (2017)]}},}\ }\href {\doibase 10.1103/PhysRevB.96.119902} {\bibfield
  {journal} {\bibinfo  {journal} {Phys. Rev. B}\ }\textbf {\bibinfo {volume}
  {96}},\ \bibinfo {pages} {119902} (\bibinfo {year}
  {2017}{\natexlab{b}})}\BibitemShut {NoStop}%
\bibitem [{\citenamefont {Mostaani}\ \emph {et~al.}(2017)\citenamefont
  {Mostaani}, \citenamefont {Szyniszewski}, \citenamefont {Price},
  \citenamefont {Maezono}, \citenamefont {Danovich}, \citenamefont {Hunt},
  \citenamefont {Drummond},\ and\ \citenamefont {Fal'ko}}]{PhysRevB.96.075431}%
  \BibitemOpen
  \bibfield  {author} {\bibinfo {author} {\bibfnamefont {E.}~\bibnamefont
  {Mostaani}}, \bibinfo {author} {\bibfnamefont {M.}~\bibnamefont
  {Szyniszewski}}, \bibinfo {author} {\bibfnamefont {C.~H.}\ \bibnamefont
  {Price}}, \bibinfo {author} {\bibfnamefont {R.}~\bibnamefont {Maezono}},
  \bibinfo {author} {\bibfnamefont {M.}~\bibnamefont {Danovich}}, \bibinfo
  {author} {\bibfnamefont {R.~J.}\ \bibnamefont {Hunt}}, \bibinfo {author}
  {\bibfnamefont {N.~D.}\ \bibnamefont {Drummond}}, \ and\ \bibinfo {author}
  {\bibfnamefont {V.~I.}\ \bibnamefont {Fal'ko}},\ }\bibfield  {title}
  {\enquote {\bibinfo {title} {{Diffusion Quantum Monte Carlo study of
  excitonic complexes in two-dimensional transition-metal dichalcogenides}},}\
  }\href {\doibase 10.1103/PhysRevB.96.075431} {\bibfield  {journal} {\bibinfo
  {journal} {Phys. Rev. B}\ }\textbf {\bibinfo {volume} {96}},\ \bibinfo
  {pages} {075431} (\bibinfo {year} {2017})}\BibitemShut {NoStop}%
\bibitem [{\citenamefont {Busemeyer}, \citenamefont {MacDougall},\ and\
  \citenamefont {Wagner}(2019)}]{PhysRevB.99.081118}%
  \BibitemOpen
  \bibfield  {author} {\bibinfo {author} {\bibfnamefont {B.}~\bibnamefont
  {Busemeyer}}, \bibinfo {author} {\bibfnamefont {G.~J.}\ \bibnamefont
  {MacDougall}}, \ and\ \bibinfo {author} {\bibfnamefont {L.~K.}\ \bibnamefont
  {Wagner}},\ }\bibfield  {title} {\enquote {\bibinfo {title} {{Prediction for
  the singlet-triplet excitation energy for the spinel
  ${\mathrm{MgTi}}_{2}{\mathrm{O}}_{4}$ using first-principles Diffusion Monte
  Carlo}},}\ }\href {\doibase 10.1103/PhysRevB.99.081118} {\bibfield  {journal}
  {\bibinfo  {journal} {Phys. Rev. B}\ }\textbf {\bibinfo {volume} {99}},\
  \bibinfo {pages} {081118} (\bibinfo {year} {2019})}\BibitemShut {NoStop}%
\bibitem [{\citenamefont {Hunt}\ \emph {et~al.}(2020)\citenamefont {Hunt},
  \citenamefont {Monserrat}, \citenamefont {Z\'olyomi},\ and\ \citenamefont
  {Drummond}}]{PhysRevB.101.205115}%
  \BibitemOpen
  \bibfield  {author} {\bibinfo {author} {\bibfnamefont {R.~J.}\ \bibnamefont
  {Hunt}}, \bibinfo {author} {\bibfnamefont {B.}~\bibnamefont {Monserrat}},
  \bibinfo {author} {\bibfnamefont {V.}~\bibnamefont {Z\'olyomi}}, \ and\
  \bibinfo {author} {\bibfnamefont {N.~D.}\ \bibnamefont {Drummond}},\
  }\bibfield  {title} {\enquote {\bibinfo {title} {{Diffusion Quantum Monte
  Carlo and GW study of the electronic properties of monolayer and bulk
  hexagonal boron nitride}},}\ }\href {\doibase 10.1103/PhysRevB.101.205115}
  {\bibfield  {journal} {\bibinfo  {journal} {Phys. Rev. B}\ }\textbf {\bibinfo
  {volume} {101}},\ \bibinfo {pages} {205115} (\bibinfo {year}
  {2020})}\BibitemShut {NoStop}%
\bibitem [{\citenamefont {Saritas}\ \emph
  {et~al.}(2019{\natexlab{a}})\citenamefont {Saritas}, \citenamefont {Krogel},
  \citenamefont {Okamoto}, \citenamefont {Lee},\ and\ \citenamefont
  {Reboredo}}]{PhysRevMaterials.3.124414}%
  \BibitemOpen
  \bibfield  {author} {\bibinfo {author} {\bibfnamefont {K.}~\bibnamefont
  {Saritas}}, \bibinfo {author} {\bibfnamefont {J.~T.}\ \bibnamefont {Krogel}},
  \bibinfo {author} {\bibfnamefont {S.}~\bibnamefont {Okamoto}}, \bibinfo
  {author} {\bibfnamefont {H.~N.}\ \bibnamefont {Lee}}, \ and\ \bibinfo
  {author} {\bibfnamefont {F.~A.}\ \bibnamefont {Reboredo}},\ }\bibfield
  {title} {\enquote {\bibinfo {title} {{Structural, electronic, and magnetic
  properties of bulk and epitaxial ${\mathrm{LaCoO}}_{3}$ through Diffusion
  Monte Carlo}},}\ }\href {\doibase 10.1103/PhysRevMaterials.3.124414}
  {\bibfield  {journal} {\bibinfo  {journal} {Phys. Rev. Materials}\ }\textbf
  {\bibinfo {volume} {3}},\ \bibinfo {pages} {124414} (\bibinfo {year}
  {2019}{\natexlab{a}})}\BibitemShut {NoStop}%
\bibitem [{\citenamefont {Horv\'athov\'a}\ \emph {et~al.}(2012)\citenamefont
  {Horv\'athov\'a}, \citenamefont {Dubeck\'y}, \citenamefont {Mitas},\ and\
  \citenamefont {\ifmmode~\check{S}\else
  \v{S}\fi{}tich}}]{PhysRevLett.109.053001}%
  \BibitemOpen
  \bibfield  {author} {\bibinfo {author} {\bibfnamefont {L.}~\bibnamefont
  {Horv\'athov\'a}}, \bibinfo {author} {\bibfnamefont {M.}~\bibnamefont
  {Dubeck\'y}}, \bibinfo {author} {\bibfnamefont {L.}~\bibnamefont {Mitas}}, \
  and\ \bibinfo {author} {\bibfnamefont {I.}~\bibnamefont
  {\ifmmode~\check{S}\else \v{S}\fi{}tich}},\ }\bibfield  {title} {\enquote
  {\bibinfo {title} {Spin multiplicity and symmetry breaking in
  vanadium-benzene complexes},}\ }\href {\doibase
  10.1103/PhysRevLett.109.053001} {\bibfield  {journal} {\bibinfo  {journal}
  {Phys. Rev. Lett.}\ }\textbf {\bibinfo {volume} {109}},\ \bibinfo {pages}
  {053001} (\bibinfo {year} {2012})}\BibitemShut {NoStop}%
\bibitem [{\citenamefont {{\v R}ez{\'a}{\v c}}, \citenamefont {Riley},\ and\
  \citenamefont {Hobza}(2012)}]{nanopart}%
  \BibitemOpen
  \bibfield  {author} {\bibinfo {author} {\bibfnamefont {J.}~\bibnamefont {{\v
  R}ez{\'a}{\v c}}}, \bibinfo {author} {\bibfnamefont {K.~E.}\ \bibnamefont
  {Riley}}, \ and\ \bibinfo {author} {\bibfnamefont {P.}~\bibnamefont
  {Hobza}},\ }\bibfield  {title} {\enquote {\bibinfo {title} {Benchmark
  calculations of noncovalent interactions of halogenated molecules},}\ }\href
  {\doibase 10.1021/ct300647k} {\bibfield  {journal} {\bibinfo  {journal}
  {Journal of Chemical Theory and Computation}\ }\textbf {\bibinfo {volume}
  {8}},\ \bibinfo {pages} {4285--4292} (\bibinfo {year} {2012})}\BibitemShut
  {NoStop}%
\bibitem [{\citenamefont {Frank}\ \emph {et~al.}(2019)\citenamefont {Frank},
  \citenamefont {Derian}, \citenamefont {Tok\'ar}, \citenamefont {Mitas},
  \citenamefont {Fabian},\ and\ \citenamefont {\ifmmode~\check{S}\else
  \v{S}\fi{}tich}}]{PhysRevX.9.011018}%
  \BibitemOpen
  \bibfield  {author} {\bibinfo {author} {\bibfnamefont {T.}~\bibnamefont
  {Frank}}, \bibinfo {author} {\bibfnamefont {R.}~\bibnamefont {Derian}},
  \bibinfo {author} {\bibfnamefont {K.}~\bibnamefont {Tok\'ar}}, \bibinfo
  {author} {\bibfnamefont {L.}~\bibnamefont {Mitas}}, \bibinfo {author}
  {\bibfnamefont {J.}~\bibnamefont {Fabian}}, \ and\ \bibinfo {author}
  {\bibfnamefont {I.}~\bibnamefont {\ifmmode~\check{S}\else \v{S}\fi{}tich}},\
  }\bibfield  {title} {\enquote {\bibinfo {title} {Many-body {Quantum Monte
  Carlo} study of {2D} materials: Cohesion and band gap in single-layer
  phosphorene},}\ }\href {\doibase 10.1103/PhysRevX.9.011018} {\bibfield
  {journal} {\bibinfo  {journal} {Phys. Rev. X}\ }\textbf {\bibinfo {volume}
  {9}},\ \bibinfo {pages} {011018} (\bibinfo {year} {2019})}\BibitemShut
  {NoStop}%
\bibitem [{\citenamefont {Shin}\ \emph {et~al.}(2020)\citenamefont {Shin},
  \citenamefont {Krogel}, \citenamefont {Kent}, \citenamefont {Benali},\ and\
  \citenamefont {Heinonen}}]{shin_krogel_kent_benali_heinonen_2020}%
  \BibitemOpen
  \bibfield  {author} {\bibinfo {author} {\bibfnamefont {H.}~\bibnamefont
  {Shin}}, \bibinfo {author} {\bibfnamefont {J.}~\bibnamefont {Krogel}},
  \bibinfo {author} {\bibfnamefont {P.}~\bibnamefont {Kent}}, \bibinfo {author}
  {\bibfnamefont {A.}~\bibnamefont {Benali}}, \ and\ \bibinfo {author}
  {\bibfnamefont {O.}~\bibnamefont {Heinonen}},\ }\bibfield  {title} {\enquote
  {\bibinfo {title} {{Structural and optical properties of bulk and monolayer
  GeSe : A Quantum Monte Carlo Study}},}\ }\href@noop {} {\bibfield  {journal}
  {\bibinfo  {journal} {APS March Meeting Presentation}\ ,\ \bibinfo {pages}
  {Abstract: R57.00008}} (\bibinfo {year} {2020})}\BibitemShut {NoStop}%
\bibitem [{\citenamefont {Koloren\ifmmode~\check{c}\else \v{c}\fi{}},
  \citenamefont {Hu},\ and\ \citenamefont {Mitas}(2010)}]{PhysRevB.82.115108}%
  \BibitemOpen
  \bibfield  {author} {\bibinfo {author} {\bibfnamefont {J.~c.~v.}\
  \bibnamefont {Koloren\ifmmode~\check{c}\else \v{c}\fi{}}}, \bibinfo {author}
  {\bibfnamefont {S.}~\bibnamefont {Hu}}, \ and\ \bibinfo {author}
  {\bibfnamefont {L.}~\bibnamefont {Mitas}},\ }\bibfield  {title} {\enquote
  {\bibinfo {title} {{Wave functions for Quantum Monte Carlo calculations in
  solids: Orbitals from density functional theory with hybrid
  exchange-correlation functionals}},}\ }\href {\doibase
  10.1103/PhysRevB.82.115108} {\bibfield  {journal} {\bibinfo  {journal} {Phys.
  Rev. B}\ }\textbf {\bibinfo {volume} {82}},\ \bibinfo {pages} {115108}
  (\bibinfo {year} {2010})}\BibitemShut {NoStop}%
\bibitem [{\citenamefont {Yu}, \citenamefont {Wagner},\ and\ \citenamefont
  {Ertekin}(2015)}]{doi:10.1063/1.4937421}%
  \BibitemOpen
  \bibfield  {author} {\bibinfo {author} {\bibfnamefont {J.}~\bibnamefont
  {Yu}}, \bibinfo {author} {\bibfnamefont {L.~K.}\ \bibnamefont {Wagner}}, \
  and\ \bibinfo {author} {\bibfnamefont {E.}~\bibnamefont {Ertekin}},\
  }\bibfield  {title} {\enquote {\bibinfo {title} {{Towards a systematic
  assessment of errors in Diffusion Monte Carlo calculations of semiconductors:
  Case study of zinc selenide and zinc oxide}},}\ }\href {\doibase
  10.1063/1.4937421} {\bibfield  {journal} {\bibinfo  {journal} {The Journal of
  Chemical Physics}\ }\textbf {\bibinfo {volume} {143}},\ \bibinfo {pages}
  {224707} (\bibinfo {year} {2015})}\BibitemShut {NoStop}%
\bibitem [{\citenamefont {Needs}\ and\ \citenamefont
  {Towler}(2003)}]{doi:10.1142/S0217979203020533}%
  \BibitemOpen
  \bibfield  {author} {\bibinfo {author} {\bibfnamefont {R.~J.}\ \bibnamefont
  {Needs}}\ and\ \bibinfo {author} {\bibfnamefont {M.~D.}\ \bibnamefont
  {Towler}},\ }\bibfield  {title} {\enquote {\bibinfo {title} {{The Diffusion
  Quantum Monte Carlo Method: Designing Trial Wave Functions for NiO}},}\
  }\href {\doibase 10.1142/S0217979203020533} {\bibfield  {journal} {\bibinfo
  {journal} {International Journal of Modern Physics B}\ }\textbf {\bibinfo
  {volume} {17}},\ \bibinfo {pages} {5425--5434} (\bibinfo {year} {2003})},\
  \Eprint {http://arxiv.org/abs/https://doi.org/10.1142/S0217979203020533}
  {https://doi.org/10.1142/S0217979203020533} \BibitemShut {NoStop}%
\bibitem [{\citenamefont {Mitra}\ \emph {et~al.}(2015)\citenamefont {Mitra},
  \citenamefont {Krogel}, \citenamefont {Santana},\ and\ \citenamefont
  {Reboredo}}]{doi:10.1063/1.4934262}%
  \BibitemOpen
  \bibfield  {author} {\bibinfo {author} {\bibfnamefont {C.}~\bibnamefont
  {Mitra}}, \bibinfo {author} {\bibfnamefont {J.~T.}\ \bibnamefont {Krogel}},
  \bibinfo {author} {\bibfnamefont {J.~A.}\ \bibnamefont {Santana}}, \ and\
  \bibinfo {author} {\bibfnamefont {F.~A.}\ \bibnamefont {Reboredo}},\
  }\bibfield  {title} {\enquote {\bibinfo {title} {{Many-body ab initio
  Diffusion Quantum Monte Carlo applied to the strongly correlated oxide
  NiO}},}\ }\href {\doibase 10.1063/1.4934262} {\bibfield  {journal} {\bibinfo
  {journal} {The Journal of Chemical Physics}\ }\textbf {\bibinfo {volume}
  {143}},\ \bibinfo {pages} {164710} (\bibinfo {year} {2015})}\BibitemShut
  {NoStop}%
\bibitem [{\citenamefont {Schiller}, \citenamefont {Wagner},\ and\
  \citenamefont {Ertekin}(2015)}]{PhysRevB.92.235209}%
  \BibitemOpen
  \bibfield  {author} {\bibinfo {author} {\bibfnamefont {J.~A.}\ \bibnamefont
  {Schiller}}, \bibinfo {author} {\bibfnamefont {L.~K.}\ \bibnamefont
  {Wagner}}, \ and\ \bibinfo {author} {\bibfnamefont {E.}~\bibnamefont
  {Ertekin}},\ }\bibfield  {title} {\enquote {\bibinfo {title} {{Phase
  stability and properties of manganese oxide polymorphs: Assessment and
  insights from Diffusion Monte Carlo}},}\ }\href {\doibase
  10.1103/PhysRevB.92.235209} {\bibfield  {journal} {\bibinfo  {journal} {Phys.
  Rev. B}\ }\textbf {\bibinfo {volume} {92}},\ \bibinfo {pages} {235209}
  (\bibinfo {year} {2015})}\BibitemShut {NoStop}%
\bibitem [{\citenamefont {Saritas}\ \emph {et~al.}(2018)\citenamefont
  {Saritas}, \citenamefont {Krogel}, \citenamefont {Kent},\ and\ \citenamefont
  {Reboredo}}]{PhysRevMaterials.2.085801}%
  \BibitemOpen
  \bibfield  {author} {\bibinfo {author} {\bibfnamefont {K.}~\bibnamefont
  {Saritas}}, \bibinfo {author} {\bibfnamefont {J.~T.}\ \bibnamefont {Krogel}},
  \bibinfo {author} {\bibfnamefont {P.~R.~C.}\ \bibnamefont {Kent}}, \ and\
  \bibinfo {author} {\bibfnamefont {F.~A.}\ \bibnamefont {Reboredo}},\
  }\bibfield  {title} {\enquote {\bibinfo {title} {{Diffusion Monte Carlo: A
  pathway towards an accurate theoretical description of manganese oxides}},}\
  }\href {\doibase 10.1103/PhysRevMaterials.2.085801} {\bibfield  {journal}
  {\bibinfo  {journal} {Phys. Rev. Materials}\ }\textbf {\bibinfo {volume}
  {2}},\ \bibinfo {pages} {085801} (\bibinfo {year} {2018})}\BibitemShut
  {NoStop}%
\bibitem [{\citenamefont {Saritas}\ \emph {et~al.}(2017)\citenamefont
  {Saritas}, \citenamefont {Mueller}, \citenamefont {Wagner},\ and\
  \citenamefont {Grossman}}]{formation-qmc}%
  \BibitemOpen
  \bibfield  {author} {\bibinfo {author} {\bibfnamefont {K.}~\bibnamefont
  {Saritas}}, \bibinfo {author} {\bibfnamefont {T.}~\bibnamefont {Mueller}},
  \bibinfo {author} {\bibfnamefont {L.}~\bibnamefont {Wagner}}, \ and\ \bibinfo
  {author} {\bibfnamefont {J.~C.}\ \bibnamefont {Grossman}},\ }\bibfield
  {title} {\enquote {\bibinfo {title} {Investigation of a {Quantum Monte Carlo}
  protocol to achieve high accuracy and high-throughput materials formation
  energies},}\ }\href {\doibase 10.1021/acs.jctc.6b01179} {\bibfield  {journal}
  {\bibinfo  {journal} {Journal of Chemical Theory and Computation}\ }\textbf
  {\bibinfo {volume} {13}},\ \bibinfo {pages} {1943--1951} (\bibinfo {year}
  {2017})}\BibitemShut {NoStop}%
\bibitem [{\citenamefont {Trail}\ \emph {et~al.}(2017)\citenamefont {Trail},
  \citenamefont {Monserrat}, \citenamefont {L\'opez~R\'{\i}os}, \citenamefont
  {Maezono},\ and\ \citenamefont {Needs}}]{PhysRevB.95.121108}%
  \BibitemOpen
  \bibfield  {author} {\bibinfo {author} {\bibfnamefont {J.}~\bibnamefont
  {Trail}}, \bibinfo {author} {\bibfnamefont {B.}~\bibnamefont {Monserrat}},
  \bibinfo {author} {\bibfnamefont {P.}~\bibnamefont {L\'opez~R\'{\i}os}},
  \bibinfo {author} {\bibfnamefont {R.}~\bibnamefont {Maezono}}, \ and\
  \bibinfo {author} {\bibfnamefont {R.~J.}\ \bibnamefont {Needs}},\ }\bibfield
  {title} {\enquote {\bibinfo {title} {{Quantum Monte Carlo study of the
  energetics of the rutile, anatase, brookite, and columbite
  ${\mathrm{TiO}}_{2}$ polymorphs}},}\ }\href {\doibase
  10.1103/PhysRevB.95.121108} {\bibfield  {journal} {\bibinfo  {journal} {Phys.
  Rev. B}\ }\textbf {\bibinfo {volume} {95}},\ \bibinfo {pages} {121108}
  (\bibinfo {year} {2017})}\BibitemShut {NoStop}%
\bibitem [{\citenamefont {Luo}\ \emph {et~al.}(2016)\citenamefont {Luo},
  \citenamefont {Benali}, \citenamefont {Shulenburger}, \citenamefont {Krogel},
  \citenamefont {Heinonen},\ and\ \citenamefont {Kent}}]{Luo_2016}%
  \BibitemOpen
  \bibfield  {author} {\bibinfo {author} {\bibfnamefont {Y.}~\bibnamefont
  {Luo}}, \bibinfo {author} {\bibfnamefont {A.}~\bibnamefont {Benali}},
  \bibinfo {author} {\bibfnamefont {L.}~\bibnamefont {Shulenburger}}, \bibinfo
  {author} {\bibfnamefont {J.~T.}\ \bibnamefont {Krogel}}, \bibinfo {author}
  {\bibfnamefont {O.}~\bibnamefont {Heinonen}}, \ and\ \bibinfo {author}
  {\bibfnamefont {P.~R.~C.}\ \bibnamefont {Kent}},\ }\bibfield  {title}
  {\enquote {\bibinfo {title} {{Phase stability of TiO$_2$ polymorphs from
  Diffusion Quantum Monte Carlo}},}\ }\href {\doibase
  10.1088/1367-2630/18/11/113049} {\bibfield  {journal} {\bibinfo  {journal}
  {New Journal of Physics}\ }\textbf {\bibinfo {volume} {18}},\ \bibinfo
  {pages} {113049} (\bibinfo {year} {2016})}\BibitemShut {NoStop}%
\bibitem [{\citenamefont {Benali}\ \emph {et~al.}(2016)\citenamefont {Benali},
  \citenamefont {Shulenburger}, \citenamefont {Krogel}, \citenamefont {Zhong},
  \citenamefont {Kent},\ and\ \citenamefont {Heinonen}}]{C6CP02067D}%
  \BibitemOpen
  \bibfield  {author} {\bibinfo {author} {\bibfnamefont {A.}~\bibnamefont
  {Benali}}, \bibinfo {author} {\bibfnamefont {L.}~\bibnamefont
  {Shulenburger}}, \bibinfo {author} {\bibfnamefont {J.~T.}\ \bibnamefont
  {Krogel}}, \bibinfo {author} {\bibfnamefont {X.}~\bibnamefont {Zhong}},
  \bibinfo {author} {\bibfnamefont {P.~R.~C.}\ \bibnamefont {Kent}}, \ and\
  \bibinfo {author} {\bibfnamefont {O.}~\bibnamefont {Heinonen}},\ }\bibfield
  {title} {\enquote {\bibinfo {title} {{Quantum Monte Carlo analysis of a
  charge ordered insulating antiferromagnet: the Ti$_4$O$_7$ Magnéli
  phase}},}\ }\href {\doibase 10.1039/C6CP02067D} {\bibfield  {journal}
  {\bibinfo  {journal} {Phys. Chem. Chem. Phys.}\ }\textbf {\bibinfo {volume}
  {18}},\ \bibinfo {pages} {18323--18335} (\bibinfo {year} {2016})}\BibitemShut
  {NoStop}%
\bibitem [{\citenamefont {Zheng}\ and\ \citenamefont
  {Wagner}(2015)}]{PhysRevLett.114.176401}%
  \BibitemOpen
  \bibfield  {author} {\bibinfo {author} {\bibfnamefont {H.}~\bibnamefont
  {Zheng}}\ and\ \bibinfo {author} {\bibfnamefont {L.~K.}\ \bibnamefont
  {Wagner}},\ }\bibfield  {title} {\enquote {\bibinfo {title} {Computation of
  the correlated metal-insulator transition in vanadium dioxide from first
  principles},}\ }\href {\doibase 10.1103/PhysRevLett.114.176401} {\bibfield
  {journal} {\bibinfo  {journal} {Phys. Rev. Lett.}\ }\textbf {\bibinfo
  {volume} {114}},\ \bibinfo {pages} {176401} (\bibinfo {year}
  {2015})}\BibitemShut {NoStop}%
\bibitem [{\citenamefont {Kyl\"anp\"a\"a}\ \emph {et~al.}(2017)\citenamefont
  {Kyl\"anp\"a\"a}, \citenamefont {Balachandran}, \citenamefont {Ganesh},
  \citenamefont {Heinonen}, \citenamefont {Kent},\ and\ \citenamefont
  {Krogel}}]{PhysRevMaterials.1.065408}%
  \BibitemOpen
  \bibfield  {author} {\bibinfo {author} {\bibfnamefont {I.}~\bibnamefont
  {Kyl\"anp\"a\"a}}, \bibinfo {author} {\bibfnamefont {J.}~\bibnamefont
  {Balachandran}}, \bibinfo {author} {\bibfnamefont {P.}~\bibnamefont
  {Ganesh}}, \bibinfo {author} {\bibfnamefont {O.}~\bibnamefont {Heinonen}},
  \bibinfo {author} {\bibfnamefont {P.~R.~C.}\ \bibnamefont {Kent}}, \ and\
  \bibinfo {author} {\bibfnamefont {J.~T.}\ \bibnamefont {Krogel}},\ }\bibfield
   {title} {\enquote {\bibinfo {title} {{Accuracy of ab initio electron
  correlation and electron densities in vanadium dioxide}},}\ }\href {\doibase
  10.1103/PhysRevMaterials.1.065408} {\bibfield  {journal} {\bibinfo  {journal}
  {Phys. Rev. Materials}\ }\textbf {\bibinfo {volume} {1}},\ \bibinfo {pages}
  {065408} (\bibinfo {year} {2017})}\BibitemShut {NoStop}%
\bibitem [{\citenamefont {Yu}, \citenamefont {Wagner},\ and\ \citenamefont
  {Ertekin}(2017)}]{PhysRevB.95.075209}%
  \BibitemOpen
  \bibfield  {author} {\bibinfo {author} {\bibfnamefont {J.}~\bibnamefont
  {Yu}}, \bibinfo {author} {\bibfnamefont {L.~K.}\ \bibnamefont {Wagner}}, \
  and\ \bibinfo {author} {\bibfnamefont {E.}~\bibnamefont {Ertekin}},\
  }\bibfield  {title} {\enquote {\bibinfo {title} {{Fixed-node Diffusion Monte
  Carlo description of nitrogen defects in zinc oxide}},}\ }\href {\doibase
  10.1103/PhysRevB.95.075209} {\bibfield  {journal} {\bibinfo  {journal} {Phys.
  Rev. B}\ }\textbf {\bibinfo {volume} {95}},\ \bibinfo {pages} {075209}
  (\bibinfo {year} {2017})}\BibitemShut {NoStop}%
\bibitem [{\citenamefont {Santana}\ \emph {et~al.}(2015)\citenamefont
  {Santana}, \citenamefont {Krogel}, \citenamefont {Kim}, \citenamefont
  {Kent},\ and\ \citenamefont {Reboredo}}]{doi:10.1063/1.4919242}%
  \BibitemOpen
  \bibfield  {author} {\bibinfo {author} {\bibfnamefont {J.~A.}\ \bibnamefont
  {Santana}}, \bibinfo {author} {\bibfnamefont {J.~T.}\ \bibnamefont {Krogel}},
  \bibinfo {author} {\bibfnamefont {J.}~\bibnamefont {Kim}}, \bibinfo {author}
  {\bibfnamefont {P.~R.~C.}\ \bibnamefont {Kent}}, \ and\ \bibinfo {author}
  {\bibfnamefont {F.~A.}\ \bibnamefont {Reboredo}},\ }\bibfield  {title}
  {\enquote {\bibinfo {title} {{Structural stability and defect energetics of
  ZnO from Diffusion quantum Monte Carlo}},}\ }\href {\doibase
  10.1063/1.4919242} {\bibfield  {journal} {\bibinfo  {journal} {The Journal of
  Chemical Physics}\ }\textbf {\bibinfo {volume} {142}},\ \bibinfo {pages}
  {164705} (\bibinfo {year} {2015})}\BibitemShut {NoStop}%
\bibitem [{\citenamefont {Shin}\ \emph {et~al.}(2017)\citenamefont {Shin},
  \citenamefont {Luo}, \citenamefont {Ganesh}, \citenamefont {Balachandran},
  \citenamefont {Krogel}, \citenamefont {Kent}, \citenamefont {Benali},\ and\
  \citenamefont {Heinonen}}]{PhysRevMaterials.1.073603}%
  \BibitemOpen
  \bibfield  {author} {\bibinfo {author} {\bibfnamefont {H.}~\bibnamefont
  {Shin}}, \bibinfo {author} {\bibfnamefont {Y.}~\bibnamefont {Luo}}, \bibinfo
  {author} {\bibfnamefont {P.}~\bibnamefont {Ganesh}}, \bibinfo {author}
  {\bibfnamefont {J.}~\bibnamefont {Balachandran}}, \bibinfo {author}
  {\bibfnamefont {J.~T.}\ \bibnamefont {Krogel}}, \bibinfo {author}
  {\bibfnamefont {P.~R.~C.}\ \bibnamefont {Kent}}, \bibinfo {author}
  {\bibfnamefont {A.}~\bibnamefont {Benali}}, \ and\ \bibinfo {author}
  {\bibfnamefont {O.}~\bibnamefont {Heinonen}},\ }\bibfield  {title} {\enquote
  {\bibinfo {title} {{Electronic properties of doped and defective NiO: A
  Quantum Monte Carlo study}},}\ }\href {\doibase
  10.1103/PhysRevMaterials.1.073603} {\bibfield  {journal} {\bibinfo  {journal}
  {Phys. Rev. Materials}\ }\textbf {\bibinfo {volume} {1}},\ \bibinfo {pages}
  {073603} (\bibinfo {year} {2017})}\BibitemShut {NoStop}%
\bibitem [{\citenamefont {Shin}\ \emph {et~al.}(2018)\citenamefont {Shin},
  \citenamefont {Benali}, \citenamefont {Luo}, \citenamefont {Crabb},
  \citenamefont {Lopez-Bezanilla}, \citenamefont {Ratcliff}, \citenamefont
  {Jokisaari},\ and\ \citenamefont {Heinonen}}]{PhysRevMaterials.2.075001}%
  \BibitemOpen
  \bibfield  {author} {\bibinfo {author} {\bibfnamefont {H.}~\bibnamefont
  {Shin}}, \bibinfo {author} {\bibfnamefont {A.}~\bibnamefont {Benali}},
  \bibinfo {author} {\bibfnamefont {Y.}~\bibnamefont {Luo}}, \bibinfo {author}
  {\bibfnamefont {E.}~\bibnamefont {Crabb}}, \bibinfo {author} {\bibfnamefont
  {A.}~\bibnamefont {Lopez-Bezanilla}}, \bibinfo {author} {\bibfnamefont
  {L.~E.}\ \bibnamefont {Ratcliff}}, \bibinfo {author} {\bibfnamefont {A.~M.}\
  \bibnamefont {Jokisaari}}, \ and\ \bibinfo {author} {\bibfnamefont
  {O.}~\bibnamefont {Heinonen}},\ }\bibfield  {title} {\enquote {\bibinfo
  {title} {{Zirconia and hafnia polymorphs: Ground-state structural properties
  from Diffusion Monte Carlo}},}\ }\href {\doibase
  10.1103/PhysRevMaterials.2.075001} {\bibfield  {journal} {\bibinfo  {journal}
  {Phys. Rev. Materials}\ }\textbf {\bibinfo {volume} {2}},\ \bibinfo {pages}
  {075001} (\bibinfo {year} {2018})}\BibitemShut {NoStop}%
\bibitem [{\citenamefont {Saritas}, \citenamefont {Krogel},\ and\ \citenamefont
  {Reboredo}(2018)}]{PhysRevB.98.155130}%
  \BibitemOpen
  \bibfield  {author} {\bibinfo {author} {\bibfnamefont {K.}~\bibnamefont
  {Saritas}}, \bibinfo {author} {\bibfnamefont {J.~T.}\ \bibnamefont {Krogel}},
  \ and\ \bibinfo {author} {\bibfnamefont {F.~A.}\ \bibnamefont {Reboredo}},\
  }\bibfield  {title} {\enquote {\bibinfo {title} {{Relative energies and
  electronic structures of CoO polymorphs through ab initio Diffusion Quantum
  Monte Carlo}},}\ }\href {\doibase 10.1103/PhysRevB.98.155130} {\bibfield
  {journal} {\bibinfo  {journal} {Phys. Rev. B}\ }\textbf {\bibinfo {volume}
  {98}},\ \bibinfo {pages} {155130} (\bibinfo {year} {2018})}\BibitemShut
  {NoStop}%
\bibitem [{\citenamefont {Saritas}\ \emph
  {et~al.}(2019{\natexlab{b}})\citenamefont {Saritas}, \citenamefont {Ming},
  \citenamefont {Du},\ and\ \citenamefont {Reboredo}}]{phosphors}%
  \BibitemOpen
  \bibfield  {author} {\bibinfo {author} {\bibfnamefont {K.}~\bibnamefont
  {Saritas}}, \bibinfo {author} {\bibfnamefont {W.}~\bibnamefont {Ming}},
  \bibinfo {author} {\bibfnamefont {M.-H.}\ \bibnamefont {Du}}, \ and\ \bibinfo
  {author} {\bibfnamefont {F.~A.}\ \bibnamefont {Reboredo}},\ }\bibfield
  {title} {\enquote {\bibinfo {title} {Excitation energies of localized
  correlated defects via {Quantum Monte Carlo}: {A} case study of
  {Mn}$^{4+}$-doped phosphors},}\ }\href {\doibase 10.1021/acs.jpclett.8b03015}
  {\bibfield  {journal} {\bibinfo  {journal} {The Journal of Physical Chemistry
  Letters}\ }\textbf {\bibinfo {volume} {10}},\ \bibinfo {pages} {67--74}
  (\bibinfo {year} {2019}{\natexlab{b}})}\BibitemShut {NoStop}%
\bibitem [{\citenamefont {Saritas}\ \emph {et~al.}(2020)\citenamefont
  {Saritas}, \citenamefont {Fadel}, \citenamefont {Kozinsky},\ and\
  \citenamefont {Grossman}}]{LiNiO2}%
  \BibitemOpen
  \bibfield  {author} {\bibinfo {author} {\bibfnamefont {K.}~\bibnamefont
  {Saritas}}, \bibinfo {author} {\bibfnamefont {E.~R.}\ \bibnamefont {Fadel}},
  \bibinfo {author} {\bibfnamefont {B.}~\bibnamefont {Kozinsky}}, \ and\
  \bibinfo {author} {\bibfnamefont {J.~C.}\ \bibnamefont {Grossman}},\
  }\bibfield  {title} {\enquote {\bibinfo {title} {Charge density and redox
  potential of {LiNiO$_2$} using ab initio {Diffusion Quantum Monte Carlo}},}\
  }\href {\doibase 10.1021/acs.jpcc.9b10372} {\bibfield  {journal} {\bibinfo
  {journal} {The Journal of Physical Chemistry C}\ }\textbf {\bibinfo {volume}
  {124}},\ \bibinfo {pages} {5893--5901} (\bibinfo {year} {2020})}\BibitemShut
  {NoStop}%
\bibitem [{\citenamefont {Kresse}\ and\ \citenamefont
  {Furthm\"uller}(1996)}]{PhysRevB.54.11169}%
  \BibitemOpen
  \bibfield  {author} {\bibinfo {author} {\bibfnamefont {G.}~\bibnamefont
  {Kresse}}\ and\ \bibinfo {author} {\bibfnamefont {J.}~\bibnamefont
  {Furthm\"uller}},\ }\bibfield  {title} {\enquote {\bibinfo {title}
  {{Efficient iterative schemes for ab initio total-energy calculations using a
  plane-wave basis set}},}\ }\href {\doibase 10.1103/PhysRevB.54.11169}
  {\bibfield  {journal} {\bibinfo  {journal} {Phys. Rev. B}\ }\textbf {\bibinfo
  {volume} {54}},\ \bibinfo {pages} {11169--11186} (\bibinfo {year}
  {1996})}\BibitemShut {NoStop}%
\bibitem [{\citenamefont {Kresse}\ and\ \citenamefont
  {Joubert}(1999)}]{PhysRevB.59.1758}%
  \BibitemOpen
  \bibfield  {author} {\bibinfo {author} {\bibfnamefont {G.}~\bibnamefont
  {Kresse}}\ and\ \bibinfo {author} {\bibfnamefont {D.}~\bibnamefont
  {Joubert}},\ }\bibfield  {title} {\enquote {\bibinfo {title} {{From ultrasoft
  pseudopotentials to the projector augmented-wave method}},}\ }\href {\doibase
  10.1103/PhysRevB.59.1758} {\bibfield  {journal} {\bibinfo  {journal} {Phys.
  Rev. B}\ }\textbf {\bibinfo {volume} {59}},\ \bibinfo {pages} {1758--1775}
  (\bibinfo {year} {1999})}\BibitemShut {NoStop}%
\bibitem [{\citenamefont {Dancoff}(1950)}]{PhysRev.78.382}%
  \BibitemOpen
  \bibfield  {author} {\bibinfo {author} {\bibfnamefont {S.~M.}\ \bibnamefont
  {Dancoff}},\ }\bibfield  {title} {\enquote {\bibinfo {title} {Non-adiabatic
  meson theory of nuclear forces},}\ }\href {\doibase 10.1103/PhysRev.78.382}
  {\bibfield  {journal} {\bibinfo  {journal} {Phys. Rev.}\ }\textbf {\bibinfo
  {volume} {78}},\ \bibinfo {pages} {382--385} (\bibinfo {year}
  {1950})}\BibitemShut {NoStop}%
\bibitem [{\citenamefont {Foulkes}\ \emph {et~al.}(2001)\citenamefont
  {Foulkes}, \citenamefont {Mitas}, \citenamefont {Needs},\ and\ \citenamefont
  {Rajagopal}}]{RevModPhys.73.33}%
  \BibitemOpen
  \bibfield  {author} {\bibinfo {author} {\bibfnamefont {W.~M.~C.}\
  \bibnamefont {Foulkes}}, \bibinfo {author} {\bibfnamefont {L.}~\bibnamefont
  {Mitas}}, \bibinfo {author} {\bibfnamefont {R.~J.}\ \bibnamefont {Needs}}, \
  and\ \bibinfo {author} {\bibfnamefont {G.}~\bibnamefont {Rajagopal}},\
  }\bibfield  {title} {\enquote {\bibinfo {title} {{Quantum Monte Carlo
  simulations of solids}},}\ }\href {\doibase 10.1103/RevModPhys.73.33}
  {\bibfield  {journal} {\bibinfo  {journal} {Rev. Mod. Phys.}\ }\textbf
  {\bibinfo {volume} {73}},\ \bibinfo {pages} {33--83} (\bibinfo {year}
  {2001})}\BibitemShut {NoStop}%
\bibitem [{\citenamefont {Needs}\ \emph {et~al.}(2009)\citenamefont {Needs},
  \citenamefont {Towler}, \citenamefont {Drummond},\ and\ \citenamefont
  {R{\'{\i}}os}}]{Needs_2009}%
  \BibitemOpen
  \bibfield  {author} {\bibinfo {author} {\bibfnamefont {R.~J.}\ \bibnamefont
  {Needs}}, \bibinfo {author} {\bibfnamefont {M.~D.}\ \bibnamefont {Towler}},
  \bibinfo {author} {\bibfnamefont {N.~D.}\ \bibnamefont {Drummond}}, \ and\
  \bibinfo {author} {\bibfnamefont {P.~L.}\ \bibnamefont {R{\'{\i}}os}},\
  }\bibfield  {title} {\enquote {\bibinfo {title} {{Continuum Variational and
  Diffusion Quantum Monte Carlo calculations}},}\ }\href {\doibase
  10.1088/0953-8984/22/2/023201} {\bibfield  {journal} {\bibinfo  {journal}
  {Journal of Physics: Condensed Matter}\ }\textbf {\bibinfo {volume} {22}},\
  \bibinfo {pages} {023201} (\bibinfo {year} {2009})}\BibitemShut {NoStop}%
\bibitem [{\citenamefont {Kim}\ \emph {et~al.}(2018)\citenamefont {Kim},
  \citenamefont {Baczewski}, \citenamefont {Beaudet}, \citenamefont {Benali},
  \citenamefont {Bennett}, \citenamefont {Berrill}, \citenamefont {Blunt},
  \citenamefont {Borda}, \citenamefont {Casula}, \citenamefont {Ceperley},
  \citenamefont {Chiesa}, \citenamefont {Clark}, \citenamefont {Clay},
  \citenamefont {Delaney}, \citenamefont {Dewing}, \citenamefont {Esler},
  \citenamefont {Hao}, \citenamefont {Heinonen}, \citenamefont {Kent},
  \citenamefont {Krogel}, \citenamefont {Kylänpää}, \citenamefont {Li},
  \citenamefont {Lopez}, \citenamefont {Luo}, \citenamefont {Malone},
  \citenamefont {Martin}, \citenamefont {Mathuriya}, \citenamefont {McMinis},
  \citenamefont {Melton}, \citenamefont {Mitas}, \citenamefont {Morales},
  \citenamefont {Neuscamman}, \citenamefont {Parker}, \citenamefont {Flores},
  \citenamefont {Romero}, \citenamefont {Rubenstein}, \citenamefont {Shea},
  \citenamefont {Shin}, \citenamefont {Shulenburger}, \citenamefont {Tillack},
  \citenamefont {Townsend}, \citenamefont {Tubman}, \citenamefont {Goetz},
  \citenamefont {Vincent}, \citenamefont {Yang}, \citenamefont {Yang},
  \citenamefont {Zhang},\ and\ \citenamefont {Zhao}}]{Kim_2018}%
  \BibitemOpen
  \bibfield  {author} {\bibinfo {author} {\bibfnamefont {J.}~\bibnamefont
  {Kim}}, \bibinfo {author} {\bibfnamefont {A.~D.}\ \bibnamefont {Baczewski}},
  \bibinfo {author} {\bibfnamefont {T.~D.}\ \bibnamefont {Beaudet}}, \bibinfo
  {author} {\bibfnamefont {A.}~\bibnamefont {Benali}}, \bibinfo {author}
  {\bibfnamefont {M.~C.}\ \bibnamefont {Bennett}}, \bibinfo {author}
  {\bibfnamefont {M.~A.}\ \bibnamefont {Berrill}}, \bibinfo {author}
  {\bibfnamefont {N.~S.}\ \bibnamefont {Blunt}}, \bibinfo {author}
  {\bibfnamefont {E.~J.~L.}\ \bibnamefont {Borda}}, \bibinfo {author}
  {\bibfnamefont {M.}~\bibnamefont {Casula}}, \bibinfo {author} {\bibfnamefont
  {D.~M.}\ \bibnamefont {Ceperley}}, \bibinfo {author} {\bibfnamefont
  {S.}~\bibnamefont {Chiesa}}, \bibinfo {author} {\bibfnamefont {B.~K.}\
  \bibnamefont {Clark}}, \bibinfo {author} {\bibfnamefont {R.~C.}\ \bibnamefont
  {Clay}}, \bibinfo {author} {\bibfnamefont {K.~T.}\ \bibnamefont {Delaney}},
  \bibinfo {author} {\bibfnamefont {M.}~\bibnamefont {Dewing}}, \bibinfo
  {author} {\bibfnamefont {K.~P.}\ \bibnamefont {Esler}}, \bibinfo {author}
  {\bibfnamefont {H.}~\bibnamefont {Hao}}, \bibinfo {author} {\bibfnamefont
  {O.}~\bibnamefont {Heinonen}}, \bibinfo {author} {\bibfnamefont {P.~R.~C.}\
  \bibnamefont {Kent}}, \bibinfo {author} {\bibfnamefont {J.~T.}\ \bibnamefont
  {Krogel}}, \bibinfo {author} {\bibfnamefont {I.}~\bibnamefont {Kylänpää}},
  \bibinfo {author} {\bibfnamefont {Y.~W.}\ \bibnamefont {Li}}, \bibinfo
  {author} {\bibfnamefont {M.~G.}\ \bibnamefont {Lopez}}, \bibinfo {author}
  {\bibfnamefont {Y.}~\bibnamefont {Luo}}, \bibinfo {author} {\bibfnamefont
  {F.~D.}\ \bibnamefont {Malone}}, \bibinfo {author} {\bibfnamefont {R.~M.}\
  \bibnamefont {Martin}}, \bibinfo {author} {\bibfnamefont {A.}~\bibnamefont
  {Mathuriya}}, \bibinfo {author} {\bibfnamefont {J.}~\bibnamefont {McMinis}},
  \bibinfo {author} {\bibfnamefont {C.~A.}\ \bibnamefont {Melton}}, \bibinfo
  {author} {\bibfnamefont {L.}~\bibnamefont {Mitas}}, \bibinfo {author}
  {\bibfnamefont {M.~A.}\ \bibnamefont {Morales}}, \bibinfo {author}
  {\bibfnamefont {E.}~\bibnamefont {Neuscamman}}, \bibinfo {author}
  {\bibfnamefont {W.~D.}\ \bibnamefont {Parker}}, \bibinfo {author}
  {\bibfnamefont {S.~D.~P.}\ \bibnamefont {Flores}}, \bibinfo {author}
  {\bibfnamefont {N.~A.}\ \bibnamefont {Romero}}, \bibinfo {author}
  {\bibfnamefont {B.~M.}\ \bibnamefont {Rubenstein}}, \bibinfo {author}
  {\bibfnamefont {J.~A.~R.}\ \bibnamefont {Shea}}, \bibinfo {author}
  {\bibfnamefont {H.}~\bibnamefont {Shin}}, \bibinfo {author} {\bibfnamefont
  {L.}~\bibnamefont {Shulenburger}}, \bibinfo {author} {\bibfnamefont {A.~F.}\
  \bibnamefont {Tillack}}, \bibinfo {author} {\bibfnamefont {J.~P.}\
  \bibnamefont {Townsend}}, \bibinfo {author} {\bibfnamefont {N.~M.}\
  \bibnamefont {Tubman}}, \bibinfo {author} {\bibfnamefont {B.~V.~D.}\
  \bibnamefont {Goetz}}, \bibinfo {author} {\bibfnamefont {J.~E.}\ \bibnamefont
  {Vincent}}, \bibinfo {author} {\bibfnamefont {D.~C.}\ \bibnamefont {Yang}},
  \bibinfo {author} {\bibfnamefont {Y.}~\bibnamefont {Yang}}, \bibinfo {author}
  {\bibfnamefont {S.}~\bibnamefont {Zhang}}, \ and\ \bibinfo {author}
  {\bibfnamefont {L.}~\bibnamefont {Zhao}},\ }\bibfield  {title} {\enquote
  {\bibinfo {title} {{QMCPACK}: an open sourceab initioquantum monte carlo
  package for the electronic structure of atoms, molecules and solids},}\
  }\href {\doibase 10.1088/1361-648x/aab9c3} {\ \textbf {\bibinfo {volume}
  {30}},\ \bibinfo {pages} {195901} (\bibinfo {year} {2018})}\BibitemShut
  {NoStop}%
\bibitem [{\citenamefont {Kent}\ \emph {et~al.}(2020)\citenamefont {Kent},
  \citenamefont {Annaberdiyev}, \citenamefont {Benali}, \citenamefont
  {Bennett}, \citenamefont {Landinez~Borda}, \citenamefont {Doak},
  \citenamefont {Hao}, \citenamefont {Jordan}, \citenamefont {Krogel},
  \citenamefont {Kylänpää}, \citenamefont {Lee}, \citenamefont {Luo},
  \citenamefont {Malone}, \citenamefont {Melton}, \citenamefont {Mitas},
  \citenamefont {Morales}, \citenamefont {Neuscamman}, \citenamefont
  {Reboredo}, \citenamefont {Rubenstein}, \citenamefont {Saritas},
  \citenamefont {Upadhyay}, \citenamefont {Wang}, \citenamefont {Zhang},\ and\
  \citenamefont {Zhao}}]{doi:10.1063/5.0004860}%
  \BibitemOpen
  \bibfield  {author} {\bibinfo {author} {\bibfnamefont {P.~R.~C.}\
  \bibnamefont {Kent}}, \bibinfo {author} {\bibfnamefont {A.}~\bibnamefont
  {Annaberdiyev}}, \bibinfo {author} {\bibfnamefont {A.}~\bibnamefont
  {Benali}}, \bibinfo {author} {\bibfnamefont {M.~C.}\ \bibnamefont {Bennett}},
  \bibinfo {author} {\bibfnamefont {E.~J.}\ \bibnamefont {Landinez~Borda}},
  \bibinfo {author} {\bibfnamefont {P.}~\bibnamefont {Doak}}, \bibinfo {author}
  {\bibfnamefont {H.}~\bibnamefont {Hao}}, \bibinfo {author} {\bibfnamefont
  {K.~D.}\ \bibnamefont {Jordan}}, \bibinfo {author} {\bibfnamefont {J.~T.}\
  \bibnamefont {Krogel}}, \bibinfo {author} {\bibfnamefont {I.}~\bibnamefont
  {Kylänpää}}, \bibinfo {author} {\bibfnamefont {J.}~\bibnamefont {Lee}},
  \bibinfo {author} {\bibfnamefont {Y.}~\bibnamefont {Luo}}, \bibinfo {author}
  {\bibfnamefont {F.~D.}\ \bibnamefont {Malone}}, \bibinfo {author}
  {\bibfnamefont {C.~A.}\ \bibnamefont {Melton}}, \bibinfo {author}
  {\bibfnamefont {L.}~\bibnamefont {Mitas}}, \bibinfo {author} {\bibfnamefont
  {M.~A.}\ \bibnamefont {Morales}}, \bibinfo {author} {\bibfnamefont
  {E.}~\bibnamefont {Neuscamman}}, \bibinfo {author} {\bibfnamefont {F.~A.}\
  \bibnamefont {Reboredo}}, \bibinfo {author} {\bibfnamefont {B.}~\bibnamefont
  {Rubenstein}}, \bibinfo {author} {\bibfnamefont {K.}~\bibnamefont {Saritas}},
  \bibinfo {author} {\bibfnamefont {S.}~\bibnamefont {Upadhyay}}, \bibinfo
  {author} {\bibfnamefont {G.}~\bibnamefont {Wang}}, \bibinfo {author}
  {\bibfnamefont {S.}~\bibnamefont {Zhang}}, \ and\ \bibinfo {author}
  {\bibfnamefont {L.}~\bibnamefont {Zhao}},\ }\bibfield  {title} {\enquote
  {\bibinfo {title} {{QMCPACK: Advances in the development, efficiency, and
  application of auxiliary field and real-space Variational and Diffusion
  Quantum Monte Carlo}},}\ }\href {\doibase 10.1063/5.0004860} {\bibfield
  {journal} {\bibinfo  {journal} {The Journal of Chemical Physics}\ }\textbf
  {\bibinfo {volume} {152}},\ \bibinfo {pages} {174105} (\bibinfo {year}
  {2020})}\BibitemShut {NoStop}%
\bibitem [{\citenamefont {Krogel}(2016)}]{nexus}%
  \BibitemOpen
  \bibfield  {author} {\bibinfo {author} {\bibfnamefont {J.~T.}\ \bibnamefont
  {Krogel}},\ }\bibfield  {title} {\enquote {\bibinfo {title} {Nexus: A modular
  workflow management system for quantum simulation codes},}\ }\href {\doibase
  https://doi.org/10.1016/j.cpc.2015.08.012} {\bibfield  {journal} {\bibinfo
  {journal} {Computer Physics Communications}\ }\textbf {\bibinfo {volume}
  {198}},\ \bibinfo {pages} {154 -- 168} (\bibinfo {year} {2016})}\BibitemShut
  {NoStop}%
\bibitem [{\citenamefont {Giannozzi}\ \emph {et~al.}(2009)\citenamefont
  {Giannozzi}, \citenamefont {Baroni}, \citenamefont {Bonini}, \citenamefont
  {Calandra}, \citenamefont {Car}, \citenamefont {Cavazzoni}, \citenamefont
  {Ceresoli}, \citenamefont {Chiarotti}, \citenamefont {Cococcioni},
  \citenamefont {Dabo}, \citenamefont {Corso}, \citenamefont {de~Gironcoli},
  \citenamefont {Fabris}, \citenamefont {Fratesi}, \citenamefont {Gebauer},
  \citenamefont {Gerstmann}, \citenamefont {Gougoussis}, \citenamefont
  {Kokalj}, \citenamefont {Lazzeri}, \citenamefont {Martin-Samos},
  \citenamefont {Marzari}, \citenamefont {Mauri}, \citenamefont {Mazzarello},
  \citenamefont {Paolini}, \citenamefont {Pasquarello}, \citenamefont
  {Paulatto}, \citenamefont {Sbraccia}, \citenamefont {Scandolo}, \citenamefont
  {Sclauzero}, \citenamefont {Seitsonen}, \citenamefont {Smogunov},
  \citenamefont {Umari},\ and\ \citenamefont {Wentzcovitch}}]{Giannozzi_2009}%
  \BibitemOpen
  \bibfield  {author} {\bibinfo {author} {\bibfnamefont {P.}~\bibnamefont
  {Giannozzi}}, \bibinfo {author} {\bibfnamefont {S.}~\bibnamefont {Baroni}},
  \bibinfo {author} {\bibfnamefont {N.}~\bibnamefont {Bonini}}, \bibinfo
  {author} {\bibfnamefont {M.}~\bibnamefont {Calandra}}, \bibinfo {author}
  {\bibfnamefont {R.}~\bibnamefont {Car}}, \bibinfo {author} {\bibfnamefont
  {C.}~\bibnamefont {Cavazzoni}}, \bibinfo {author} {\bibfnamefont
  {D.}~\bibnamefont {Ceresoli}}, \bibinfo {author} {\bibfnamefont {G.~L.}\
  \bibnamefont {Chiarotti}}, \bibinfo {author} {\bibfnamefont {M.}~\bibnamefont
  {Cococcioni}}, \bibinfo {author} {\bibfnamefont {I.}~\bibnamefont {Dabo}},
  \bibinfo {author} {\bibfnamefont {A.~D.}\ \bibnamefont {Corso}}, \bibinfo
  {author} {\bibfnamefont {S.}~\bibnamefont {de~Gironcoli}}, \bibinfo {author}
  {\bibfnamefont {S.}~\bibnamefont {Fabris}}, \bibinfo {author} {\bibfnamefont
  {G.}~\bibnamefont {Fratesi}}, \bibinfo {author} {\bibfnamefont
  {R.}~\bibnamefont {Gebauer}}, \bibinfo {author} {\bibfnamefont
  {U.}~\bibnamefont {Gerstmann}}, \bibinfo {author} {\bibfnamefont
  {C.}~\bibnamefont {Gougoussis}}, \bibinfo {author} {\bibfnamefont
  {A.}~\bibnamefont {Kokalj}}, \bibinfo {author} {\bibfnamefont
  {M.}~\bibnamefont {Lazzeri}}, \bibinfo {author} {\bibfnamefont
  {L.}~\bibnamefont {Martin-Samos}}, \bibinfo {author} {\bibfnamefont
  {N.}~\bibnamefont {Marzari}}, \bibinfo {author} {\bibfnamefont
  {F.}~\bibnamefont {Mauri}}, \bibinfo {author} {\bibfnamefont
  {R.}~\bibnamefont {Mazzarello}}, \bibinfo {author} {\bibfnamefont
  {S.}~\bibnamefont {Paolini}}, \bibinfo {author} {\bibfnamefont
  {A.}~\bibnamefont {Pasquarello}}, \bibinfo {author} {\bibfnamefont
  {L.}~\bibnamefont {Paulatto}}, \bibinfo {author} {\bibfnamefont
  {C.}~\bibnamefont {Sbraccia}}, \bibinfo {author} {\bibfnamefont
  {S.}~\bibnamefont {Scandolo}}, \bibinfo {author} {\bibfnamefont
  {G.}~\bibnamefont {Sclauzero}}, \bibinfo {author} {\bibfnamefont {A.~P.}\
  \bibnamefont {Seitsonen}}, \bibinfo {author} {\bibfnamefont {A.}~\bibnamefont
  {Smogunov}}, \bibinfo {author} {\bibfnamefont {P.}~\bibnamefont {Umari}}, \
  and\ \bibinfo {author} {\bibfnamefont {R.~M.}\ \bibnamefont {Wentzcovitch}},\
  }\bibfield  {title} {\enquote {\bibinfo {title} {{{QUANTUM} {ESPRESSO}: a
  modular and open-source software project for quantum simulations of
  materials}},}\ }\href {\doibase 10.1088/0953-8984/21/39/395502} {\bibfield
  {journal} {\bibinfo  {journal} {Journal of Physics: Condensed Matter}\
  }\textbf {\bibinfo {volume} {21}},\ \bibinfo {pages} {395502} (\bibinfo
  {year} {2009})}\BibitemShut {NoStop}%
\bibitem [{\citenamefont {Slater}(1929)}]{PhysRev.34.1293}%
  \BibitemOpen
  \bibfield  {author} {\bibinfo {author} {\bibfnamefont {J.~C.}\ \bibnamefont
  {Slater}},\ }\bibfield  {title} {\enquote {\bibinfo {title} {The theory of
  complex spectra},}\ }\href {\doibase 10.1103/PhysRev.34.1293} {\bibfield
  {journal} {\bibinfo  {journal} {Phys. Rev.}\ }\textbf {\bibinfo {volume}
  {34}},\ \bibinfo {pages} {1293--1322} (\bibinfo {year} {1929})}\BibitemShut
  {NoStop}%
\bibitem [{\citenamefont {Jastrow}(1955)}]{PhysRev.98.1479}%
  \BibitemOpen
  \bibfield  {author} {\bibinfo {author} {\bibfnamefont {R.}~\bibnamefont
  {Jastrow}},\ }\bibfield  {title} {\enquote {\bibinfo {title} {Many-body
  problem with strong forces},}\ }\href {\doibase 10.1103/PhysRev.98.1479}
  {\bibfield  {journal} {\bibinfo  {journal} {Phys. Rev.}\ }\textbf {\bibinfo
  {volume} {98}},\ \bibinfo {pages} {1479--1484} (\bibinfo {year}
  {1955})}\BibitemShut {NoStop}%
\bibitem [{\citenamefont {Drummond}, \citenamefont {Towler},\ and\
  \citenamefont {Needs}(2004)}]{PhysRevB.70.235119}%
  \BibitemOpen
  \bibfield  {author} {\bibinfo {author} {\bibfnamefont {N.~D.}\ \bibnamefont
  {Drummond}}, \bibinfo {author} {\bibfnamefont {M.~D.}\ \bibnamefont
  {Towler}}, \ and\ \bibinfo {author} {\bibfnamefont {R.~J.}\ \bibnamefont
  {Needs}},\ }\bibfield  {title} {\enquote {\bibinfo {title} {{Jastrow
  correlation factor for atoms, molecules, and solids}},}\ }\href {\doibase
  10.1103/PhysRevB.70.235119} {\bibfield  {journal} {\bibinfo  {journal} {Phys.
  Rev. B}\ }\textbf {\bibinfo {volume} {70}},\ \bibinfo {pages} {235119}
  (\bibinfo {year} {2004})}\BibitemShut {NoStop}%
\bibitem [{\citenamefont {Umrigar}\ \emph {et~al.}(2007)\citenamefont
  {Umrigar}, \citenamefont {Toulouse}, \citenamefont {Filippi}, \citenamefont
  {Sorella},\ and\ \citenamefont {Hennig}}]{PhysRevLett.98.110201}%
  \BibitemOpen
  \bibfield  {author} {\bibinfo {author} {\bibfnamefont {C.~J.}\ \bibnamefont
  {Umrigar}}, \bibinfo {author} {\bibfnamefont {J.}~\bibnamefont {Toulouse}},
  \bibinfo {author} {\bibfnamefont {C.}~\bibnamefont {Filippi}}, \bibinfo
  {author} {\bibfnamefont {S.}~\bibnamefont {Sorella}}, \ and\ \bibinfo
  {author} {\bibfnamefont {R.~G.}\ \bibnamefont {Hennig}},\ }\bibfield  {title}
  {\enquote {\bibinfo {title} {Alleviation of the fermion-sign problem by
  optimization of many-body wave functions},}\ }\href {\doibase
  10.1103/PhysRevLett.98.110201} {\bibfield  {journal} {\bibinfo  {journal}
  {Phys. Rev. Lett.}\ }\textbf {\bibinfo {volume} {98}},\ \bibinfo {pages}
  {110201} (\bibinfo {year} {2007})}\BibitemShut {NoStop}%
\bibitem [{\citenamefont {Umrigar}\ and\ \citenamefont
  {Filippi}(2005)}]{PhysRevLett.94.150201}%
  \BibitemOpen
  \bibfield  {author} {\bibinfo {author} {\bibfnamefont {C.~J.}\ \bibnamefont
  {Umrigar}}\ and\ \bibinfo {author} {\bibfnamefont {C.}~\bibnamefont
  {Filippi}},\ }\bibfield  {title} {\enquote {\bibinfo {title} {Energy and
  variance optimization of many-body wave functions},}\ }\href {\doibase
  10.1103/PhysRevLett.94.150201} {\bibfield  {journal} {\bibinfo  {journal}
  {Phys. Rev. Lett.}\ }\textbf {\bibinfo {volume} {94}},\ \bibinfo {pages}
  {150201} (\bibinfo {year} {2005})}\BibitemShut {NoStop}%
\bibitem [{\citenamefont {Mitas}, \citenamefont {Shirley},\ and\ \citenamefont
  {Ceperley}(1991)}]{doi:10.1063/1.460849}%
  \BibitemOpen
  \bibfield  {author} {\bibinfo {author} {\bibfnamefont {L.}~\bibnamefont
  {Mitas}}, \bibinfo {author} {\bibfnamefont {E.~L.}\ \bibnamefont {Shirley}},
  \ and\ \bibinfo {author} {\bibfnamefont {D.~M.}\ \bibnamefont {Ceperley}},\
  }\bibfield  {title} {\enquote {\bibinfo {title} {{Nonlocal pseudopotentials
  and Diffusion Monte Carlo}},}\ }\href {\doibase 10.1063/1.460849} {\bibfield
  {journal} {\bibinfo  {journal} {The Journal of Chemical Physics}\ }\textbf
  {\bibinfo {volume} {95}},\ \bibinfo {pages} {3467--3475} (\bibinfo {year}
  {1991})}\BibitemShut {NoStop}%
\bibitem [{\citenamefont {Burkatzki}, \citenamefont {Filippi},\ and\
  \citenamefont {Dolg}(2007)}]{doi:10.1063/1.2741534}%
  \BibitemOpen
  \bibfield  {author} {\bibinfo {author} {\bibfnamefont {M.}~\bibnamefont
  {Burkatzki}}, \bibinfo {author} {\bibfnamefont {C.}~\bibnamefont {Filippi}},
  \ and\ \bibinfo {author} {\bibfnamefont {M.}~\bibnamefont {Dolg}},\
  }\bibfield  {title} {\enquote {\bibinfo {title} {{Energy-consistent
  pseudopotentials for Quantum Monte Carlo calculations}},}\ }\href {\doibase
  10.1063/1.2741534} {\bibfield  {journal} {\bibinfo  {journal} {The Journal of
  Chemical Physics}\ }\textbf {\bibinfo {volume} {126}},\ \bibinfo {pages}
  {234105} (\bibinfo {year} {2007})}\BibitemShut {NoStop}%
\bibitem [{\citenamefont {Burkatzki}, \citenamefont {Filippi},\ and\
  \citenamefont {Dolg}(2008)}]{doi:10.1063/1.2987872}%
  \BibitemOpen
  \bibfield  {author} {\bibinfo {author} {\bibfnamefont {M.}~\bibnamefont
  {Burkatzki}}, \bibinfo {author} {\bibfnamefont {C.}~\bibnamefont {Filippi}},
  \ and\ \bibinfo {author} {\bibfnamefont {M.}~\bibnamefont {Dolg}},\
  }\bibfield  {title} {\enquote {\bibinfo {title} {{Energy-consistent
  small-core pseudopotentials for 3d-transition metals adapted to Quantum Monte
  Carlo calculations}},}\ }\href {\doibase 10.1063/1.2987872} {\bibfield
  {journal} {\bibinfo  {journal} {The Journal of Chemical Physics}\ }\textbf
  {\bibinfo {volume} {129}},\ \bibinfo {pages} {164115} (\bibinfo {year}
  {2008})}\BibitemShut {NoStop}%
\bibitem [{\citenamefont {Wang}\ \emph
  {et~al.}(2019{\natexlab{b}})\citenamefont {Wang}, \citenamefont
  {Annaberdiyev}, \citenamefont {Melton}, \citenamefont {Bennett},
  \citenamefont {Shulenburger},\ and\ \citenamefont
  {Mitas}}]{doi:10.1063/1.5121006}%
  \BibitemOpen
  \bibfield  {author} {\bibinfo {author} {\bibfnamefont {G.}~\bibnamefont
  {Wang}}, \bibinfo {author} {\bibfnamefont {A.}~\bibnamefont {Annaberdiyev}},
  \bibinfo {author} {\bibfnamefont {C.~A.}\ \bibnamefont {Melton}}, \bibinfo
  {author} {\bibfnamefont {M.~C.}\ \bibnamefont {Bennett}}, \bibinfo {author}
  {\bibfnamefont {L.}~\bibnamefont {Shulenburger}}, \ and\ \bibinfo {author}
  {\bibfnamefont {L.}~\bibnamefont {Mitas}},\ }\bibfield  {title} {\enquote
  {\bibinfo {title} {{A new generation of effective core potentials from
  correlated calculations: 4s and 4p main group elements and first row
  additions}},}\ }\href {\doibase 10.1063/1.5121006} {\bibfield  {journal}
  {\bibinfo  {journal} {The Journal of Chemical Physics}\ }\textbf {\bibinfo
  {volume} {151}},\ \bibinfo {pages} {144110} (\bibinfo {year}
  {2019}{\natexlab{b}})}\BibitemShut {NoStop}%
\bibitem [{\citenamefont {Togo}\ and\ \citenamefont
  {Tanaka}(2018)}]{togo2018textttspglib}%
  \BibitemOpen
  \bibfield  {author} {\bibinfo {author} {\bibfnamefont {A.}~\bibnamefont
  {Togo}}\ and\ \bibinfo {author} {\bibfnamefont {I.}~\bibnamefont {Tanaka}},\
  }\href@noop {} {\enquote {\bibinfo {title} {{$\texttt{Spglib}$: a software
  library for crystal symmetry search}},}\ } (\bibinfo {year} {2018}),\ \Eprint
  {http://arxiv.org/abs/1808.01590} {arXiv:1808.01590 [cond-mat.mtrl-sci]}
  \BibitemShut {NoStop}%
\bibitem [{\citenamefont {Hinuma}\ \emph {et~al.}(2017)\citenamefont {Hinuma},
  \citenamefont {Pizzi}, \citenamefont {Kumagai}, \citenamefont {Oba},\ and\
  \citenamefont {Tanaka}}]{HINUMA2017140}%
  \BibitemOpen
  \bibfield  {author} {\bibinfo {author} {\bibfnamefont {Y.}~\bibnamefont
  {Hinuma}}, \bibinfo {author} {\bibfnamefont {G.}~\bibnamefont {Pizzi}},
  \bibinfo {author} {\bibfnamefont {Y.}~\bibnamefont {Kumagai}}, \bibinfo
  {author} {\bibfnamefont {F.}~\bibnamefont {Oba}}, \ and\ \bibinfo {author}
  {\bibfnamefont {I.}~\bibnamefont {Tanaka}},\ }\bibfield  {title} {\enquote
  {\bibinfo {title} {Band structure diagram paths based on crystallography},}\
  }\href {\doibase https://doi.org/10.1016/j.commatsci.2016.10.015} {\bibfield
  {journal} {\bibinfo  {journal} {Computational Materials Science}\ }\textbf
  {\bibinfo {volume} {128}},\ \bibinfo {pages} {140 -- 184} (\bibinfo {year}
  {2017})}\BibitemShut {NoStop}%
\bibitem [{\citenamefont {Archibald}, \citenamefont {Krogel},\ and\
  \citenamefont {Kent}(2018)}]{doi:10.1063/1.5040584}%
  \BibitemOpen
  \bibfield  {author} {\bibinfo {author} {\bibfnamefont {R.}~\bibnamefont
  {Archibald}}, \bibinfo {author} {\bibfnamefont {J.~T.}\ \bibnamefont
  {Krogel}}, \ and\ \bibinfo {author} {\bibfnamefont {P.~R.~C.}\ \bibnamefont
  {Kent}},\ }\bibfield  {title} {\enquote {\bibinfo {title} {{Gaussian process
  based optimization of molecular geometries using statistically sampled energy
  surfaces from Quantum Monte Carlo}},}\ }\href {\doibase 10.1063/1.5040584}
  {\bibfield  {journal} {\bibinfo  {journal} {The Journal of Chemical Physics}\
  }\textbf {\bibinfo {volume} {149}},\ \bibinfo {pages} {164116} (\bibinfo
  {year} {2018})}\BibitemShut {NoStop}%
\bibitem [{\citenamefont {Sorella}\ and\ \citenamefont
  {Capriotti}(2010)}]{doi:10.1063/1.3516208}%
  \BibitemOpen
  \bibfield  {author} {\bibinfo {author} {\bibfnamefont {S.}~\bibnamefont
  {Sorella}}\ and\ \bibinfo {author} {\bibfnamefont {L.}~\bibnamefont
  {Capriotti}},\ }\bibfield  {title} {\enquote {\bibinfo {title} {{Algorithmic
  differentiation and the calculation of forces by Quantum Monte Carlo}},}\
  }\href {\doibase 10.1063/1.3516208} {\bibfield  {journal} {\bibinfo
  {journal} {The Journal of Chemical Physics}\ }\textbf {\bibinfo {volume}
  {133}},\ \bibinfo {pages} {234111} (\bibinfo {year} {2010})}\BibitemShut
  {NoStop}%
\bibitem [{\citenamefont {Buda}\ \emph {et~al.}(2017)\citenamefont {Buda},
  \citenamefont {Lane}, \citenamefont {Barbiellini}, \citenamefont
  {Ruzsinszky}, \citenamefont {Sun},\ and\ \citenamefont {Bansil}}]{scan-2d}%
  \BibitemOpen
  \bibfield  {author} {\bibinfo {author} {\bibfnamefont {I.~G.}\ \bibnamefont
  {Buda}}, \bibinfo {author} {\bibfnamefont {C.}~\bibnamefont {Lane}}, \bibinfo
  {author} {\bibfnamefont {B.}~\bibnamefont {Barbiellini}}, \bibinfo {author}
  {\bibfnamefont {A.}~\bibnamefont {Ruzsinszky}}, \bibinfo {author}
  {\bibfnamefont {J.}~\bibnamefont {Sun}}, \ and\ \bibinfo {author}
  {\bibfnamefont {A.}~\bibnamefont {Bansil}},\ }\bibfield  {title} {\enquote
  {\bibinfo {title} {Characterization of thin film materials using {SCAN}
  meta-{GGA}, an accurate nonempirical density functional},}\ }\href {\doibase
  10.1038/srep44766} {\bibfield  {journal} {\bibinfo  {journal} {Scientific
  Reports}\ }\textbf {\bibinfo {volume} {7}},\ \bibinfo {pages} {44766}
  (\bibinfo {year} {2017})}\BibitemShut {NoStop}%
\bibitem [{\citenamefont {Annaberdiyev}\ \emph {et~al.}(2020)\citenamefont
  {Annaberdiyev}, \citenamefont {Melton}, \citenamefont {Bennett},
  \citenamefont {Wang},\ and\ \citenamefont {Mitas}}]{atomcalcs}%
  \BibitemOpen
  \bibfield  {author} {\bibinfo {author} {\bibfnamefont {A.}~\bibnamefont
  {Annaberdiyev}}, \bibinfo {author} {\bibfnamefont {C.~A.}\ \bibnamefont
  {Melton}}, \bibinfo {author} {\bibfnamefont {M.~C.}\ \bibnamefont {Bennett}},
  \bibinfo {author} {\bibfnamefont {G.}~\bibnamefont {Wang}}, \ and\ \bibinfo
  {author} {\bibfnamefont {L.}~\bibnamefont {Mitas}},\ }\bibfield  {title}
  {\enquote {\bibinfo {title} {Accurate atomic correlation and total energies
  for correlation consistent effective core potentials},}\ }\bibfield
  {booktitle} {\emph {\bibinfo {booktitle} {Journal of Chemical Theory and
  Computation}},\ }\href {\doibase 10.1021/acs.jctc.9b00962} {\bibfield
  {journal} {\bibinfo  {journal} {Journal of Chemical Theory and Computation}\
  }\textbf {\bibinfo {volume} {16}},\ \bibinfo {pages} {1482--1502} (\bibinfo
  {year} {2020})}\BibitemShut {NoStop}%
\bibitem [{\citenamefont {Grimme}(2006)}]{doi:10.1002/jcc.20495}%
  \BibitemOpen
  \bibfield  {author} {\bibinfo {author} {\bibfnamefont {S.}~\bibnamefont
  {Grimme}},\ }\bibfield  {title} {\enquote {\bibinfo {title} {Semiempirical
  {GGA}-type density functional constructed with a long-range dispersion
  correction},}\ }\href {\doibase 10.1002/jcc.20495} {\bibfield  {journal}
  {\bibinfo  {journal} {Journal of Computational Chemistry}\ }\textbf {\bibinfo
  {volume} {27}},\ \bibinfo {pages} {1787--1799} (\bibinfo {year}
  {2006})}\BibitemShut {NoStop}%
\bibitem [{\citenamefont {Peng}\ \emph {et~al.}(2016)\citenamefont {Peng},
  \citenamefont {Yang}, \citenamefont {Perdew},\ and\ \citenamefont
  {Sun}}]{PhysRevX.6.041005}%
  \BibitemOpen
  \bibfield  {author} {\bibinfo {author} {\bibfnamefont {H.}~\bibnamefont
  {Peng}}, \bibinfo {author} {\bibfnamefont {Z.-H.}\ \bibnamefont {Yang}},
  \bibinfo {author} {\bibfnamefont {J.~P.}\ \bibnamefont {Perdew}}, \ and\
  \bibinfo {author} {\bibfnamefont {J.}~\bibnamefont {Sun}},\ }\bibfield
  {title} {\enquote {\bibinfo {title} {Versatile van der {Waals} density
  functional based on a meta-generalized gradient approximation},}\ }\href
  {\doibase 10.1103/PhysRevX.6.041005} {\bibfield  {journal} {\bibinfo
  {journal} {Phys. Rev. X}\ }\textbf {\bibinfo {volume} {6}},\ \bibinfo {pages}
  {041005} (\bibinfo {year} {2016})}\BibitemShut {NoStop}%
\end{thebibliography}%

\end{document}